\documentstyle[12pt]{article}
\def\NPB{{\em Nucl. Phys.} B}
\def\PLB{{\em Phys. Lett.}  B}
\def\PRL{{\em Phys. Rev. Lett.}}
\def\PRD{{\em Phys. Rev.} D}
\def\ZPC{{\em Z. Phys.} C}
\def\PRC{{\em Phys. Rep.} C}

\def\ra{\rightarrow}

\def\be{\begin{equation}}
\def\ee{\end{equation}}
\def\bea{\begin{eqnarray}}
\def\eea{\end{eqnarray}}
\def\lsim{\:\raisebox{-0.5ex}{$\stackrel{\textstyle<}{\sim}$}\:}
\def\gsim{\:\raisebox{-0.5ex}{$\stackrel{\textstyle>}{\sim}$}\:}
\def\gev{GeV }
\def\eg{{\it{e.g. }}}
\def\ie{{\it{i.e. }}}
\def\g2{GeV$^2$}
\def\psr{{pseudorapidity }}
\def\bi{\bibitem}
\begin{document}
\titlepage
\begin{flushright}
DESY 98-013\\[1.5ex]
IFT 97/14\\[1.5ex]
{\large \bf hep-ph/9806291} \\
February 1998\\
\end{flushright}
\vskip 0.5cm
\centerline{\bf {\huge {{Survey of recent data on  photon structure
 functions}}}}
\vskip 0.5cm 
\centerline{\bf {\huge {and resolved photon processes}}}
\vskip 2.5cm
\centerline{\Large Maria Krawczyk$^a$ and Andrzej Zembrzuski}                 
\centerline{Institute of Theoretical Physics, Warsaw University, Poland}

\centerline{$^a$ and Deutsches Electronen-Synchrotron DESY, Hamburg, Germany} 
\vskip 0.5cm
\centerline{and}
\vskip 0.5cm
\centerline{\Large Magdalena Staszel}
\centerline{Institute of Experimental Physics, Warsaw University, Poland}

%%%%%%%%%%%%%%%%%%%%%%%%%%%%%%%%%%%%%%%%%%%%%%%%%%%%%%%%%%%%%%


%%%%%%%%%%%%%%%%%%%%%%%%%%%%%%%%%%%%%%%%%%%%%%%%%%%%%%%%%%%%%%
%\maketitle

\vskip 3cm
\begin{abstract}
Present data on the partonic content  of the photon
are reviewed. The results on the unpolarized structure functions
from DIS experiments and on large $p_T$  
jet production processes in $\gamma \gamma$ and 
$\gamma p$ collisions are discussed for both  real and  virtual 
photons. A few
related topics like the QED structure functions of the photon
and the structure function of the electron are also shortly discussed.
\end{abstract}
\newpage
\titlepage
\tableofcontents
%%%%%%%%%%%%%%%%%%%%%%%%%%%%%%%%%%%%%%%%%%%%%%%%%%
%
%
%%%%%%%%%%%%%%%%%%%%%%%%%%%%%%%%%%%%%%%%%%%%%%%%%%
\newpage
\section{Introduction}
The concept of the hadronic (partonic) structure of photon is used  in 
describing high energy photon - hadron interactions 
(for a general discussion see \eg review articles \cite{rev}).
There are two basic types of 
inclusive processes where the structure of photon is tested in
existing experiments:
\begin{itemize}
\item the deep inelastic scattering (DIS$_{\gamma}$), $e{\gamma}
\rightarrow e$ $hadrons $, where the structure functions of photon $F^{\gamma}
_{1,2,...}$ are measured,
\item
the large $p_T$ jet production in ${\gamma}p$
and ${\gamma}{\gamma}$ collisions
(one of the so called resolved photon processes),
where individual quark and gluon densities in the photon
may be probed.
\end{itemize}

The early experiments of the first type were performed
at PETRA and  PEP $e^+e^-$ colliders \cite{early}
\footnote{The first measurement was done in 1981 by the PLUTO
Collaboration
\cite{early}.}. 
Final analyses of LEP 1 
data on $F^{\gamma}_2$ taken at CM energy $\sim M_Z$ appeared recently,
new data are being collected and analyzed at $e^+e^-$ LEP collider running
at higher energies (CM energy: 130-136 and 161-172 GeV), 
and  at  KEK collider (energy $\sim$ 60 GeV). 
\footnote{In principle the measurement could be
performed also at the SLC collider (at energy $\sim$~91~GeV).} 
All together they cover a 
wide range of $Q^2$ from 0.24 to 390 GeV$^2$;
the range of the $x_{Bj}$ variable extends 
from $\sim$ 0.002 to 0.98.

The latter type of measurements, \ie measurement of the 
jets 
in  the resolved photon processes, started  a decade later
{\footnote {First evidence for the resolved process was found in 1990 
by the AMY collaboration at KEK \cite{amy2}.}}.
Data have been and are still being taken   
in  photoproduction processes at the $ep$ collider
HERA (with the CM energy 300 GeV) 
and in ${\gamma}{\gamma}$ collisions at the above mentioned $e^+e^-$ machines. 
They  have just started to give first results on the light 
quark (the effective parton density) and gluonic content of the photon.

In this paper  the status of recent measurements of
the structure function of unpolarized photons for both real and 
virtual photons
in ``DIS$_{\gamma}$'' experiments and   
in large $p_T$ jet processes involving resolved photons is discussed
\footnote{So far there is no data for the polarized
photon; for recent discussion of   the 
polarized structure functions of photons see \eg \cite{mr,sv}.}.
In the presentation we focus on recent results  
(data collected in years 1990 to 1996)\footnote{The 
exception is made for the structure of the virtual photon
and the electromagnetic structure function for the photon,
where old experimental data are also reviewed.},
where qualitative change has appeared in both types of measurements.
Final results based on few years' 
runs at LEP 1 are being published 
and a summing - up can be done. 
On the other hand, impressive progress has been obtained in 
pinning down the individual parton contributions in 
resolved photon processes with large $p_{T}$ jets at HERA.
These two methods of studying the ``structure'' of photon
are becoming recently even more closely related, as
 it turned out  that in the reconstruction of $F_2^{\gamma}$
by the unfolding procedure the detailed analysis of  the final 
hadronic state, with the contribution
due to   resolved photon processes,  has to be performed.

The discrepancies observed by many collaborations
in the description of the hadronic final 
states in the DIS$_{\gamma}$ experiments have
enforced the advanced study of various aspects of hadron production in
this experimental setup. 
 We found it important to collect these results together
and in close relation to the large $p_T$ jet production
results, where   problems with  the proper description
 of the underlying events appear as well.
We hope that this way of presentation 
may help to clarify the situation with the production of hadrons
in photon induced processes.

A very comprehensive paper
"A compilation of  data on two-photon reactions 
leading to hadron final states" by D. Morgan, M. R. Pennington 
and M. R. Whalley  
\cite{wal}  contains data published up to March '94. 
Beside the inclusive hadron production  it contains also the
data on the exclusive processes and on the total cross sections, 
which are beyond the scope of the present paper.

Our survey is based on the published data, with few very recent  
preliminary results presented at the 1997 conferences.
The short descriptions of results
obtained by the experimental group
 together with  some representative
figures as well as  comments/conclusions quoted 
from original publications are given.
We describe the data by the name of the collaboration in alphabetic order,
the publication date and the reference listed at the end of the paper.
Each of the  different topics is introduced by a short review of the
basic theoretical ideas and notation. 

In Sec.2, data for the real photon are discussed. 
Sec.2.1 deals with  DIS$_{\gamma}$  
results (structure function $F_2^{\gamma}$),  
Sec.2.2 with  data on  hadronic final states in DIS$_{\gamma}$ experiments
 for the real photon and Sec.2.3 with large $p_T$ jet production 
in $\gamma \gamma$ and $\gamma p$ collisions. 
 In Sec.3 existing data for the virtual photon are presented.
In Sec.4 related topics are shortly discussed, namely 
 leptonic structure functions of the photon
(Sec.4.1) and the structure function of the electron (Sec.4.2). 
Sec.5 contains the summary and outlook. 
In the Appendix, the existing parton parametrizations are listed.
%%%%%%%%%%%%%%%%%%%%%%%%%%%%%%%%%%%%%%%%%%%%%%%%%%
%
%
%%%%%%%%%%%%%%%%%%%%%%%%%%%%%%%%%%%%%%%%%%%%%%%%%%
\newpage
\section{Partonic content of the real photon}
\subsection{DIS$_{\gamma}$ experiments}
\begin{figure}[ht]
\vskip -1.5cm\relax\noindent\hskip 2.cm
       \relax{\includegraphics{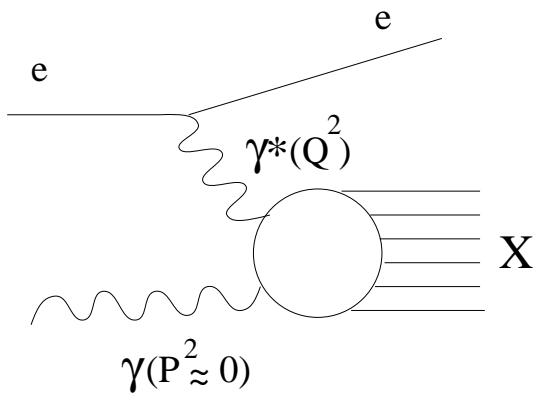}}
\vspace{6.2cm }
\caption{{\small\sl Deep inelastic scattering
$e{\gamma}~(P^2\approx0)\ra eX$. The photon 
probe has virtuality $q^2=-Q^2$.}}
\label{fig:jerzy1}
\end{figure}

In this section we consider the standard DIS$_{\gamma}$ measurements 
in unpolarized $e{\gamma}$ collisions based on the process (Fig.~\ref{fig:jerzy1}):
\be
{e\gamma} \ra e {\ }hadrons, \label{eg}
\ee
at $e^{+}e^-$ colliders (so called single - tagged events) \cite{rev}.
Here the target is a real (in practice  almost real) photon, coming 
from the initial electron (positron). 

The flux of these real photons in present experimental setups can be 
approximated by  the Weizs\"acker - Williams formula (called also 
Equivalent Photon Approximation EPA) \cite{ww,rev}.
\footnote{Recent critical discussion of this approach for the 
${\gamma}{\gamma}$ processes in $e^+e^-$ collisions at LEP 2
can be found in \cite{schu}.}
It corresponds to the typical soft bremsstrahlung spectrum.
\footnote{Note that at $e\gamma$ or $\gamma \gamma$ 
options planned at NL Colliders, where the real photons
would be obtained in the Compton backscattering process, 
the corresponding flux is expected to be much harder \cite{tel}.} 
There is a possibility to introduce here the structure function of 
the electron, which would contain the convolution of the 
Weizs\"acker - Williams flux for virtual photons  and the structure 
function of ${\gamma}^*$ (see discussion in Sec.3 and 4).

The cross section for process (\ref{eg}), where the real photon
with four momentum $p$ is probed by the virtual photon with four 
momentum $q=k-k'$ ($q^2=-Q^2<0$), is given by:
\begin{eqnarray}
\label{si}
{{d\sigma^{e { \gamma} \ra e X}}\over{dx_{Bj}dy}}=
{{4\pi \alpha^2 (2p\cdot k)}\over
{Q^4}}{[(1-y)F_2^{\gamma}+x_{B_j}y^2F_1^{\gamma}]} \\
={{2\pi \alpha^2 (2p\cdot k)}\over
{Q^4}}{[(1+(1-y)^2)F_2^{\gamma}-y^2F_L^{\gamma}]}, \label{si2} 
\end{eqnarray}
with the relation among the transverse $F_T^{\gamma}$ 
(= $F_1^{\gamma}$) and longitudinal $F_L^{\gamma}$  structure 
functions: $F_2^{\gamma}=F_L^{\gamma}+2x_{Bj}F_1^{\gamma}$.
$k(k')$ are four momenta of initial (final) electron
and the standard DIS variables are defined by  
\be
x_{Bj}={Q^2 \over {2p \cdot q}}     
,~~~~~~y={{p\cdot q}\over {p\cdot k}}. \label{xy} 
\ee
Note that in the LAB frame $y$ describes the scaled energy of the 
exchanged photon. In practice, the variable $y$ is small and 
the cross section (2) is effectively saturated by $F_2^{\gamma}$. 
Note that  in $e^+e^-$  collisions there is always a small 
off-shellness of the photon-target, $p^2=-P^2\neq0$, and there may 
appear in addition the third structure function, $F_3^{\gamma}$.
It  disappears however from the cross section in the single-tag 
measurements after integration over the azimuthal angle (since it 
enters the cross section as the term $F_3^{\gamma} \cos (2\phi)$).
See Sec. 4.1 for the discussion on the additional structure functions.

\subsubsection{Theoretical description}
In the Quark Parton Model (QPM or PM) one assumes that the hadronic 
final state in Eq.~(\ref{eg}) is due to the production of quark pairs:
$q_i$ and $\bar q_i$ ($i$=1,2...$N_f$ - number of flavours) with the 
fractional charge $Q_i$. The  $F_2^{\gamma}$ is obtained by the 
integration over the transverse momentum of the outgoing quark with 
respect to the target photon direction. The full (PM or the lowest 
order) expression for $F_2^{\gamma}$, keeping the terms with quark 
mass $m_{q_i}$, is given by the Bethe-Heitler formula \cite{rev}: 
\begin{eqnarray}
\nonumber{F}_2^{\gamma}={{{\alpha}}\over{{\pi}}}N_c\sum_{i=1}
^{N_f} Q_i^4 x_{Bj}
[ (-1+8x_{Bj}(1-x_{Bj})-x_{Bj}(1-x_{Bj}){{4m_{q_i}^2}\over{Q^2}}){\beta}\\
{+[x_{Bj}^2+(1-x_{Bj})^2+x_{Bj}(1-3x_{Bj}){{4m_{q_i}^2}\over{Q^2}}
-x_{Bj}^2{{8m_{q_i}^2}\over{Q^2}}]\ln{{1+{\beta}}\over{1-{\beta}}} ] }
\label{mh},
\end{eqnarray}
where the quark velocity $\beta$ is given by
\begin{eqnarray}
{\beta}=\sqrt{1-{{4m_{q_i}^2}\over s}}=\sqrt{1-{{4m_{q_i}^2x_{Bj}}\over {Q^2(1-x_{Bj})}}}.
\label{beta}
\end{eqnarray}
The energy of the $\gamma^* \gamma$ collision, $s$, is equal to the
the square of the invariant mass of the hadronic system $W^2$ and is given by
\begin{eqnarray}
s=W^2={{Q^2}\over x_{Bj}}(1-x_{Bj}).
\label{w2}
\end{eqnarray}
In the limit of $s$ well above threshold, \ie for $\beta \approx 1$, 
which for fixed $x_{Bj}$ (not too small and not too close to 1) 
corresponds to the  Bjorken limit, one gets
\be
\ln{{1+{\beta}}\over{1-{\beta}}}=\ln{{(1+{\beta})^2}\over{1-{\beta}^2}}
\approx\ln{{Q^2(1-x_{Bj})}\over{m_{q_i}^2x_{Bj}}}.
\ee
The structure function $F_2^{\gamma}$ can be approximated in such 
case by
\begin{eqnarray} 
F_2^{\gamma}= 
{{\alpha }\over{\pi}}N_c\sum^{N_f}_{i=1} 
Q_i^4 x_{Bj}[ [x_{Bj}^2+(1-x_{Bj})^2]\ln{W^2\over m^2_{q_i}}
+8x_{Bj}(1-x_{Bj})-1]
\label{f2}
\end{eqnarray}
and it can be used to define the quark densities in the photon:
\begin{eqnarray} 
F_2^{\gamma} =x_{Bj}\sum^{2N_f}_{i=1} Q_i^2 q_i^{\gamma}(x_{Bj},Q^2).
\label{f2q}
\end{eqnarray}  
In the above formulae $N_c$=3 denotes the number of colors. In the 
last  equation a (natural) assumption, that quark and antiquark 
distributions in photon  are the same, has been  introduced. Note that
\begin{itemize}
\item
$ F_2^{\gamma}$ is calculable in the PM, in contrast to the structure 
function of hadrons, \eg the nucleon structure function $ F_2^{N}$.
\item
$F_2^{\gamma}$ is proportional to $\alpha$, the fine structure 
coupling constant ($\approx$ 1/137). There is an overall logarithmic 
dependence on the energy scale squared $W^2$ (or $Q^2$, see 
Eq.~(\ref{w2})). A large value of $F_2^{\gamma}$, or in other words 
a large quark density, is predicted  at large $x_{Bj}$.
\item  $F_2^{\gamma}$ is a sum of the quark densities with the factors 
$Q_i^4$, where $Q_i$ is fractional charge of quark, so the individual
quark density is proportional to $Q^2_i$ (see Eqs. (\ref{f2}, \ref{f2q})).
\end{itemize}
The longitudinal structure function $F_L^{\gamma}$ is not zero in 
PM, in contrast to the corresponding function for hadrons, and it 
is scale invariant:
\be 
F_L^{\gamma}={{\alpha}\over{\pi}}N_c\sum_{i=1}^{N_f}Q^4_i
x_{Bj}[4x_{Bj}
(1-x_{Bj})].
\ee
Note that also the third structure function is scale invariant:  
\be
F_3^{\gamma}={{{\alpha}}\over{{\pi}}}N_c\sum_{i=1}^{N_f}Q^4_i
x_{Bj}[-x_{Bj}^2].
\ee
In the leading logarithmic approximation (LLA) the Callan-Gross 
relation $F_2^{\gamma}=2x_{Bj} F_1^{\gamma}$ holds as in the case 
of hadrons. 

In this approximation the PM formula for the quark density 
is given by
\be
q_i^{\gamma}(x_{Bj},Q^2)|_{PM}^{LL}
={{\alpha }\over {2\pi}}N_c Q_i^2 [ [x_{Bj}^2+
(1-x_{Bj})^2]\ln{{Q^2}\over {\Lambda_{QCD}^2}}]. 
\ee
Since the LL contribution corresponds to the on-shell quarks one 
can treat $x_{Bj}$ in the above formula as equal to the part of four 
momentum of the initial photon carried by the quark;  this latter 
variable is usually  denoted by $x_{\gamma}$ (or simply $x$). Note 
also that in the above formula instead of a quark mass there appears 
the QCD scale $\Lambda_{QCD}$. Therefore this expression allows to 
describe all the light quark (light as compared to the scale $W^2$)
contributions to $F_2^{\gamma}$ in a universal way. Heavy quark 
contributions should be treated separately, according to the 
Eqs.~(\ref{mh}, \ref{beta}) \cite{rev}.

An additional $Q^2$ dependence will appear in $F_2^{\gamma}$ and in 
$q^{\gamma}_i$ due to the QCD corrections, which can be described by 
the Dokshitzer-Gribov-Lipatov-Altarelli-Parisi equations or by other 
techniques \cite{rev,witten,dewitt,bb,uw}. In this framework the 
gluonic content of photon, $g^{\gamma}$, appears as well.

The inhomogeneous DGLAP equations for the real photon 
can be represented in the following way 
(below $q_i^{\gamma}$ is  used for quarks and antiquarks):
\begin{eqnarray}
\nonumber
{{{\partial q_i^{\gamma}}\over{\partial lnQ^2}}=
{{{\alpha}}\over{2{\pi}}}Q^2_i P_{q \gamma}
{+{{{\alpha}_s}\over{2{\pi}}}\int^1_x{{dy}\over y}
[P_{qq}({x\over y})q_i^{\gamma}(y)+
P_{qg}({x\over y})g^{\gamma}(y)]}}
\end{eqnarray}
\begin{eqnarray}
{{\partial g^{\gamma}}\over{\partial lnQ^2}}=
0+{{{{\alpha}_s}\over{2{\pi}}}\int^1_x{{dy}\over y}
[P_{gq}({x\over y})\sum_{i=1}^{2N_f} q_i^{\gamma}(y)+
P_{gg}({x\over y})g^{\gamma}(y)]}
\end{eqnarray}
with the standard splitting functions $P_{qq},P_{qg},P_{gq},P_{gg}$
and in addition with the function 
\be
P_{q{\gamma}}
=N_c[x^2+(1-x)^2],
\ee
describing the splitting of the photon into quarks.

Note that it is possible to solve the above equations without the initial
conditions \cite{witten}, assuming the LL or NLL behaviour of
the solution. Obtained in this way the, so called, asymptotic 
solutions, have a singular behaviour at small $x$ \cite{uw}.
Therefore in practice while solving the  equations (14) 
the initial conditions have to be assumed from a model or 
taken from measurements at some (low) 
$Q^2_0$ scale \cite{grg}. On this basis the parton parametrizations
for the photon can be  constructed, see the Appendix for the details
of existing parton parametrizations.

In pre-QCD times, the hadronic structure of the photon has been solely
attributed to the vector meson component ($\rho, \omega, \phi$) 
in the (real) photon. Therefore for the matrix element  of the photon 
between \eg the nucleon states,  the following representation 
by the corresponding matrix elements for the $\rho$ current  
was assumed \cite{bernstein}:
\be
<N\mid J_{\mu}\mid N>=-{{m_{\rho}^2}\over{g_{\rho}}}{1\over{q^2-m^2_{\rho}}}
<N\mid {\rho}_{\mu}\mid N>+...
\ee

That was the basic assumption of the 
Vector Dominance Model (VDM) or Generalized VDM (GVDM) 
if higher vector mesons states were included \cite{bernstein,vdm}. 
This non-perturbative (or hadronic)
contribution to the structure of the photon
is  present of course in the measured structure function $F_2^{\gamma}$,
as well as in $q^{\gamma}$ or   $g^{\gamma}$
extracted from the experimental data.

\subsubsection{Data on $F^{\gamma}_2$}
Here we discuss the latest results for
$F_2^{\gamma}$ for a real photon based on single-tagged events 
at $e^+e^-$ colliders (\ie DIS$_{\gamma}$ events, see Eq. (\ref{eg})
and Fig. \ref{fig:jerzy1}).
Note that in practice not $x_{Bj}$ but the quantity $x_{vis}$ is measured,
\be
x_{vis}={Q^2 \over {Q^2+W_{vis}^2}},
\ee
where $W_{vis}$ is the invariant mass of the visible hadronic system.
For the nonzero target-photon virtuality $P^2$, the measured quantity
is 
\be
x_{vis}={Q^2 \over {Q^2+W_{vis}^2+P^2}}.
\ee
From a measured value for $x_{vis}$ one can reconstruct 
$x_{Bj}$ by an unfolding procedure.

Recently, in order to perform reliable unfolding, the events are  
grouped into classes of similar topology (see \eg {\bf DELPHI 96b}),
which roughly  coincide with the contributions due to QPM, VDM 
or RPC (RPC - Resolved Photon Contribution,
where photons interact through their partonic contents).

In the early analyses the separation between these contributions 
(QPM with QCD corrections  and VDM, basically) was obtained 
by introducing the parameter of the minimal transverse momentum 
of the produced hadrons, $p_T^0$.
The VDM (or hadronic) contribution should be
negligible for the production of particles with $p_T$ 
larger than $p_T^0$,
where, on the other hand, other  contributions should be included.
The FKP approach \cite{fkp}, where this kind of cutoff parameter is built in,
was natural here.
Note that in generation of   two quark final state
(in \eg F2GEN event generators)  two extreme types of angular 
dependence for final quarks are assumed:
point-like (PL)
with angular dependence as in the lepton pair production 
in {\underline {real}} photon-photon collision and 
a ``peripheral'' one with the  distribution 
corresponding to an exponential form of the $p_T$ dependence,
with a mean of 300 MeV (as if the photon interacted as 
a hadron)~\footnote{also a 
mixed 'perimiss' dependence was studied \cite{opal2}.}.
Nowadays the Monte Carlo generators
HERWIG and PYTHIA, adapted for DIS$_{\gamma}$ in 1995,
 can describe all these types of contributions
with any of the existing  parton parametrizations.
It is no longer necessary to fit an empirical $p_T^0$ parameter
to the data before unfolding.

It is worth noticing here, that at present energies
(or $Q^2$ scales) light quark ($u,d,s$)  and heavy quark
distributions are treated differently. 
For $c,b$ quarks the QPM formula (Eq. (\ref{mh})) is applied 
(at least close to the threshold),
while for the lighter quarks the QCD corrections are necessary. 
Note also that in the earlier measurements the $c$ quark contribution 
was usually subtracted from the structure function $F_2^{\gamma}$.

Unfolding  can be performed
using the  traditional method based on linear scale 
or using the new approach with the logarithmic scale,
 especially useful for extracting $F_2^{\gamma}$
results in the small $x_{Bj}$ region.

As we have already pointed out, the study of the hadronic final state 
became part of the measurements of the structure function 
for the photon at $e^+e^-$ colliders.
The details of the studies of the hadronic final state will be presented
in the next section. Here we would only like to stress 
that some discrepancies have been found for certain distributions,
like the pseudorapidity $\eta$ of final hadrons  distributions.
This fact is included in the estimation of 
the uncertainty of the measured function $F_2^{\gamma}$.

The general features of $F_2^{\gamma}$, as far as $x_{Bj}$ and 
$\log Q^2$
dependences are concerned, agree with the theoretical expectations,
although the precision of the data does not allow
in many cases 
to distinguish between existing parton parametrizations
and/or clarify the small $x_{Bj}$ behaviour of $F_2^{\gamma}$.  

We start the presentation  of the data from  the LEP collider, 
then TRISTAN data  are discussed. Collective figures of
$F_2^{\gamma}$ versus $x_{Bj}$ and $F_2^{\gamma}$ versus $Q^2$
are presented at  the end of this section.

The sets of data are described by the name of the collaboration
(in alphabetic order), 
the publication date,
and reference listed at the end of the paper.
In comments we  quote statements from the original papers 
(for abbreviations used for the parton parametrizations see 
the Appendix).
\newline\newline
\centerline{\bf \huge DATA}
\newline\newline
$\bullet${\bf {ALEPH 97a \cite{finch} (LEP 1)}}\\
Data on $F_2^{\gamma}$ at LEP 1 energy for $x_{Bj}$ from
0.002 to 0.9 and $Q^2$ between 6 and 44~GeV$^2$ were  collected in 
the period 1991-94. 
An analysis of the hadronic final state was performed
using the QPM+VMD model and the
HERWIG simulations (with GRV, GS and LAC1 parton parametrizations)
 (see next section).
\vspace*{0.3cm}
\begin{figure}[ht]
\vskip 9.5in\relax\noindent\hskip -3.2cm
       \relax{\includegraphics{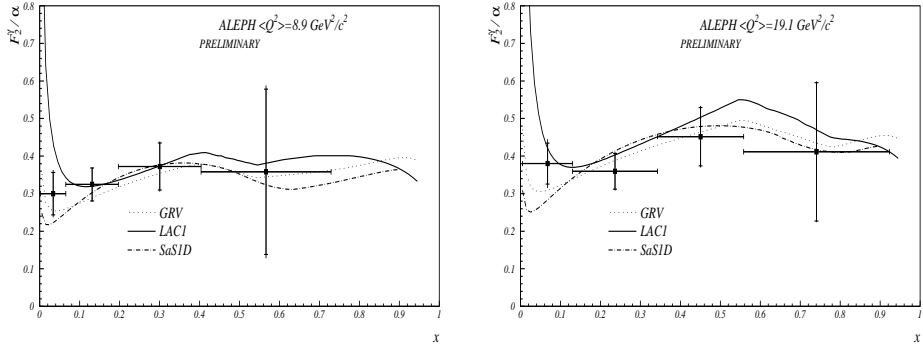}}
\vspace{-20.3cm}
\caption{ {\small\sl The unfolded $F_2^{\gamma}/\alpha$ for $<Q^2>$=8.9
and 19.1 GeV$^2$
compared with predictions of GRV, LAC1, and SaS1D 
parametrizations.}}
\label{fig:finch3}
\end{figure}

\begin{figure}[ht]
\vskip 0.cm\relax\noindent\hskip -3.9cm
       \relax{\includegraphics{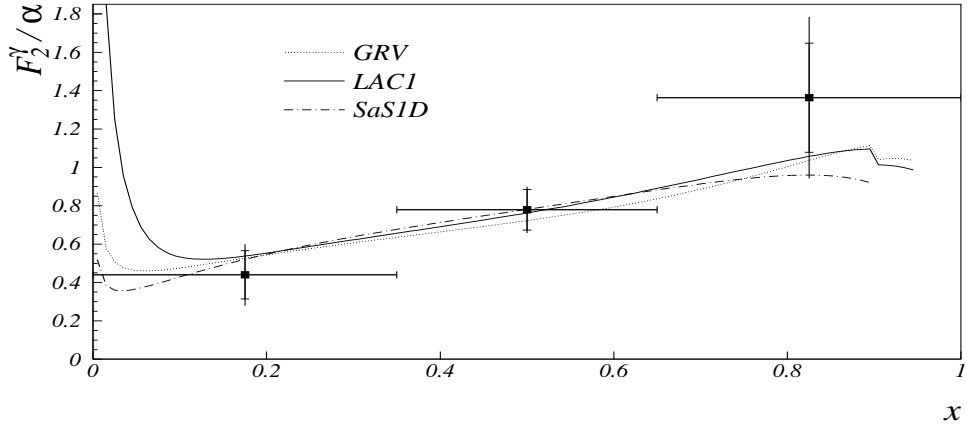}}
\vspace{0.5cm}
\caption{\small\sl The structure function $F_2^{\gamma}/\alpha$
as a function of $x_{Bj}$ (from \cite{LP315}).}
\label{fig:LP3154}
\end{figure}

\begin{figure}[ht]
\vskip 0.cm\relax\noindent\hskip -0.5cm
       \relax{\includegraphics{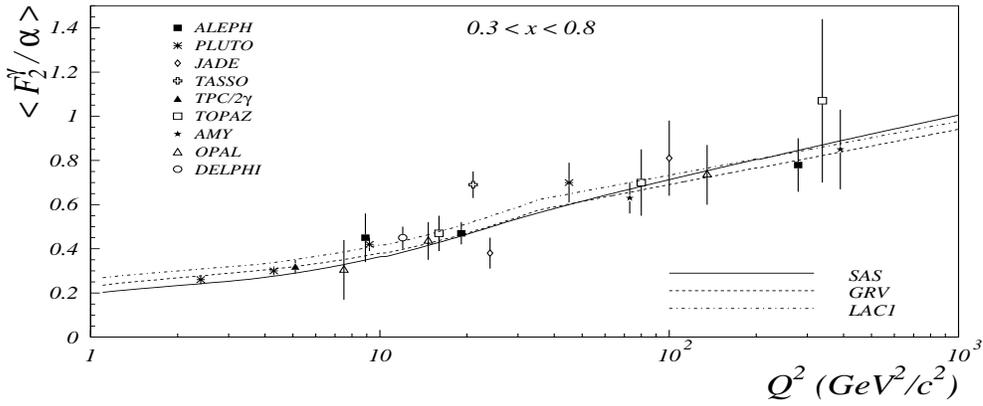}}
\vspace{0.1cm}
\caption{\small\sl The $Q^2$ dependence of $F_2^{\gamma}/\alpha$ 
for the range 0.3$<x_{Bj}<$0.8 (from \cite{LP315}).}
\label{fig:LP3155b}
\end{figure}

The unfolded results for $F_2^{\gamma}$ are presented in
Fig.~\ref{fig:finch3}. 
$F_2^{\gamma}$ averaged over $x_{Bj}$ and $Q^2$ in two
bins of $Q^2$ is given in the  table:
$$
\begin{array}{|c|c|c|}
\hline
<Q^2>&x_{Bj}&F_2^{\gamma}/\alpha\\
~[GeV^2]~&&(stat.+syst.)\\
\hline
8.9 &0.3-0.8&0.36\pm 0.16\pm 0.06\\
\hline
19.1&0.3-0.8&0.44\pm 0.08\pm 0.02\\
\hline
\end{array}
$$
\newline
Comment: {\it Problem with the proper description of  hadronic 
energy flow as a function  of pseudorapidity or azimuthal
separation angle was found (see next section).}
~\newline\newline   
$\bullet${\bf {ALEPH 97b \cite{LP315} (LEP I) }}\\
Data taken from 1991 to 1995 at average $Q^2$ equal to 279 GeV$^2$.
The hadronic final state was also studied using 
four QPM+VDM models and HERWIG Monte Carlo program (with 
GS2, GRV LO, LAC1 and SaS1d parton 
pa\-ra\-me\-tri\-za\-tions, see next section).
The results for $F_2^{\gamma}$ are presented in Figs.~\ref{fig:LP3154}
and \ref{fig:LP3155b}.

~\newline
Comment: {\it Problem with the proper description of the 
pseudorapidity distribution of final hadrons (see next section).}\\
\vspace*{4.7cm}
\vspace*{4.6cm}

\begin{figure}[hc]
\vskip 0.0in\relax\noindent\hskip 0.cm
       \relax{\includegraphics{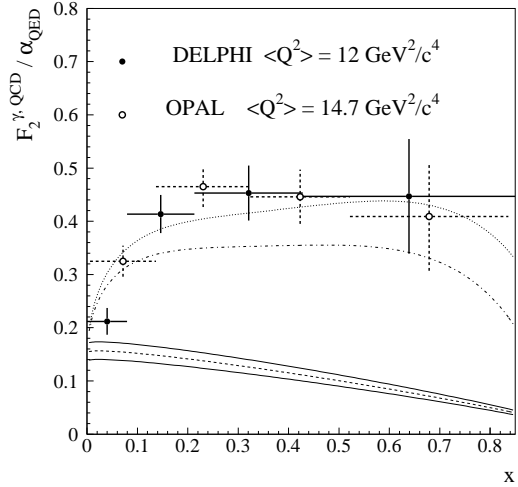}}
\vspace{0.cm}
\caption{\small\sl Unfolded $F_2^{\gamma}/\alpha$ 
for the light quarks (DELPHI and OPAL data).
The curves show the  
sum of the GVMD model prediction multiplied  
by the threshold factor 1-$x_{Bj}$ and the prediction
of the FKP parametrization for the point-like 
part of $F_2^{\gamma}$,
with different values of parameter $p_T^0$:
0.1 GeV (upper line) and 0.5 GeV (lower line). The bottom
curves show the GVDM contribution with different target masses
(from \cite{delphi2}).}
\label{fig:delphi7}
\end{figure}

~\newline
$\bullet${\bf {DELPHI 96a \cite{delphi2} (LEP 1) }}\\
Data on $F_2^{\gamma}$ were taken in the period 1991-93 for $Q^2$ between  
4 and 30 GeV$^2$ and for $x_{Bj}$ down to 0.003.
The so called   $F_2^{\gamma (QED)}$ was  also measured
and  compared with QED prediction, 
also for nonzero averaged virtuality $P^2$ of the
target photon (see Sec. 4.1 for  details).
Estimated  target photon 
virtuality $P^2$   was used in the unfolding
of  $F_2^{\gamma}$ 
\footnote{
It was found that although  $<P^2>$=0.13
GeV$^2$, a fixed value of 0.04 GeV$^2$ fits the data better.}.
The TWOGAM event generator  was used to simulate QPM events
and another event generator was 
used to  obtained the QCD correction (LL) to the point-like
contribution for light quarks in the FKP approach.
 The  GVDM  and the point-like (FKP)  contributions were studied, 
with $p_T^0$
=0.1 and 0.5 GeV. 
Results for $F_2^{\gamma}$ at $<Q^2>$=12 GeV$^2$ 
are presented in Fig.~\ref{fig:delphi7} and in the  table below: 
$$
\begin{array}{|c|c|c|}
\hline
<Q^2>&x_{Bj}&F_2^{\gamma}/\alpha\\
~[GeV^2]~&&(stat. + syst.)\\
\hline
~~~12~~~&0.003 - 0.080&0.21\pm0.03\pm0.06\\
&0.080 - 0.213&0.41\pm0.04\pm0.05\\
&0.213 - 0.428&0.45\pm0.05\pm0.05\\
&0.428 - 0.847&0.45\pm0.11\pm0.10\\
\hline
\end{array}
$$
\vspace*{5.6cm}

\begin{figure}[ht]
\vskip 7.in\relax\noindent\hskip -2.4cm
       \relax{\includegraphics{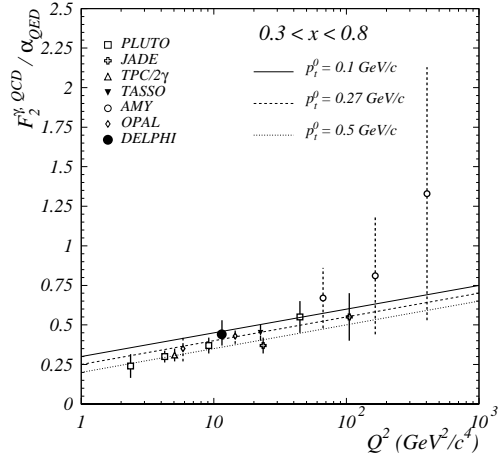}}
\vspace{-12.cm}
\caption{\small\sl $F_2^{\gamma}/\alpha$ averaged over $x_{Bj}$ 
between 0.3 and 0.8. The curves 
show the FKP parametrization 
predictions for different values of 
the parameter $p_T^0$  (from \cite{delphi2}).}
\label{fig:delphi9}
\end{figure}

The averaged value of  $F_2^{\gamma}/\alpha$ over the $x_{Bj}$ 
range between 0.3 and 0.8 was extracted:
$$
\begin{array}{|c|c|}
\hline
<Q^2>&<F_2^{\gamma}/\alpha>\\
~[GeV^2]~& \\
\hline
~12~&0.45 \pm0.08\\
\hline
\end{array}
$$
For comparison with other measurements of the $Q^2$ dependence of
the $F_2^{\gamma}$, see Fig.\ref{fig:delphi9}.
\newpage

\begin{figure}[hb]
\vskip 0cm\relax\noindent\hskip -3.cm
       \relax{\includegraphics{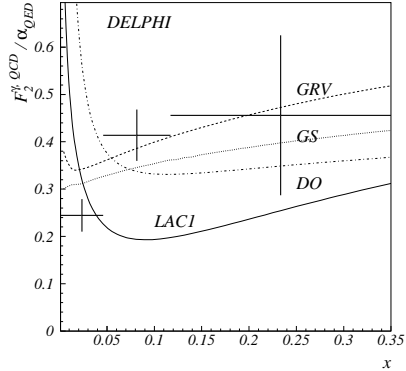}}
\vspace{0cm}
\caption{\small\sl Unfolded $F_2^{\gamma}/\alpha$ for $<Q^2>=$   
12 GeV$^2$ from the DELPHI experiment with   
the LO parametrizations LAC1, GS, DO, 
and GRV (from \cite{delphi2}).}
\label{fig:delphi10}
\end{figure}

Study of the $F_2^{\gamma}$ behaviour at $<Q^2>$=12 GeV$^2$ in 
the low $x_{Bj}$ domain leads to
following results (Fig.~\ref{fig:delphi10}):
$$
\begin{array}{|c|c|c|}
\hline
<Q^2>&x_{Bj}&F_2^{\gamma}/\alpha\\
~[GeV^2]~&&(stat. + syst.)\\
\hline
12&0.003 - 0.046&0.24\pm0.03\pm0.07\\
&0.046 - 0.117&0.41\pm0.05\pm0.08\\
&0.117 - 0.350&0.46\pm0.17\pm0.09\\
\hline
\end{array}
$$
\vspace*{4.3cm}

~\newline
Comment: {\it No rise of $F_2^{\gamma}$ at small {$x_{Bj}$} has been found.
GRV  and GS leading order parametrizations of the 
quark density in the photon are in agreement with data.}
\newline\newline
$\bullet${\bf {DELPHI 96b \cite{delphi} (LEP 1)}}\\
Measurement  of the photon structure function
 $F_2^{\gamma}$ (data collected in the years 1991-95)
together with a study  of the hadronic final state, in 
particular the resolved target 
photon contribution (see next section), is reported.

Two types of models for the extraction of the 
VDM contribution were used for the $<Q^2>$=$13$ GeV$^2$ data
(GVDM and "TPC/2$\gamma$"-type, for details see Ref.\cite{delphi}).
QPM, and RPC contribution with the GS2, SaS4 and GRV3 parton parametrizations
were studied.
The results for the $<Q^2>$=106 GeV$^2$ were also obtained.
The unfolding was done in a linear and logarithmic scale in $x_{Bj}$.

\begin{figure}[ht]
\vskip 0.in\relax\noindent\hskip 1.5cm
       \relax{\includegraphics{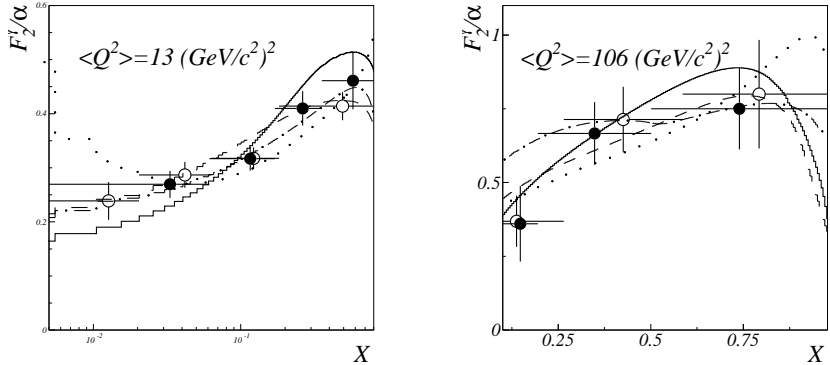}}
\vspace{0.ex}
\caption{\small\sl Unfolded $F_2^{\gamma}/\alpha$ compared with 
QPM+GVDM+RPC (GS2) (solid line) and
predictions of   SaS4 (dashed line), GS2 (dashed-dotted)
and GRV3 (dotted) parametrizations (from \cite{delphi}).} 
\label{fig:tyap96}
\end{figure}

Fig.~\ref{fig:tyap96} 
shows the $x_{Bj}$ dependence for both $Q^2$ samples.
\vspace*{4.5cm}

The averaged value of $F_2^{\gamma}/\alpha$ for $0.3 <x_{Bj}<0.8$
as a function of $Q^2$ was extracted:
$$
\begin{array}{|c|c|}
\hline
<Q^2>&<F_2^{\gamma}/\alpha>\\
~[GeV^2]~& (stat.+syst.)\\
\hline
~13&0.38~\,\pm0.031\pm0.016\\
106\,&0.576\pm0.081\pm0.076\\
\hline
\end{array}
$$
~\newline
Comment:{\it  The importance of the final hadronic 
state topology was noticed and the  study of a
linear and logarithmic unfolding performed.}
\newline\newline
$\bullet${\bf {DELPHI 97 \cite{tyapkin} (LEP 1, LEP 2)}}\\
This paper reports a 
recent study of $F_2^{\gamma}$ in the $Q^2$ range between
3 and 150 GeV$^2$, based on data from the 1994-95 runs for energies
around the $Z^0$ mass and from 1996 for energies between 161
and 172 GeV. An analysis of the hadronic final state is performed
and compared with predictions of the TWOGAM Monte Carlo
program, where QPM, VDM and RPC(GS2) parts are included (see next section).

\begin{figure}[hc]
\vskip 0in\relax\noindent\hskip -3.5cm
       \relax{\includegraphics{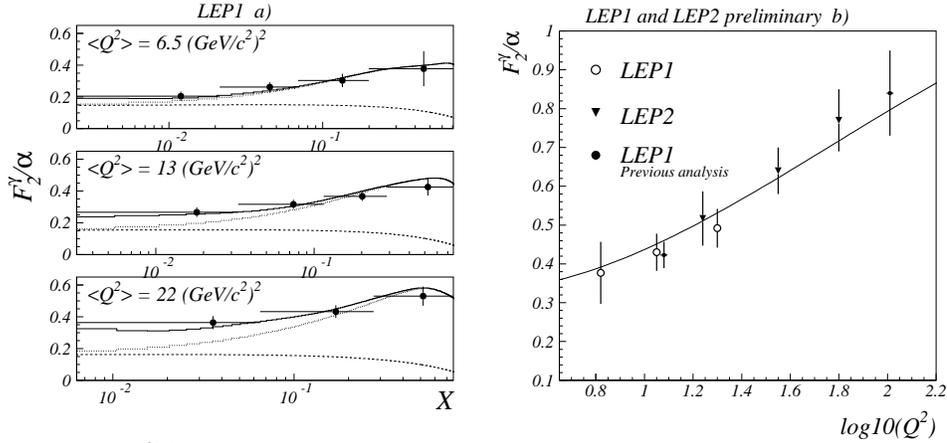}}
\vspace{0cm}
\caption{\small\sl a) $F_2^{\gamma}/\alpha$ versus $x_{Bj}$ 
from the DELPHI experiment 
based on LEP 1 data for three values of $<Q^2>$: 6.5, 13, and
22 GeV$^2$. The solid line corresponds to QPM+GVDM+RPC(GS2), dotted - 
QPM+GVDM and 
dashed - GVDM.
 b) $F_2^{\gamma}/\alpha$ 
averaged over $x_{Bj}$ from 0.3 to 0.8.
Results obtained from LEP1 and LEP2 data are shown together with
results of a previous analysis of LEP1 data
(from \cite{tyapkin}).}
\label{fig:tyapkin4}
\end{figure}

Unfolded results for $F_2^{\gamma}$ as a
function of $x_{Bj}$ and $Q^2$ are presented in 
Figs.~\ref{fig:tyapkin4}a and~\ref{fig:tyapkin4}b, respectively.

~\newline
Comment:{\it The data for hadronic final state distributions,
including energy flow versus  pseudorapidity, agree
with predictions of the TWOGAM simulations.}\\
\vspace*{5.cm}

\begin{figure}[ht]
\vskip 2.in\relax\noindent\hskip 1.45cm
       \relax{\includegraphics{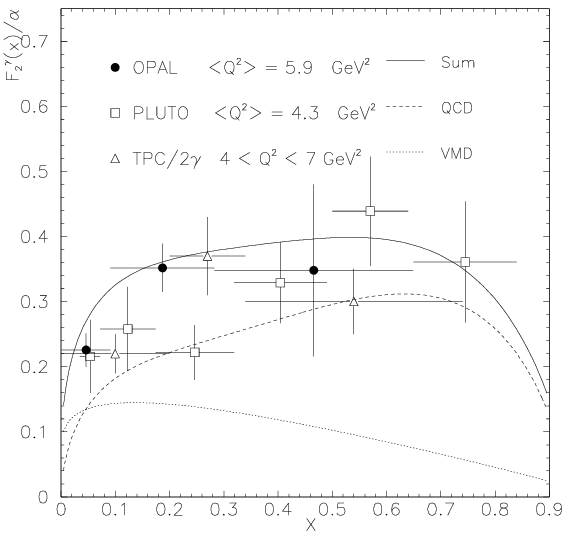}}
\vskip -0.2in\relax\noindent\hskip 7.85cm       
       \relax{\includegraphics{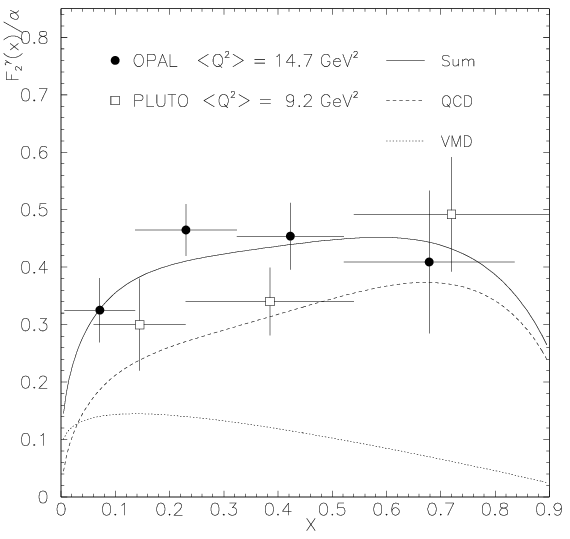}}
\vspace{-0.4cm}
\caption{\small\sl Unfolded $F_2^{\gamma}/\alpha$ shown with previous
measurements in other experiments at similar $<Q^2>$ values. 
The curves show contributions 
of VDM (dots), QCD-based model (dashes) and their sum (line) for 
a) $<Q^2>$=5.9 GeV$^2$ b) $<Q^2>$=14.7 GeV$^2$ 
(from \cite{opal94}).}
\label{fig:opal94a}
\end{figure}

~\newline
$\bullet${\bf {OPAL 94 \cite{opal94} (LEP 1)}}\\
Measurement of $F_2^{\gamma}$ at $<Q^2>$=5.9 and 14.7 GeV$^2$
was performed using data from the period 1990-92.
The VDM and pointlike  contributions (in the FKP approach)
 are separated by the cutoff parameter 
$p_T^0$ found to be 0.27$\pm 0.10$ GeV. 
\newline
Results compared with 
PLUTO and TPC/2$\gamma$
data are presented in Fig.~\ref{fig:opal94a}. 
\vspace*{0.5cm}

~\newline
$\bullet${\bf {OPAL 97a \cite{opal2} (LEP 1) }}\\
The measurement of $F_2^{\gamma}$ was done 
for $6<Q^2<30$ GeV$^2$ and  $60<Q^2<400$~GeV$^2$, 
using the full sample of data at the $Z$ peak (years 1990-95).
The detailed 
analysis of the hadronic final states was performed and sizeable 
discrepancies with the expectations 
 were found especially at low $x_{Bj}$ (see next section).
The influence of the choice of different MC generators (HERWIG, PYTHIA)
on $F_2^{\gamma}$
was studied. The estimation of the $P^2$ values was done based on 
the SaS1 parametrization.

\begin{figure}[ht]
\vskip 0.in\relax\noindent\hskip 0.in
       \relax{\includegraphics{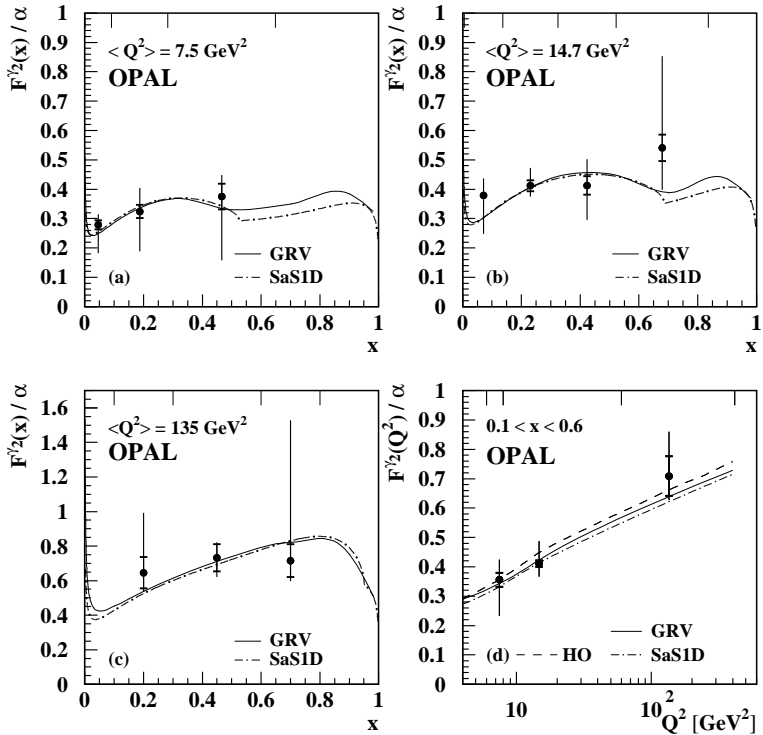}}
\vspace{0.ex}
\caption{\small\sl $F_2^{\gamma}/\alpha$ data from the OPAL experiment for
the number of flavours $N_f$=4. Curves in (a)-(d) show predictions
of GRV and SaS 1D parametrizations (from \cite{opal2}).}
\label{fig:scan1}
\end{figure}

The results for the $F_2^{\gamma}$, presented in  
Fig.~\ref{fig:scan1}a,b,c,
are as 
follows (the value of $F_2^{\gamma}/{\alpha}$
is given at the centre of the $x_{Bj}$ bin):
$$
\begin{array}{|r|r|c|}
\hline
\,\,<Q^2>~~\!\!&x_{Bj}~~~~~~&F_2^{\gamma}/{\alpha}\\  
~[GeV^2]~~~&&(stat. + syst.)\\
\hline
~~~~7.5~~~~~&0.001-0.091&0.28\pm0.02^{+0.03}_{-0.10}\\
&0.091-0.283&0.32\pm0.02^{+0.08}_{-0.13}\\
&0.283-0.649&0.38\pm0.04^{+0.06}_{-0.21}\\
\hline
~~~14.7~~~~~&0.006-0.137&0.38\pm0.01^{+0.06}_{-0.13}\\
&0.137-0.324&0.41\pm0.02^{+0.06}_{-0.03}\\
&0.324-0.522&0.41\pm0.03^{+0.08}_{-0.11}\\
&0.522-0.836&0.54\pm0.05^{+0.31}_{-0.13}\\
\hline
~~135.0~~~~~&0.100-0.300&0.65\pm0.09^{+0.33}_{-0.06}\\
&0.300-0.600&0.73\pm0.08^{+0.04}_{-0.08}\\
&0.600-0.800&0.72\pm0.10^{+0.81}_{-0.07}\\
\hline
\end{array} 
$$
\vspace*{9cm}

The measurement of $F_2^{\gamma}/{\alpha}$ as a function of $Q^2$
averaged over $x_{Bj}$ range, 
0.1$<x_{Bj}<$0.6, leads to 
following results (Fig.~\ref{fig:scan1}d):
$$
\begin{array}{|r|c|}
\hline
\,\,<Q^2>~~\!\!&<F_2^{\gamma}/{\alpha}>\\ 
~[GeV^2]~~~&(stat. + syst.)\\
\hline
~~~~7.5~~~~~ &0.36\pm0.02^{+0.06}_{-0.12}\\
~~~14.7~~~~~&0.41\pm0.01^{+0.08}_{-0.04}\\
~~135.0~~~~~&0.71\pm0.07^{+0.14}_{-0.05}\\
\hline
\end{array} 
$$
and the slope $d(F_2^{\gamma}/{\alpha})/dlnQ^2$ is measured to be 
$0.13^{+0.06}_{-0.04}$.
~\newline\newline
Comment: {\it No correction for $P^2 \neq$ 0 was made.
Large discrepancies between the hadronic energy flow
data and Monte Carlo simulations are observed at low x$_{vis}$,
when the results are presented versus pseudorapidity or 
azimuthal angle (see next section).
}
\newline\newline
$\bullet${\bf{OPAL 97b \cite{bech} (LEP 1)}}
~\newline
Measurement of $F_2^{\gamma}$ at LEP 1 was done for $<Q^2>$=1.86 GeV$^2$
and $<Q^2>$=3.76 GeV$^2$ as a function of $x_{Bj}$, reaching
the lowest ever measured (center of bin) value: $x_{Bj}$=0.0025. 
For a better sensitivity on the  low $x_{Bj}$ 
region the unfolding procedure was performed
on a logarithmic scale. Final state topology
was analysed as well (using the HERWIG, PYTHIA, F2GEN generators 
- both with the pointlike and the peripheral distributions).
Discrepancy between the data (hadron energy flow)
and results from the Monte Carlo generators, as well as
between different models  was found for low $x_{vis}<$0.05 
(see Sec.2.2).

\begin{figure}[ht]
\vskip 0.in\relax\noindent\hskip 1.2cm
       \relax{\includegraphics{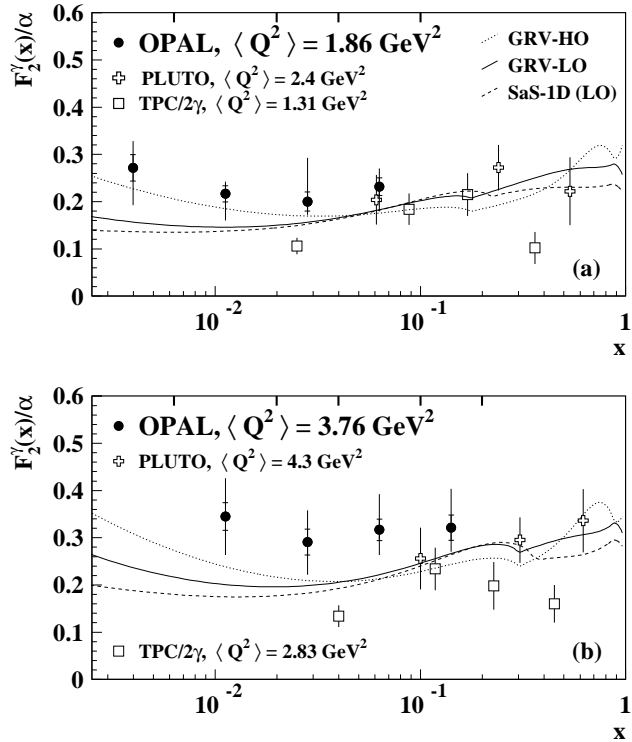}}
\vspace{0.1cm}
\caption{\small\sl OPAL $F_2^{\gamma}/{\alpha}$ data
(circles) as a function of $x_{Bj}$ for 
 $<Q^2>$= 1.86 GeV$^2$ (a) and $<Q^2>$=
3.76 GeV$^2$ (b). Also data from 
PLUTO (crosses) for $<Q^2>$= 2.4, 4.3 GeV$^2$ 
and TPC/2$\gamma$ (squares) for $<Q^2>$=1.31, 
2.83 GeV$^2$ are shown. The curves show predictions of
GRV-HO (dots), GRV-LO (line), and SaS1D (dashed) parametrizations.
The range of the $x_{Bj}$-bins of the OPAL results are marked at the tops of 
figures (from \cite{bech}).}
\label{fig:bech2}
\end{figure}

Obtained values of $F_2^{\gamma}/{\alpha}$  
are shown in Fig.~\ref{fig:bech2} together with measurements from 
PLUTO and TPC/2$\gamma$. 

~\newline
Comment: {\it ``No correction for $P^2\neq$0 was made. GRV-HO is consistent 
with the low-x OPAL results in the lower 
$Q^2$ bin, but at  higher $Q^2$
it underestimates the low x OPAL
data. GRV-LO and SaS1D describe the unfolded
results worse than GRV-HO. Shapes of measured $F_2^{\gamma}$
are flat within the errors, but a small rise in the low x region is 
not excluded.''}\\
\vspace*{9.cm}

~\newline
$\bullet${\bf{OPAL 97c \cite{LP291} (LEP 2)}}
\newline
New data on $F_2^{\gamma}$
from LEP 2 at the CM energies 161-172 GeV were
collected in 1996. 
The $<Q^2>$ range lies between 9 and 59 GeV$^2$. Also distribution
of the final hadronic energy flow was studied
(HERWIG, PYTHIA and F2GEN, see next section).

\begin{figure}[ht]
\vskip 0.in\relax\noindent\hskip 2.5cm
       \relax{\includegraphics{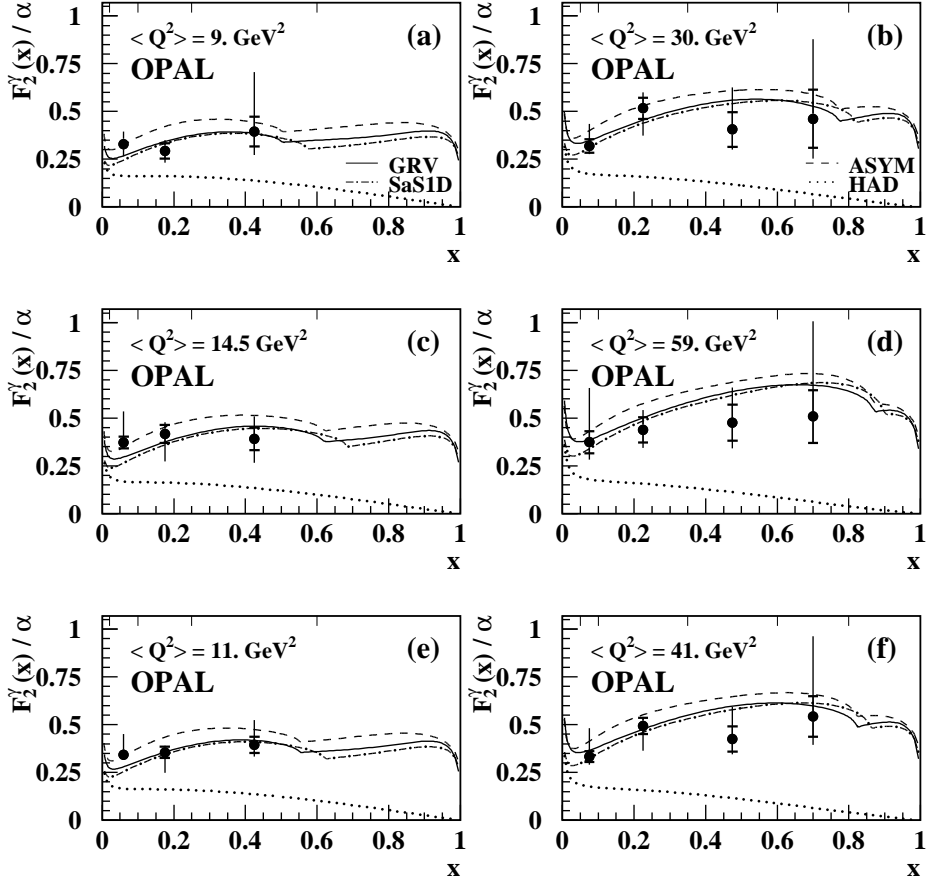}}
\vspace{0.ex}
\caption{\small\sl $x_{Bj}$-dependence
of $F_2^{\gamma}/{\alpha}$ measured by OPAL for different value of
$<Q^2>$: 9, 30, 14.5, 59, 11, and 41 GeV$^2$. For comparison
the GRV LO (solid line) and SaS1D (dashed dotted line)
parametrizations are shown;
dotted line represents the hadronic component (VDM?) (HAD)  
and the dashed one the asymptotic solution (ASYM)  
(from \cite{LP291}).}
\label{fig:nisius1}
\end{figure}

\begin{figure}[hb]
\vskip 0.cm\relax\noindent\hskip 0.5cm
       \relax{\includegraphics{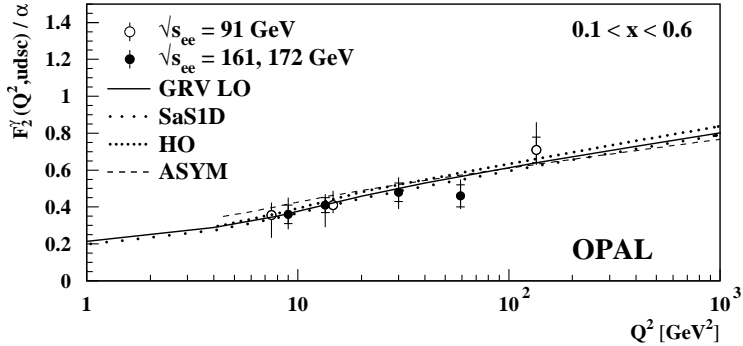}}
\vspace{0.1cm}
\caption{\small\sl $F_2^{\gamma}/{\alpha}$  averaged over 0.1$<x_{Bj}<$0.6, 
for the energy 91 GeV (open circles) and  for
the energies 161, 172 GeV (full circles). Predictions of the QCD
calculation are shown by the lines: 
solid (GRV LO), dotted (SaS1D), and double-dotted
(HO, based on GRV HO parametrization for light quarks);  dashed
line corresponds to the asymptotic solution (ASYM).
The charm contribution is calculated separately
(from \cite{LP291}).} 
\label{fig:nisius2}
\end{figure}

\begin{figure}[ht]
\vskip 0.cm\relax\noindent\hskip 1.cm
       \relax{\includegraphics{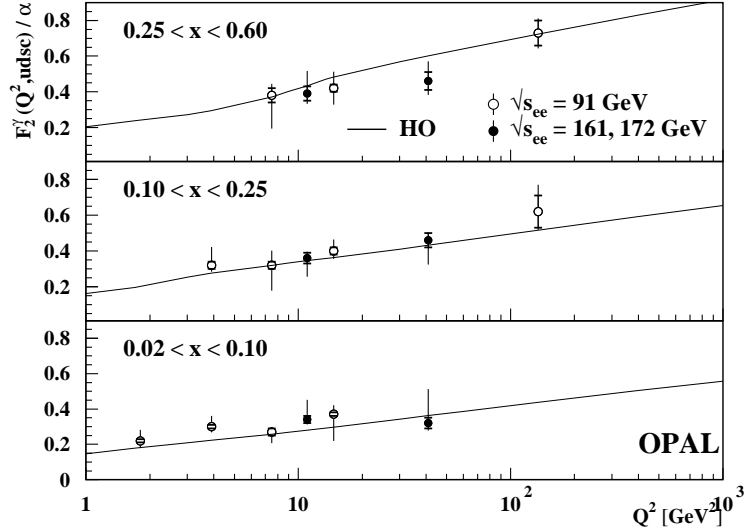}}
\vspace{0.1cm}
\caption{\small\sl $F_2^{\gamma}/{\alpha}$ for
the $0.1<x_{Bj}<0.6$ range from Fig.~\ref{fig:nisius2}
has been subdivided into three smaller bins. The data are 
compared here only to HO predictions (from \cite{LP291}).}
\label{fig:LP2914b}
\end{figure}

The unfolded $F_2^{\gamma}/{\alpha}$ as a function
of $x_{Bj}$ and $Q^2$ are presented in  
Figs~\ref{fig:nisius1}, 
\ref{fig:nisius2}. A special study of the $Q^2$ dependence of $F_2^{\gamma}$
is performed for different $x_{Bj}$ ranges, see 
Fig.~\ref{fig:LP2914b}.

A fit to the new data at the energies 161-172 GeV and the 
previous OPAL set at 91 GeV for $Q^2$ from 7.5 to 135
GeV$^2$, averaged over the $x_{Bj}$ range of 0.1-0.6
(see Fig.~\ref{fig:nisius2}), has the form:
\begin{eqnarray}
\nonumber
F_2^{\gamma}(Q^2)/\alpha =(0.16\pm 0.05^{+0.17}_{-0.16})
+(0.10\pm 0.02^{+0.05}_{-0.02})\ln (Q^2/\rm {GeV^2}). 
\end{eqnarray}
~\newline
Comment: {\it No correction for P$^2\neq$0 was made. Discrepancies
are observed in the hadronic energy flow between the data and
the HERWIG and PYTHIA simulations, especially
at x$_{vis}<$0.1. Accuracy of the data does not allow to see
the expected  different slope of $F_2^{\gamma}$
versus $log$Q$^2$ for different x$_{Bj}$ ranges.}\\
%\newpage
\vspace*{11.cm}
\vspace*{4.1cm}
\vspace*{6.1cm}

\begin{figure}[hb]
\vskip 0.in\relax\noindent\hskip 1.5cm
       \relax{\includegraphics{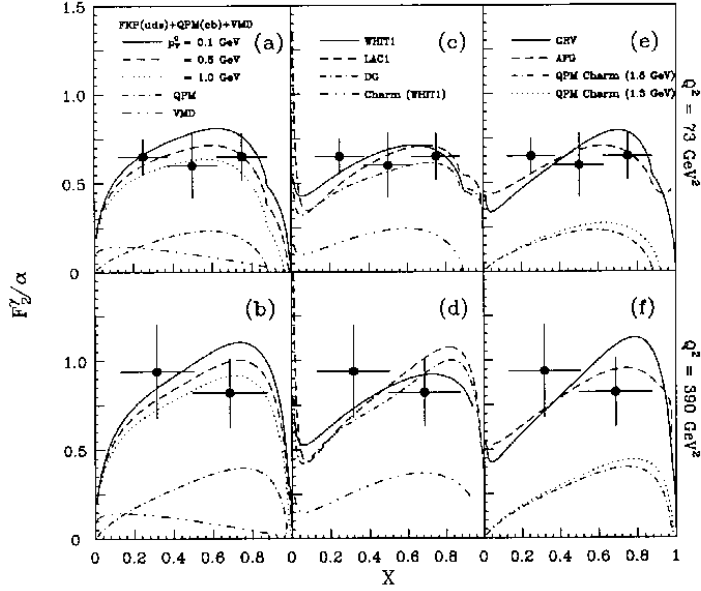}}
\vspace{0.ex}
\caption{\small\sl $F_2^{\gamma}/{\alpha}$ data from 
the AMY collaboration compared
with a few parametrizations of the photon structure:
FKP, QPM (quark parton model), VMD model (a,b),
WHIT1, LAC1, DG (c,d), GRV, AFG (e,f).
Upper (a,c,e) and lower (b,d,f) figures correspond to averaged
$Q^2$=73 GeV$^2$ and $Q^2$=390 GeV$^2$, respectively 
(from \cite{amy}). }
\label{fig:scan2}
\end{figure}

~\newline    
$\bullet${\bf{AMY 95 \cite{amy} (TRISTAN) }}\\
A high $Q^2$ measurement of the 
photon structure function $F_2^{\gamma}$ 
was performed.
Unfolded $F_2^{\gamma}$
is presented in Fig.~\ref{fig:scan2} (below are the corresponding
numbers).
$$
\begin{array}{|r|c|c|}
\hline
\,\,<Q^2>~\!&<x_{Bj}>&F_2^{\gamma}/{\alpha}\\
~[GeV^2]~~\!&&(stat. + syst.)\\
\hline
~~~~73~~~~~&0.25&0.65\pm0.08\pm0.06\\
&0.50&0.60\pm0.16\pm0.03\\
&0.75&0.65\pm0.11\pm0.08\\
\hline
~~~390~~~~~&0.31&0.94\pm0.23\pm0.10\\
&0.69&0.82\pm0.16\pm0.11\\ 
\hline
\end{array}
$$
The values of $F_2^{\gamma}$ averaged over 0.3$<x_{Bj}<$0.8
are as follows:
$$
\begin{array}{|r|c|}
\hline

\,\,<Q^2>~\!&<F_2^{\gamma}/{\alpha}>\\ 
~[GeV^2]~~\!&\\
\hline
~73~~~~~&0.63\pm0.07\\
135~~~~~&0.85\pm0.18\\
\hline
\end{array} 
$$
~\newline
Comment: {\it The observed 
$x_{Bj}$-behaviour of $F_2^{\gamma}$ is consistent with 
the GRV parametrization and with the FKP one 
(fitted parameter $p_T^0$ is equal to 0.51$\pm$0.39 GeV,
if only AMY data are included, and 0.45$\pm$0.07 GeV if all available, 
at that time, $F_2^{\gamma}$ data are taken).}\\
\vspace*{7.3cm}
%~\newline
%Comment: {\it The observed 
%$x_{Bj}$-behaviour of $F_2^{\gamma}$ is consistent with 
%the GRV parametrization and with the FKP one 
%(fitted parameter $p_T^0$ is equal to 0.51$\pm$0.39 GeV,
%if only AMY data are included, and 0.45$\pm$0.07 GeV if all available, 
%at that time, $F_2^{\gamma}$ data are taken).}

\begin{figure}[ht]
\vskip 0.in\relax\noindent\hskip 1.5cm
       \relax{\includegraphics{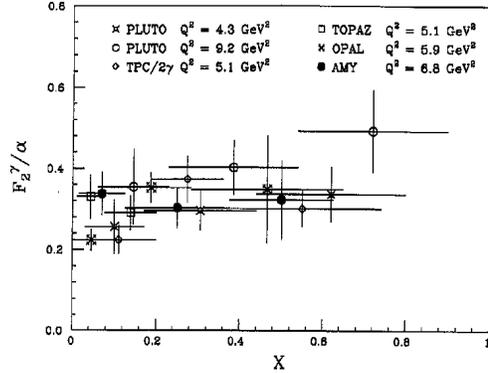}}
\vspace{0.ex}
\caption{\small\sl The measured $F_2^{\gamma}/{\alpha}$ values compared with results
of other experiments at $Q^2\sim$ 4-9 GeV$^2$ 
(from \cite{amy96}).}
\label{fig:amy971}
\end{figure}

\begin{figure}[hb]
\vskip 0.in\relax\noindent\hskip 1.5cm
       \relax{\includegraphics{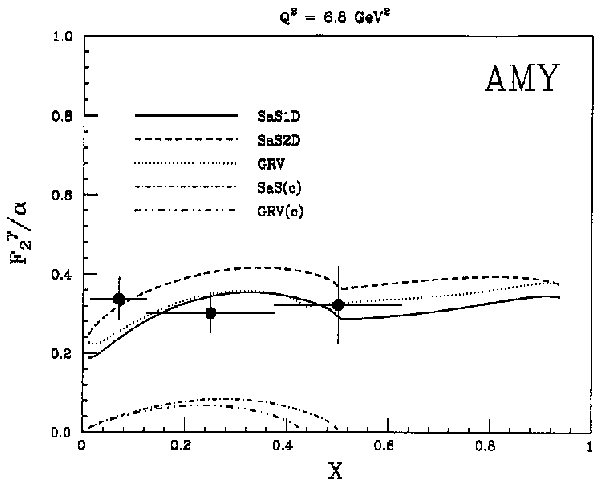}}
\vskip 0.in\relax\noindent\hskip 1.5cm
       \relax{\includegraphics{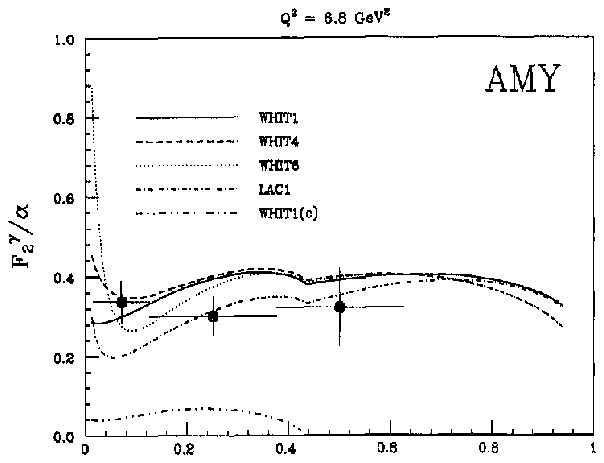}}
\vspace{0.ex}
\caption{\small\sl  The $F_2^{\gamma}/{\alpha}$ values as a function of 
$x_{Bj}$. a) Comparison with the SaS1D, SaS2D and 
GRV  predictions. Also shown are the heavy quark contributions 
from the SaS and GRV parametrizations. b) Comparison with WHIT1, WHIT4,
WHIT6 and LAC1 predictions. The heavy quark contribution is shown 
for WHIT1 parametrization (from \cite{amy96}).}
\label{fig:amy972}
\end{figure}

~\newline
$\bullet${\bf {AMY 97 \cite{amy96} (TRISTAN)}}
\newline
The measurement of $F_2^{\gamma}$ at $<Q^2>$=6.8 GeV$^2$
was performed;
the results for 
$F_2^{\gamma}$ versus $x_{Bj}$, together with earlier data,
are presented in Fig.~\ref{fig:amy971}. The 
comparison with various parametrizations is given in
Fig.~\ref{fig:amy972}.\\ 
\vspace*{4.cm}

\vspace*{4cm}
~\newline
Comment: {\it ``The $x_{Bj}$ behaviour of the measured $F_2^{\gamma}$
is consistent with the QCD-based predictions such as SaS, GRV-LO and
WHIT parametrizations, but inconsistent with LAC1 predictions for $x_{Bj}$
value around 0.07.''}
~\newline\newline 
$\bullet${\bf {TOPAZ 94 \cite{topaz} (TRISTAN)}}\\
The photon structure function $F_2^{\gamma}$ 
has been measured 
for averaged $Q^2$ values from 5.1 to 338 GeV$^2$ (see table below).
The one and two jet events in this sample are studied as well (see also 
Sec.2.2 and 2.3).

$$
\begin{array}{|l|r|l|l|c|}
\hline
~\!\!\!<Q^2>&~Q^2\!\!\!~~~~~~&<x_{Bj}>&~~~~~~x_{Bj}&F_2^{\gamma}/{\alpha}\\
~~\!\!\![GeV^2]&[GeV^2]\!\!~~~&&&(stat. + syst.)\\
\hline
~~~~~~5.1&~~3-~~10~&\!~0.043~&0.010-0.076&0.33\pm0.02\pm0.05\\
&&~\!0.138~&0.076-0.20&0.29\pm0.03\pm0.03\\
\hline
~~~~\,16&10-~~30~&~\!0.085~&0.02\,~-0.15&0.60\pm0.08\pm0.06\\        
&&~\!0.24~~&0.15\,~-0.33&0.56\pm0.09\pm0.04\\
&&~\!0.555~&0.33\,~-0.78&0.46\pm0.15\pm0.06\\
\hline
~~~~\,80&45-130~&~\!0.19~~&0.06\,~-0.32&0.68\pm0.26\pm0.05\\
&&~\!0.455~&0.32\,~-0.59&0.83\pm0.22\pm0.05\\
&&~\!0.785~&0.59\,~-0.98&0.53\pm0.21\pm0.05\\
\hline
\end{array}
$$
\vspace*{10cm}
\begin{figure}[ht]
\vskip 0.in\relax\noindent\hskip 0.cm
       \relax{\includegraphics{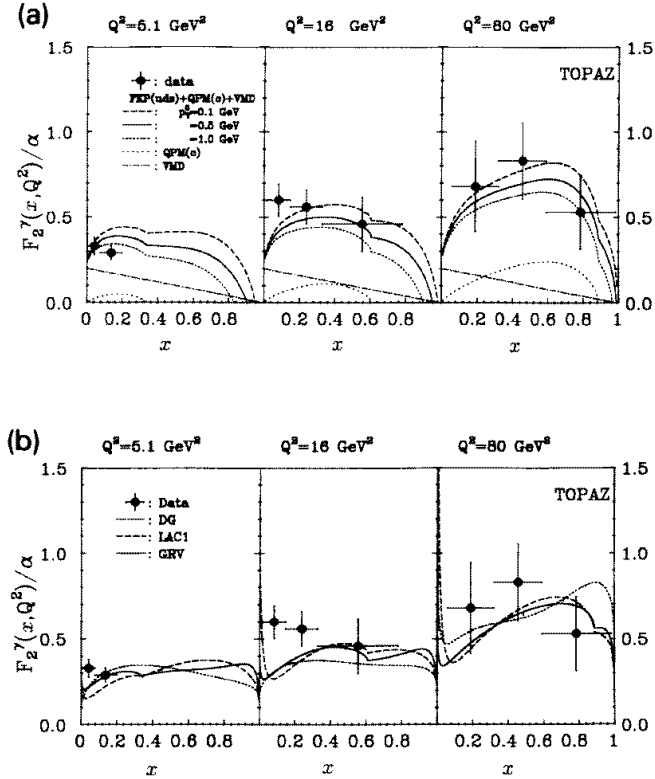}}
\vspace{0ex}
\caption{\small\sl TOPAZ 94 data for $F_2^{\gamma}/\alpha$ versus  
$x_{Bj}$ for $Q^2$=5.1, 16, 80 GeV$^2$ and a comparison with
parton parametrizations: a) FKP, QPM (parton model), VMD model
and b) DG, LAC1, GRV (from \cite{topaz}).}
\label{fig:marysia3}
\end{figure}

\begin{figure}[ht]
\vskip 0.5in\relax\noindent\hskip 0.in
       \relax{\includegraphics{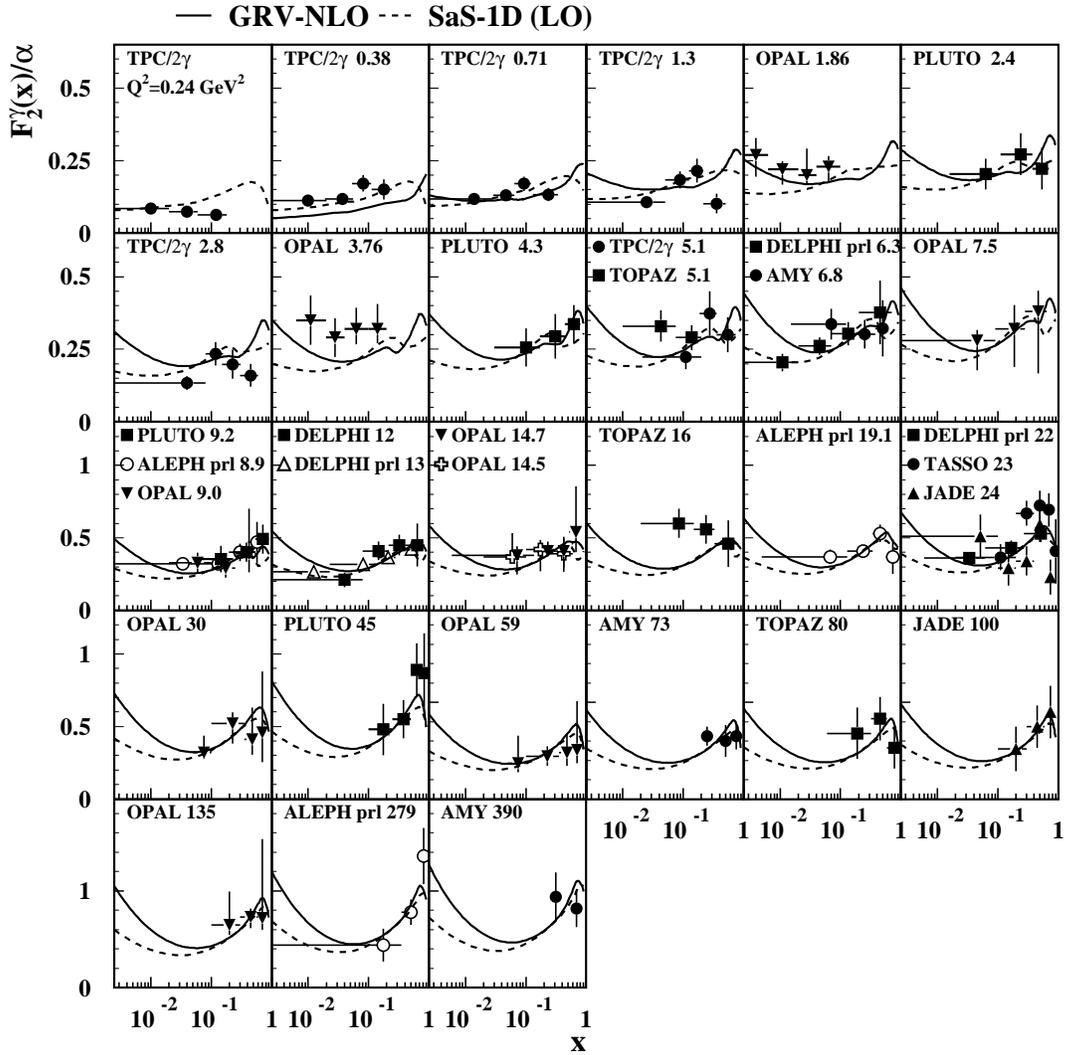}}
\vspace{-27.ex}
\caption{ {\small\sl The photon structure function 
$F_2^{\gamma}/\alpha$ as a function of $x_{Bj}$ in bins of
$Q^2$ compared to the GRV NLO (solid line) and 
SaS-1D (LO) (dashed line) parametrizations of parton distributions in
the photon (from \cite{stef}).}}
\label{fig:mary12}
\end{figure}

\begin{figure}[ht]
\vskip 0.cm\relax\noindent\hskip 0cm
       \relax{\includegraphics{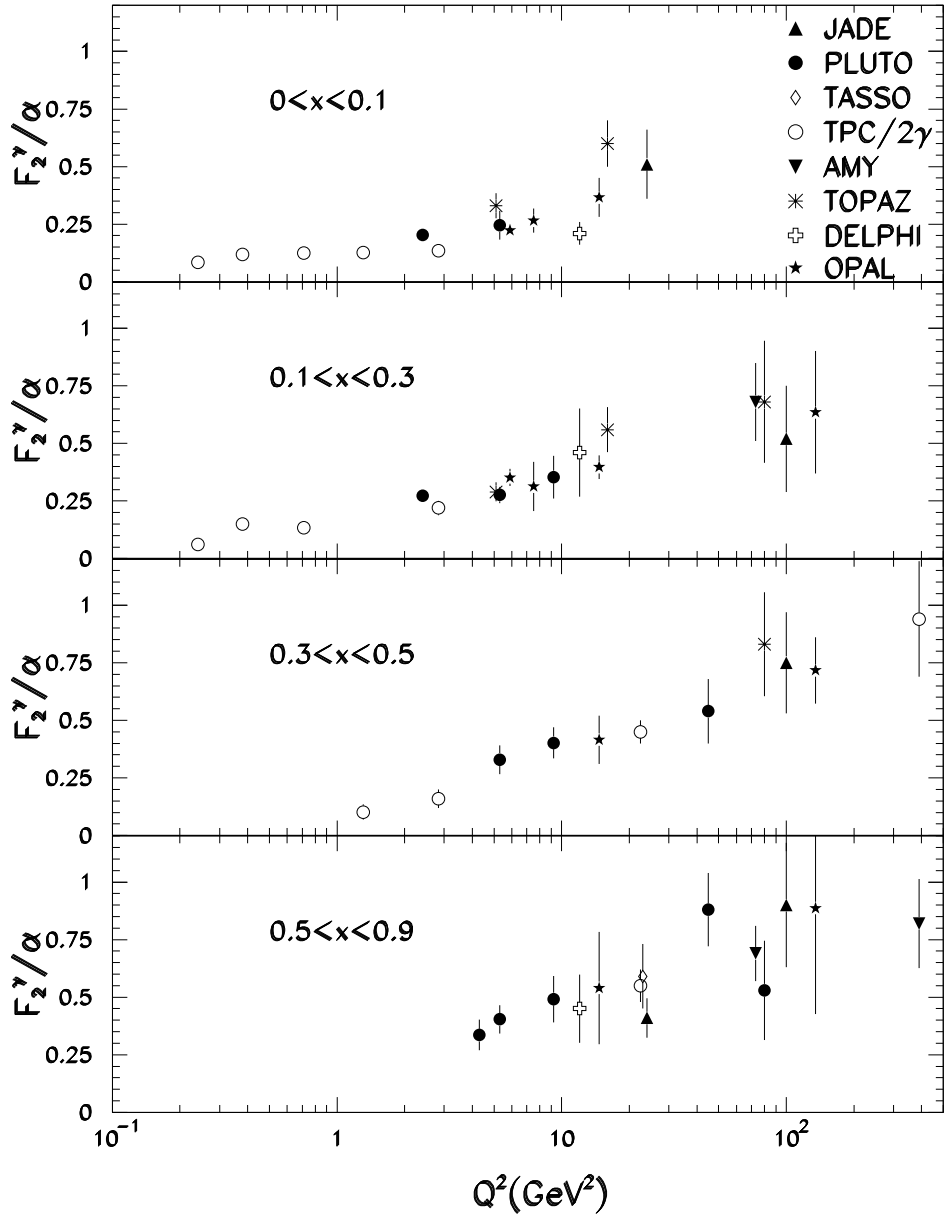}}
\vspace{0ex}
\caption{ {\small\sl The photon structure function
$F_2^{\gamma}/\alpha$ (data from various experiments)
as a function of $Q^2$
(from \cite{abr}).}}
\label{fig:habram}
\end{figure}

\begin{figure}[ht]
\vskip 0.cm\relax\noindent\hskip 0.8cm
       \relax{\includegraphics{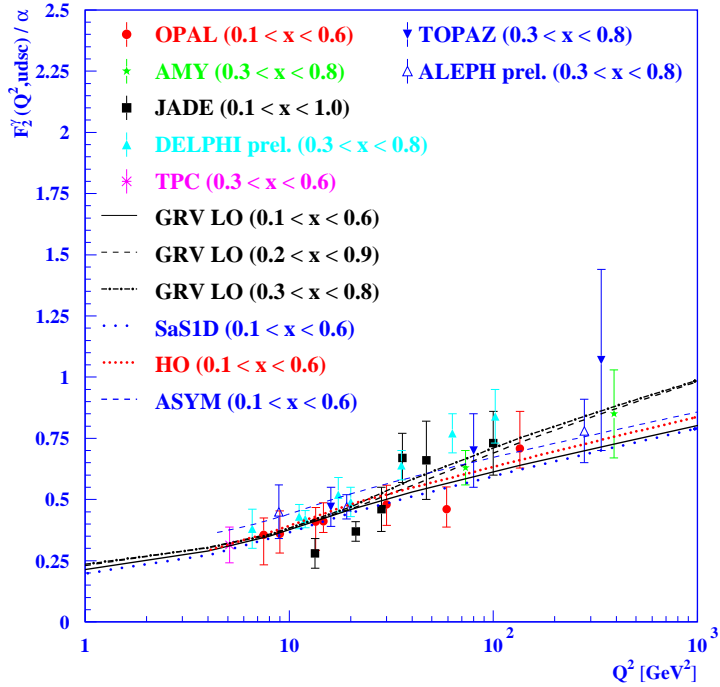}}
\vspace{0.2cm}
\caption{\small\sl The $Q^2$ evolution of $F_2^{\gamma}/{\alpha}$ 
measured by OPAL ($0.1<x_{Bj}<0.6$), AMY ($0.3<$
$x_{Bj}<0.8$), JADE ($0.1<x_{Bj}<1.$),  
DELPHI ($0.3<x_{Bj}<0.8$), TPC ($0.3<$
$x_{Bj}<0.6$), TOPAZ ($0.3<x_{Bj}<0.8$). 
The lines are predictions of parametrizations 
averaged over different $x_{Bj}$ ranges, 
HO calculation being a sum of
GRV HO parametrization for light quarks and the charm
contribution calculated independently from GRV;
also the asymptotic solution is shown (ASYM, dashed line)
  (from \cite{tnis}). }
\label{fig:nisius3}
\end{figure}

The results are presented in Fig.~\ref{fig:marysia3}.
To study the $Q^2$ dependence of $F_2^{\gamma}$,
the averaged values of  $F_2^{\gamma}/\alpha$ in the $x_{Bj}$-range
from 0.3 to 0.8 were extracted at $<Q^2>$= 16, 80 
and 338 GeV$^2$ (the numbers in parentheses are
the results for light quarks alone): 
$$
\begin{array}{|r|c|}
\hline
\,\,<Q^2>~~~~~\!\!&F_2^{\gamma}/{\alpha}\\ 
~[GeV^2]~~~~~~&\\
\hline
~16&0.47\pm0.08~(0.38\pm0.08)\\
~80&0.70\pm0.15~(0.49\pm0.15)\\
338&1.07\pm0.37~(0.72\pm0.37)\\
\hline
\end{array} 
$$
~\newline
Comment: {\it  The final hadronic state described by QPM(c)+VDM+FKP(u,d,s)
with $p_T^0$=0.1,0.5 and 1  GeV, also by GRV, 
DG and LAC1 parton parametrizations, was studied.} 
\vskip 1.cm
\centerline{*****}
\vskip 1.cm
For an overall comparison  
the collective figures of $F_2^{\gamma}/{\alpha}$ versus $x_{Bj}$
(Fig.~\ref{fig:mary12} \cite{stef}) and $F_2^{\gamma}/{\alpha}$ 
versus $Q^2$ 
(Figs.~\ref{fig:habram} \cite{abr}, \ref{fig:nisius3} \cite{tnis}),
 containing  also earlier data
not discussed here, are presented.
In Fig.~\ref{fig:mary12}
the comparison with theoretical predictions for $x_{Bj}$ 
dependence of $F_2^{\gamma}$ based
on SaS-1D (LO) and GRV NLO parton parametrizations are shown
(note the logarithmic scale).
For comparison with theoretical predictions of
the $Q^2$ evolution for different $x_{Bj}$ ranges see Fig.
 \ref{fig:nisius3},
 and also Figs. \ref{fig:LP3155b}, \ref{fig:delphi9}, 
\ref{fig:tyapkin4}b, \ref{fig:scan1}d, 
\ref{fig:nisius2}, \ref{fig:LP2914b}.

\begin{figure}[ht]
\vskip 0.cm\relax\noindent\hskip 1.5cm
       \relax{\includegraphics{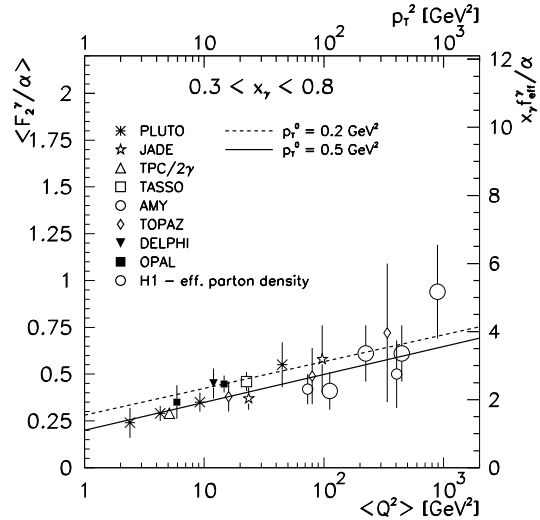}}
\vspace{0.cm}
\caption{\small\sl The scaling violation of the parton densities
compared with the $Q^2$ dependence of the $F_2^{\gamma}$
(averaged over the range 0.3$<x_{Bj}<$0.8)
(from \cite{t307}).}
\label{fig:t3074}
\end{figure}

The effective parton density, as measured at HERA collider 
by H1 group (see {\bf H1 97a} \cite{t307})
in the jet production from resolved photons,
is compared with the $F_2^{\gamma}$ data in 
Fig.~\ref{fig:t3074} as well.
\vspace*{17.4cm}
\newpage
\vspace*{15.5cm}
\newpage
\vspace*{8.2cm}
\vspace*{6.2cm}
%%%%%%%%%%%%%%%%%%%%%%%%%%%%%%%%%%%%%%%%%%%%%%%%%%
%
%
%%%%%%%%%%%%%%%%%%%%%%%%%%%%%%%%%%%%%%%%%%%%%%%%%%
\newpage
\subsection{Hadronic final states in the DIS$_{\gamma}$ experiments}
Although a detailed analysis of hadronic final states, including  
those involving large $p_T$ particles/jets, has not been the main 
aim of DIS$_{\gamma}$ experiments, nowadays it has proved crucial 
in extracting the unfolded $F_2^{\gamma}$. The obtained experimental 
results, and problems that appear while describing hadronic energy 
flow, transverse energy, pseudorapidity and other distributions 
within QCD, deserve close attention and a separate treatment - this 
section is devoted to this subject.

It is worth noticing that as $F_2^{\gamma}$ corresponds to the total 
cross section for hadron production in ${\gamma}^*{\gamma}$ 
collisions, all the regions: $Q^2\ll p^2_T, ~Q^2 \sim p^2_T$ and 
$Q^2\gg p^2_T$ contribute, with the bulk of the cross section  being 
due to the soft production, \ie with a relatively small $p_T$.
In the next section  dedicated  analyses of the {\underline {large}} 
$p_T$ jet production in real $\gamma \gamma$ and $\gamma p$ processes
will be presented. Some of the problems that will appear there are 
common with these presented here.

\subsubsection{Modeling of the hadronic final state in $\gamma^*\gamma $
collision}
In this section the ${\gamma}^*{\gamma}$ collisions resulting in 
the hadronic state will be discussed. As we have already mentioned 
in Sec.2.1.2, the QPM, VDM and RPC contributions are introduced 
to model the hadronic final state. In recent analyses the MC 
generators HERWIG and  PYTHIA are being used with chosen parton 
parametrizations. In the OPAL analysis, the generator F2GEN (based 
on TWOGEN used in older analyses), with assumed two  quarks in the 
final state is used in addition, for the comparison and systematic 
checks. In this generator one can easily put a specific assumption
on the final states. For example different angular distributions may 
be assumed in the $\gamma^* \gamma$ CM system: pointlike (PL), \ie 
as for lepton pair from two real photons, peripheral with an 
exponential distribution of the transverse momentum, or the 
``perimiss'' combination \cite{opal2}. In a recent DELPHI analysis, 
the TWOGAM generator (``fixed'' in 1993) was successfully used, 
with QPM, VDM and RPC contributions included \cite{delphi, tyapkin}.

As far as data on global hadronic variables like 
$W_{vis},\ Q^2,\ M_{ij}$ 
(invariant mass of two jets)
  and other  distributions  
are concerned,  there is a fairly good description of the data 
by existing MC generators.
The problems arise for the 
transverse energy out of the  tag plane 
(the plane defined by the initial and tagged electrons)  or for the
energy flow per event as a function of
 pseudorapidity, $\eta=-\ln(\tan(\theta/2))$,
 where the polar angle is measured  from the direction 
of the target photon.
The discrepancies are very pronounced especially for small $x_{Bj}$.
  
The first observation of the disagreement between the data and MC
models was made by the OPAL collaboration \cite{opal2}. 
``The serious discrepancies between the 
data and any of the available Monte Carlo
models are seen both within the central region of the detector
($|\eta|<2.3$), where the energy flow is well measured, 
and in the forward region, where the energy
can only be sampled.''
It is clear that the unfolding of $F_2^{\gamma}$
``will have large errors as long as the energy 
flow from different models remains in clear disagreement with the energy
flow in the data, in particular in the region of $x_{vis}<$0.1
and $Q^2<$30 GeV$^2$''
(from \cite{opal2}).
A similar effect has been seen by now by other groups.

The method used recently to improve the agreement with the 
data is the so called ``HERWIG + power law $p_t$'' generator,
where the additional power law
 spectrum of the form $dk_T^2/(k^2_T+k^2_0)$ was introduced
(it seems to be needed at HERA (ZEUS) as well, see next section).

One should be aware that in hadronic final state in DIS$_{\gamma}$
experiments two types of large scale may appear: $Q^2$ and $p_T^2$, 
describing the transverse momentum of the final hadrons/jets. The 
bulk of the data corresponds to events  with not very large $p_T$;
if for these events the relation $Q^2\gg p^2_T \gg P^2$ holds, the 
interpretation in terms of the  photon interaction between  one 
direct ($\gamma^*$) and  one  resolved real ($\gamma $) photon may 
be introduced. Then it is not clear what scale should be used in the 
parton density for a real photon $f^{\gamma}(x,\tilde Q^2)$: $p_T$
or $Q^2$. Moreover the processes corresponding to  $Q^2 \ll p^2_T$
should be treated as being resolved from the point of view of both:
a real and a virtual photon.

Note, that in  recent analyses the dependence on the number of jets 
in the final state is studied. 

\subsubsection{Data on the hadronic final state}
The analyses of the hadronic final state accompanying the 
measurements of $F_2^{\gamma}$ discussed in Sec.2.1.2 as well as the 
results from independent analyses of large $p_T$ hadron production 
in single tagged events are presented below.

\begin{figure}[ht]
\vskip 0.in\relax\noindent\hskip -3.cm
       \relax{\includegraphics{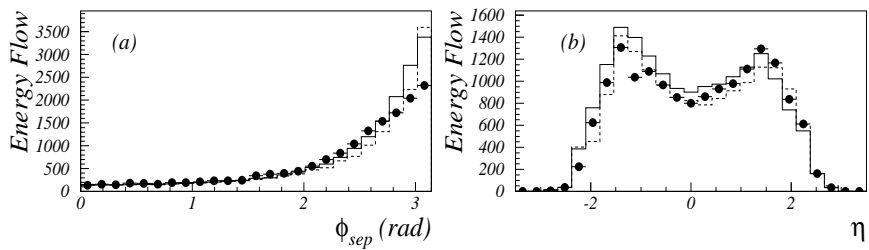}}
\vspace{-0.5cm}
\caption{\small\sl Energy flow as a function of $\phi_{sep}$ (a)
and $\eta$ (b). Predictions: QPM+VDM (solid line) and HERWIG generator with
GRV (dashed line) (from \cite{finch}).}
\label{fig:had1}
\end{figure}

For the two  photons involved, we  introduce the following notation
for their squared (positive) virtualities:  $P^2_1=Q^2$
and $P^2_2=P^2$ (with $P_1^2 \ge P_2^2$),
where the corresponding variables
used in the discussed DIS scattering are also given. 
\newline\newline
\centerline{\huge DATA} 
\newline\newline
$\bullet${\bf{ALEPH 97a \cite{finch} (LEP 1) }}\\
Hadronic energy flow in the azimuthal angle separation $\phi_{sep}$
and pseudorapidity $\eta$ was measured for $Q^2$ between 6 and 
44~GeV$^2$. The results are presented in Fig.~\ref{fig:had1}.\\
\vspace*{3cm}
~\newline
Comment:{\sl "Discrepancy with MC models (QPM + VDM and HERWIG (with 
parton parametrizations GS, GRV, LAC1)) 
when energy flow is plotted as a function of rapidity 
and azimuthal separation."}
\newline\newline  
$\bullet${\bf {ALEPH 97b \cite{LP315} (LEP 1) }}\\
Data taken from 1991 to 1995 for the average $Q^2$=279 GeV$^2$.
Four different QPM+VDM models were used and in addition
the HERWIG 5.9   generator
with the GS, GRV, LAC1 and  SaS1d  parton parametrizations.

\begin{figure}[hb]
\vskip 0.cm\relax\noindent\hskip 2.5cm
       \relax{\includegraphics{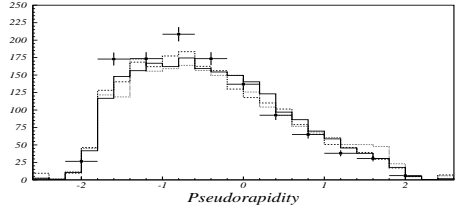}}
\vspace{-4.2cm}
\caption{\small\sl 
Pseudorapidity of all charged tracks and neutral calorimeter objects.
 Histograms are predictions of three of the models 
 (dotted line from HERWIG, others - QPM+VDM type) (from \cite{LP315}).}
\label{fig:LP3151}
\end{figure}

The pseudorapidity distribution compared with the model predictions 
is presented in Fig.~\ref{fig:LP3151}. A disagreement is found, 
similar to the one observed previously by ALEPH at lower $Q^2$
({\bf {ALEPH 97a}}), but now at the negative pseudorapidity.\\
\vspace*{6.2cm}

\begin{figure}[ht]
\vskip 0.cm\relax\noindent\hskip 0.5cm
       \relax{\includegraphics{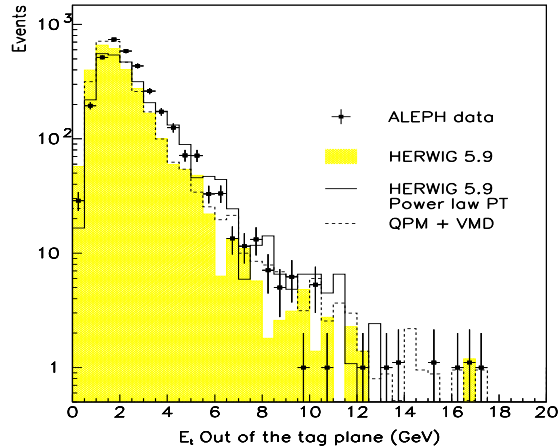}}
\vspace{0.cm}
\caption{\small\sl The energy (out of plane of the tag 
and the beam) distribution (from \cite{LP253}).}
\label{fig:LP2532b}
\end{figure}

\begin{figure}[ht]
\vskip 0.cm\relax\noindent\hskip 0.5cm
       \relax{\includegraphics{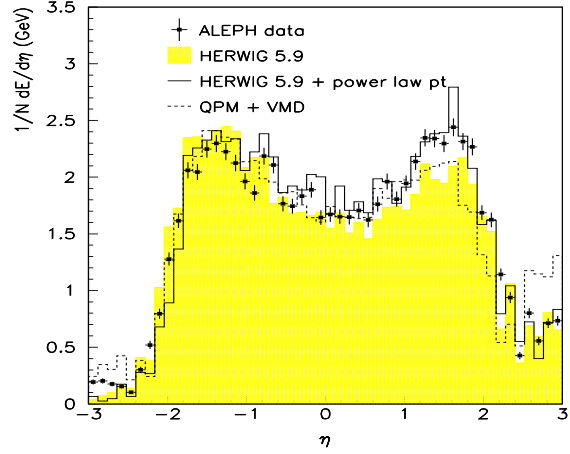}}
\vspace{0.cm}
\caption{\small\sl The energy flow versus 
pseudorapidity of the final hadrons (from \cite{LP253}).}
\label{fig:LP2532a}
\end{figure}

\begin{figure}[ht]
\vskip 0.cm\relax\noindent\hskip 0.5cm
       \relax{\includegraphics{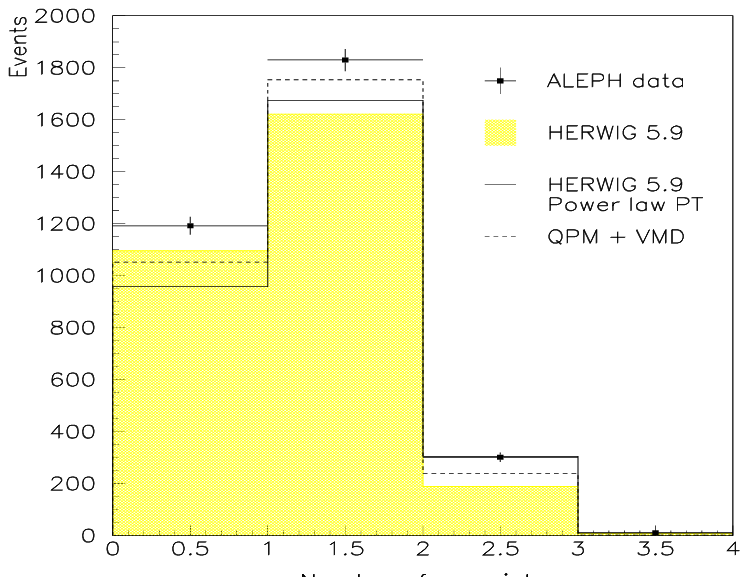}}
\vspace{0.cm}
\caption{\small\sl (from \cite{LP253}).}
\label{fig:LP2533}
\end{figure}

~\newline
Comment: {\sl ``The pseudorapidity distribution can not be described by the models
as observed previously at lower $Q^2$,
note however that the excess is now in the opposite hemisphere 
than at lower $Q^2$.''} 
~\newline\newline
$\bullet${\bf{ALEPH 97c \cite{LP253} (LEP 1) }}\\
Single tagged events collected in the years 1992-1994 
at the average $Q^2$=14.2 GeV$^2$.
A dedicated study of the final hadronic state assuming  the
QPM+VDM for $u$,$d$ and $c$ quarks and using  HERWIG 5.9 
generator (with  GRV parametrizations). Also  the modified 
HERWIG generator was used which corresponds to the 
``HERWIG + power law $p_t$'', with $k_{t0}=0.66$ GeV 
(see also {\bf OPAL 97e} where this model was introduced).
 Results are presented in  Figs.~\ref{fig:LP2532b},
\ref{fig:LP2532a}, \ref{fig:LP2533}.
The approach ``HERWIG + power law $p_t$'' 
leads everywhere to better description of the data.
\vspace*{5.7cm}

In order to pin down the source of observed discrepancy 
the analysis of the number of cone jets was performed 
in the final hadronic state
(see {\bf OPAL 97d} for the first analysis of this type).
HERWIG model where resolved photon processes
 are included should in principle give larger number of two jet events than
QPM+VDM model, having additional production mechanisms.
On the contrary, it gives less (see Fig.~\ref{fig:LP2533}).

Also the energy not assigned to jets in two jet events, 
$E_2^{non-jet}$, was studied. To check the presence of the single 
and double  resolved photon events, $x_{\gamma}$ and $x_{tag}$
distributions were measured. The ``HERWIG + power law $p_t$'' 
approach was used sucessfully to describe these data as well.
\vspace*{6.cm}

\vspace*{5.5cm}

\begin{figure}[ht]
\vskip 0.in\relax\noindent\hskip -0.6cm
       \relax{\includegraphics{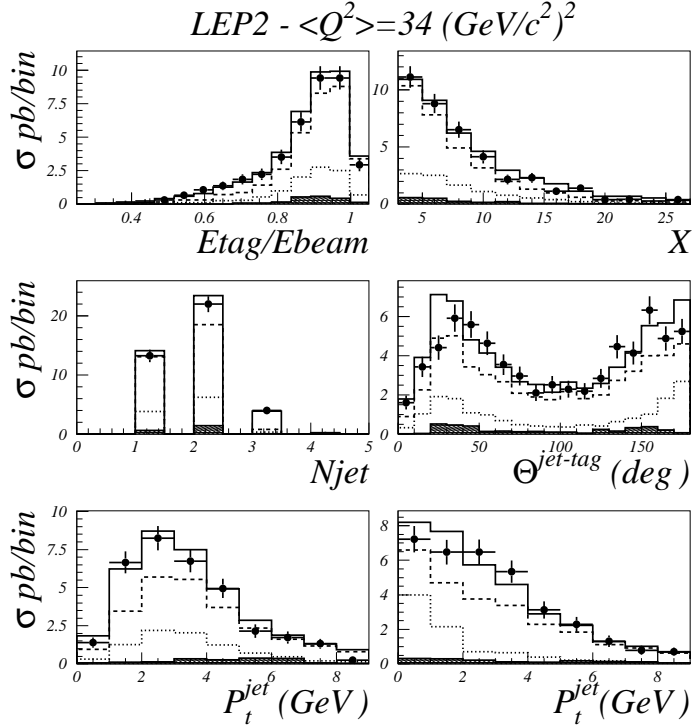}}
\vspace{0ex}
\caption{\small\sl The comparison of data and the MC prediction for 
$<Q^2>$ = 34 GeV$^2$ (LEP2). a) Energy of tagged particle; b) The 
invariant mass; c) The number of jets; d) The jet angle with respect 
to the tagged particle; e) $p_T^{jet}$ for jets in the same
hemisphere as the tagged particle; f) $p_T^{jet}$ for jets
in the opposite hemisphere. Curves show the MC predictions:
GVDM+QPM+RPC (GS2) (solid line), GVDM+QPM (dots), GVDM (dashes)
(from \cite{tyapkin}).}
\label{fig:had2}
\end{figure}

\begin{figure}[hb]
\vskip 0in\relax\noindent\hskip -0.9cm
       \relax{\includegraphics{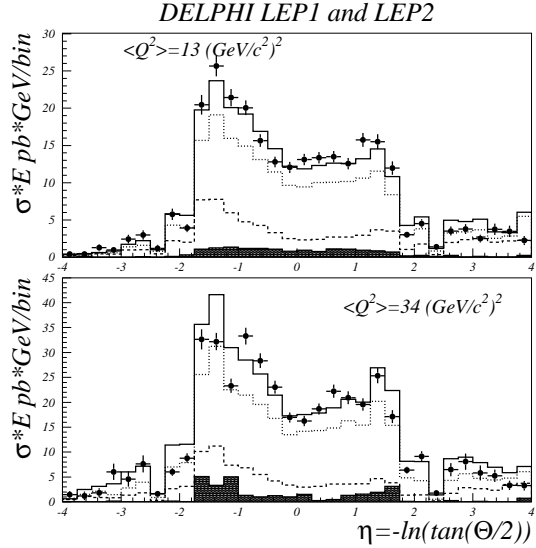}}
\vspace{0ex}
\caption{\small\sl The event energy flow as a function of the pseudorapidity.
The notations as in Fig.~\ref{fig:had2} (from \cite{tyapkin}).}
\label{fig:had3}
\end{figure}

~\newline
Comment: {\sl "``HERWIG + power law $p_t$'' is 
better in modeling the region
of large $E_{t, out}$ and the peak at positive rapidity".}  
\newpage
~\newline\newline
$\bullet${\bf{DELPHI 95 \cite{delphi3} (LEP 1) }}\\  
First evidence of hard scattering process in the  single-tagged 
$e{\gamma}$ collision in the data from the 1991-1992 run
is reported.
The values of  $E_T$ of observed jets were larger than  1.5 GeV, 
while the magnitude of the mass of virtual
photon was equal to $<P_1^2>\approx$ 0.06 GeV$^2$. This corresponds 
to the standard  resolved (almost real) photon
process, from the point of view of both photons. 
~\newline\newline
Comment: {\it This is not typical DIS experiment since the 
photon probe
is almost real. The analysis is not typical, either, 
for the standard large $p_T$
jet study. 
``The data are consistent with the predictions 
for quark and gluon density functions 
in the GS parametrization.
The sum of the contributions: VDM + QPM (Quark Parton Model) 
+ [QCD - RPC(Resolved Photon Contribution)]
is needed in order to describe the data; the DO and LAC3 parametrizations
do not adequately describe the data.''}
\newline\newline
$\bullet${\bf{DELPHI 96b \cite{delphi} (LEP 1) }}\\  
The results for the averaged $Q^2$=13 and 106 GeV$^2$
were considered and compared with TWOGAM generator. The QPM, 
GVDM contributions and, for the first time at DELPHI, RPC with GS2 
parton parametrization were included.
~\newline\newline
Comment: {\sl The resolved photon contributions are needed 
to obtain the 
description of the data. ``...the correct unfolding 
procedure leads to the
pointlike plus hadronic part of $F_2^{\gamma}$ 
in a shape that is inconsistent 
with QPM (FKP) + GVDM in the region of low-$x_{Bj}$.''}
%\item{\bf{L3 93 \cite{} (LEP)}}\\
%O. Adriani PL B 318 (1993) 575
\newline\newline
$\bullet${\bf{DELPHI 97 \cite{tyapkin} (LEP1, LEP2)}} 
\newline
Analysis of hadronic final state in the $F_2^{\gamma}$ measurement 
(see Sec.2.1.2) was performed for energies around the $Z^0$ mass and 
for 161-172 GeV (1996 run). The $Q^2$ ranges: 6, 13, 22 and 17, 34, 
63~GeV$^2$ were studied, respectively. Hadronic final state topology 
with events containg jets was studied using TWOGAM generator. For 
results see Fig.~\ref{fig:had2} for LEP2 data and Fig.~\ref{fig:had3} 
for event energy flow at LEP1 and LEP2.

~\newline
Comment: {\it ``All variables are in good agreement with 
TWOGAM predictions''.\\
Note that the $p_T$ range of jets may be of order of $<Q^2>$
for small $Q^2$ samples.}\\
\vspace*{8.5cm}

\vspace*{6.5cm} 

~\newline
$\bullet${\bf{OPAL 94 \cite{opal94} (LEP 1)}}\\
%ZP C61 (94)199\\
Data were collected in 1990-92 for averaged $Q^2$=5.9 and 14.7 GeV$^2$.
 Early analysis of the final hadronic state
(the TWOGEN generator with the contribution based on QPM, 
VDM and on the FKP approach to 
the QCD contribution) was performed.
For the estimation of the $p_T^0$ see Sec.2.1.2.
\newline\newline
$\bullet${\bf {OPAL 97a \cite{opal2} (LEP 1)}}\\    
%ZP C74 (1997) 33\\
The hadronic final state was analysed in the measurement of 
the $F_2^{\gamma}$ with
one photon highly virtual ($Q^2$ between 6-30 \g2  - 
{\it low $Q^2$ sample}
and between 60-400 \g2  - {\it high $Q^2$ sample}), 
the other being almost real. 

For generating the hadronic final state MC programs HERWIG,
PYTHIA, and for comparison F2GEN were used 
with GRV and SaS1D parton parametrizations.

\begin{figure}[hb]
\vskip 3.7cm\relax\noindent\hskip 1.8cm
       \relax{\includegraphics{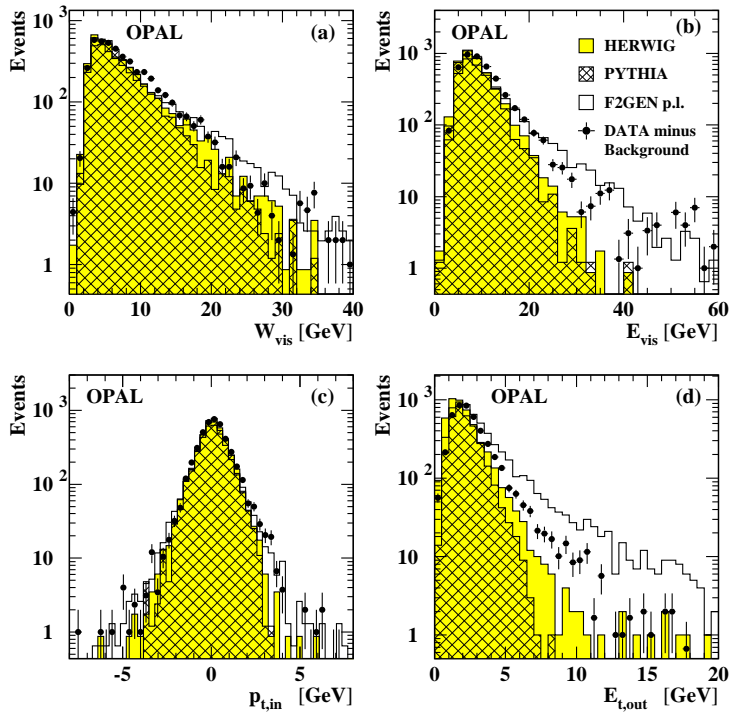}}
\vspace{-3.7cm}
\caption{\small\sl Comparison of data event quantities in the 
low-$Q^2$ sample with HERWIG, PYTHIA and F2GEN (PL)
Monte Carlo simulations. a) the distribution of the visible
invariant mass; b) the total visible energy of the event;
c) the transverse momentum of the event in the tag plane;
d) the energy out of the tag plane (from \cite{opal2}).}
\label{fig:had4}
\end{figure}

The energy ($E_{t,out}$)
transverse to the plane defined  by the beam axis and the 
tag direction, and other quantities
for the {\it low $Q^2$ sample} are presented in 
Fig.~\ref{fig:had4}.
The discrepancy found for the $E_{vis}$ and $E_{t,out}$
distributions in this sample 
(Figs.~\ref{fig:had4} b and d, respectively)
is absent in the {\it high $Q^2$ events} (not presented).
\vspace*{9cm}

\begin{figure}[ht]
\vskip 0.cm\relax\noindent\hskip 1.5cm
       \relax{\includegraphics{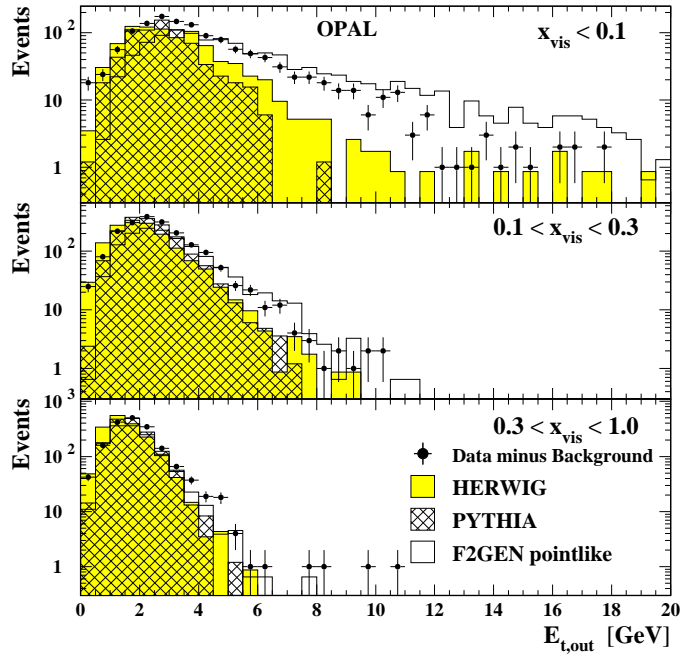}}
\vspace{0cm}
\caption{\small\sl The energy transverse to the tag plane for 
three $x_{vis}$ bins for the {\it low $Q^2$ sample}. 
MC generators as in Fig.~\ref{fig:had4} (from \cite{opal2}).}
\label{fig:had5}
\end{figure}

The hadronic energy $E_{t,out}$ distributions 
in the three $x_{Bj}$ bins are shown in Fig.~\ref{fig:had5}.
The failure of the models in the low $Q^2$ region is 
most visible at low $x_{Bj}$.

\begin{figure}[ht]
\vskip 0.cm\relax\noindent\hskip 2.cm
       \relax{\includegraphics{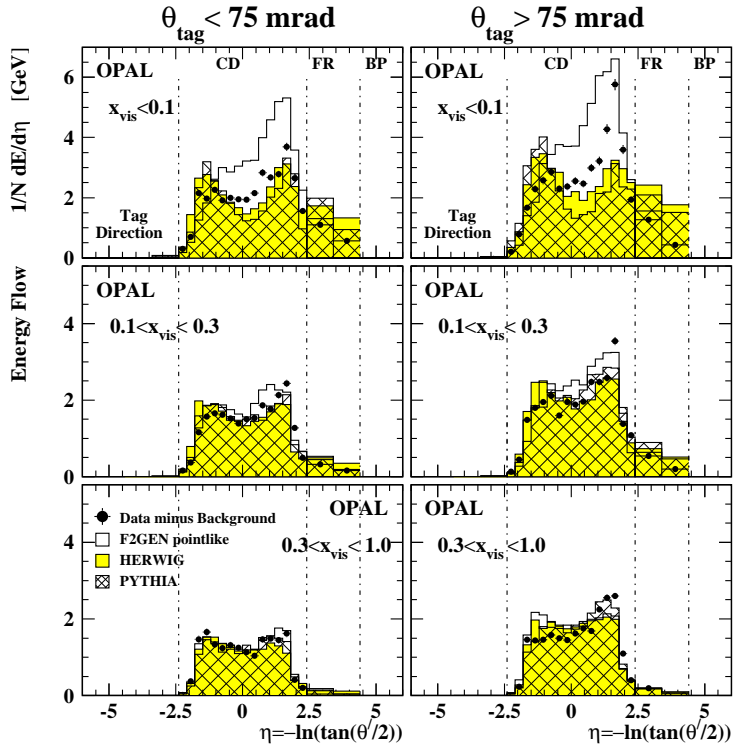}}
\vspace{0.cm}
\caption{\small\sl The hadronic energy flow per event as a function 
of the pseudorapidity $\eta$ for the data and various MC simulations,
in various ranges of $x_{vis}$ and $\theta_{tag}$ for the 
{\it low $Q^2$ sample} (from \cite{opal2}.}
\label{fig:had6}
\end{figure}

\begin{figure}[ht]
\vskip 4.3cm\relax\noindent\hskip 3cm
       \relax{\includegraphics{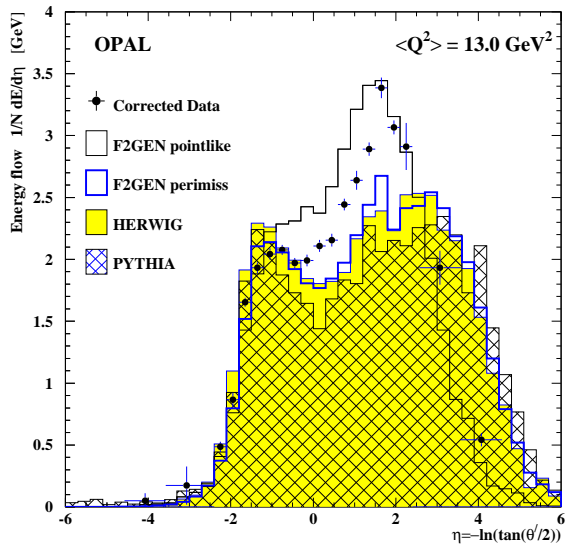}}
\vspace{-3.cm}
\caption{\small\sl The measured energy flow in the low-$Q^2$ sample
corrected for the detector inefficiencies, as a function of the 
pseudorapidity $\eta$, compared to the values generated by various MC models
 (from \cite{opal2}). }
\label{fig:had7}
\end{figure}

To establish the source of the discrepancy 
the energy flow per event 
in the {\it low $Q^2$ sample} was also studied as a function of the
pseudorapidity (see Fig.~\ref{fig:had6}).
The  
distribution of pseudorapidity $\eta$ for the 
{\it low $Q^2$ sample}, corrected for the experimental effects,
 is shown in Fig.~\ref{fig:had7}.\\ 
\vspace*{7.7cm}
\vspace*{9.5cm}
%\newpage
\vspace*{6cm}

\begin{figure}[hb]
\vskip 0.in\relax\noindent\hskip 1.5cm
       \relax{\includegraphics{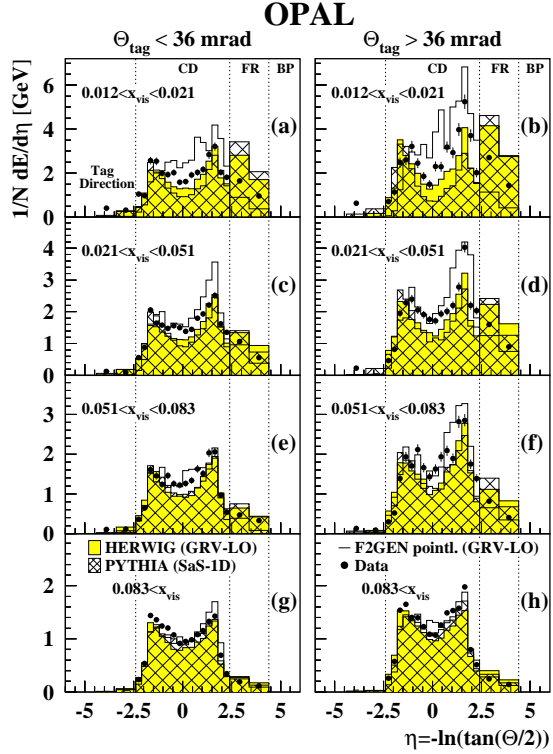}}
\vspace{0ex}
\caption{\small\sl The measured hadronic energy flow as a function of 
the pseudorapidity: a comparison of data with various MC predictions
for three bins in $x_{vis}$ and two bins in $\theta_{tag}$ 
(from \cite{bech}).}
\label{fig:had10}
\end{figure}

~\newline
Comment: {\it 
None of the generators represents the final state accurately; 
$E_{vis}, E_{t,out}$ - distributions as well as the  hadronic flow 
per event show a clear discrepancy. 
The failure of the models in the low $Q^2$ region is most marked at
low $x_{Bj}$. The differences between MC models and data in the low 
$Q^2$ region are apparent, when the energy flow per
event is plotted as a function of the pseudorapidity 
and the azimuthal angle (not shown).
``Particular attention will need to be given to the 
angular distribution of partons in $\gamma^* \gamma$ system''.

The relation between $p_T^2$ and $Q^2$ in the low $Q^2$ sample may
indicate a need
 to  take into account the structure of the virtual
photon.} 
~\newline\newline
$\bullet${\bf{OPAL 97b \cite{bech} (LEP 1)}}\\ 
The measurement of $F_2^{\gamma}$ and the modeling of the $\gamma^* \gamma$ 
fragmentation into hadrons at low $Q^2$ region (1.1 to 6.6 GeV$^2$)
and very small $x_{Bj}$ - bins from 0.0025 to 0.2 is reported. 
The hadronic energy flow as a function of $\eta$ for 
three  ${x_{Bj}}$
regions is plotted in Fig.~\ref{fig:had10}. 
~\newline\newline
Comment: {\sl Differences in the energy flow distributions 
versus the pseudorapidity (Fig.~\ref{fig:had10}) and in the summed 
 energy transverse to the tag plane are found
for $x_{vis}<0.05$ between the data and MC models
(HERWIG, PYTHIA, F2GEN).}
~\newline\newline
$\bullet${\bf{OPAL 97d \cite{rooke} (LEP 1)}}\\
An analysis of the hadronic final state was done,
in which the discrepancies between the data and predictions,
reported in {\bf OPAL 97a} \cite{opal2}, were examined from the point
of view of the number of produced jets. The data for
$Q^2\approx 6-30$ GeV$^2$ taken in the years 1994-95 were compared
to the results of the HERWIG, PYTHIA and F2GEN generators.\\
\vspace*{8.8cm}

\begin{figure}[ht]
\vskip 0.cm\relax\noindent\hskip 1.5cm
       \relax{\includegraphics{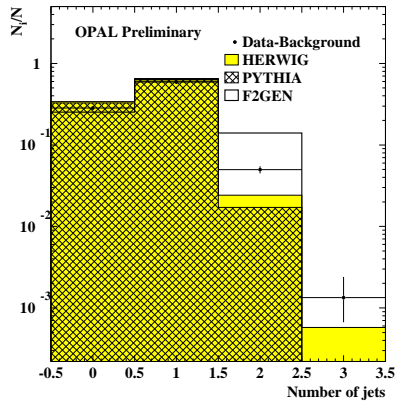}}
\vspace{0.cm}
\caption{\small\sl The fraction of events with 0-3 jets.
The points show the data with the background subtracted, with
statistical errors; histograms obtained with HERWIG, PYTHIA
and F2GEN generators
(from \cite{rooke}).}
\label{fig:LP1592}
\end{figure}

The numbers of events for different number of jets, divided by
the sum of all events, are
presented in Fig.~\ref{fig:LP1592}.\\
\vspace*{4.3cm}

\begin{figure}[hc]
\vskip 0.cm\relax\noindent\hskip 1.5cm
       \relax{\includegraphics{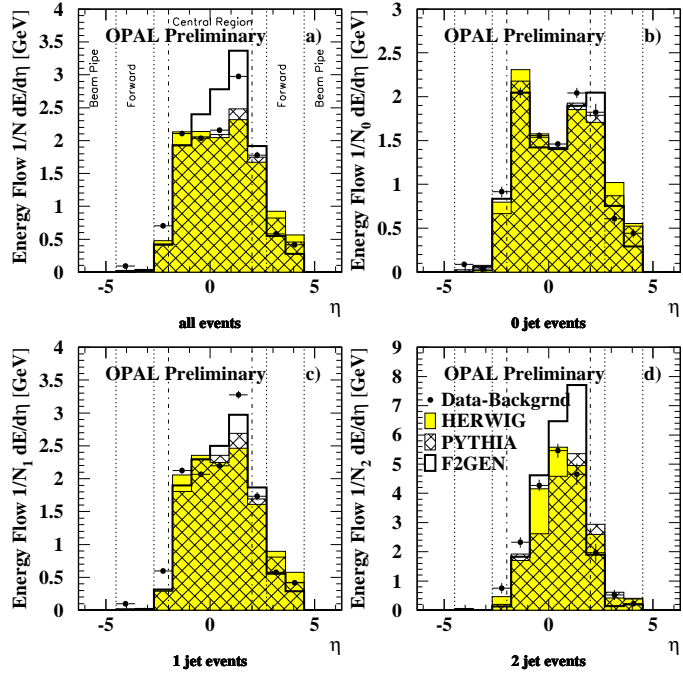}}
\vspace{0.cm}
\caption{\small\sl The hadronic energy flow per event in bins of 
the pseudorapidity $\eta$. Plots a), b), c) and d) show the average 
energy flow per event for summed events, and for events with 0, 1 
and 2 jets, respectively. The samples are represented as in 
Fig.~\ref{fig:LP1592} (from \cite{rooke}).}
\label{fig:LP1594}
\end{figure}

The results of further studies of the energy flow versus 
the pseudorapidity $\eta$ for events with different number of jets 
are shown in Fig.~\ref{fig:LP1594}.\\
\vspace*{7.4cm}

\begin{figure}[ht]
\vskip 0.cm\relax\noindent\hskip 1.5cm
       \relax{\includegraphics{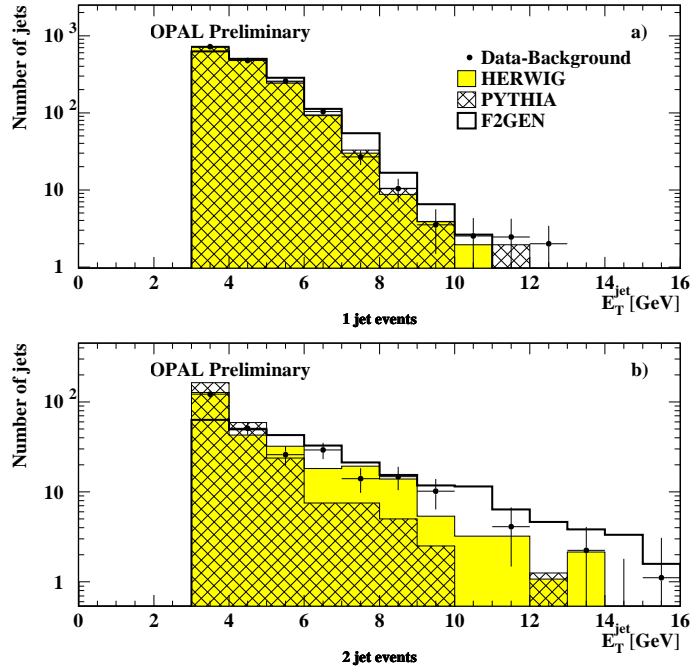}}
\vspace{-0.1cm}
\caption{\small\sl The number of a) 1-jet and b) 2-jet
events as a function of the jet transverse energy, 
$E_T^{jet}$. The HERWIG, PYTHIA and F2GEN events are 
normalized to the a) 1-jet and b) 2-jet data events 
(from \cite{rooke}).}
\label{fig:LP1597}
\end{figure}

The number of events versus the energy transverse to the tag plane, 
obtained by the different generators and observed in the experiment, 
are compared in the Fig.~\ref{fig:LP1597}.\\
\vspace*{8.2cm}

~\newline
Comment:{\it ``There is a marked difference between the data
and the Monte Carlo samples in the number of events with 
2 jets''. ``While the F2GEN sample, generated using the 
pointlike approximation, overestimates the 2 jet rate by a
factor 2.4, HERWIG and PYTHIA are too low by factors of 2.4
and 3.6 respectively.''\\
Disagreement between the data and each of generators 
for the hadronic energy flow versus the rapidity is seen for all
types of events: 0, 1 and 2-jet.\\   
``All of the Monte Carlo samples model well the $E_{t,out}$
distributions for events with 1 jet, but PYTHIA underestimates 
the $E_{t,out}$ of the events with 2 jets.''}\\
~\newline
$\bullet${\bf{OPAL 97e \cite{lauber} (LEP 1)}}\\
An analysis of OPAL data taken in 1993-95 was performed at 
$<Q^2>$ = 7.5 and 14.5 GeV$^2$. The possible improvement of 
the description of the transverse energy flow in DIS$_{\gamma}$ 
experiment by introduction of the additional intrinsic $k_t$ 
smearing, influencing the angular distributions of the hadronic final state
in the HERWIG, PYTHIA and ARIADNE generators
was studied for the first time.

\begin{figure}[ht]
\vskip 0.in\relax\noindent\hskip -2.3cm
       \relax{\includegraphics{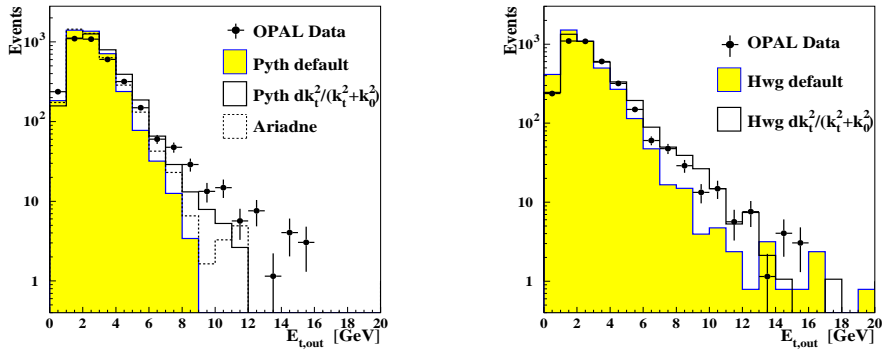}}
\vspace{0ex}
\caption{\small\sl The transverse energy out of the tag plane compared to 
PYTHIA (left) and HERWIG (right) simulations (from \cite{lauber}).}
\label{fig:had9}
\end{figure}

In  Fig.~\ref{fig:had9} the 
transverse energy out of the tag plane is presented
together with the results of the simulation by
 PYTHIA and HERWIG generators.
The energy flow as a function of $\eta$ was also studied 
for different $x$ bins and
$Q^2$ values (not shown).\\
\vspace*{4.cm}

~\newline 
Comment: {\it "Both the $E_{t,out}$ and the energy flow per event are 
greatly improved with the inclusion of the power-like distributions 
of the intrinsic $k_t$, with the exception of the peak in the energy 
flow at low $x_{vis}$ and $Q^2$ = 14.7 GeV$^2$, which still falls
short of the data." }
~\newline\newline  
$\bullet${\bf {TOPAZ 94 \cite{topaz} (TRISTAN)}}\\ 
The jet production (one and two jets) has been studied in the deep 
inelastic $e{\gamma}$ scattering
(3.0 GeV$^2<Q^2<$30 GeV$^2$), see also Sec.2.1.2.
 Events with 
the transverse momentum  of jets between 2 and 8 GeV and for 
the $|\eta|<$
0.7 were studied
using the jet cone algorithm with R=1 (see next section for the 
definition of the jets).

 The point-like
and hadron-like configurations resulting in the different final-state
topologies were studied.
~\newline
Comment: {\it The two high $p_T$ jet events are consistent with the point-like
perturbative part. "In the one - jet sample an excess over the point like 
component is observed, which is direct experimental evidence for the existence
of the hadron-like component in DIS $e\gamma$".}
%%%%%%%%%%%%%%%%%%%%%%%%%%%%%%%%%%%%%%%%%%%%%%%%%%
%
%
%%%%%%%%%%%%%%%%%%%%%%%%%%%%%%%%%%%%%%%%%%%%%%%%%%
\newpage
\subsection{Large $p_T$  processes in $\gamma \gamma$ 
and $\gamma p$ collision}
The structure function $F_2^{\gamma}$ considered 
 in Sec.2.1 is sensitive mainly to the 
combination of quark densities, 
and moreover, due to strong  dependence on the charge,
 to the $up$-type quark
distributions.
Therefore it is of great importance to have an additional and very different
from DIS$_{\gamma}$-experiments source of information. Hard 
processes involving resolved real photons provide such a source,
with the cross section being
a combination of the different parton densities convoluted with
the cross sections for basic partonic subprocesses (see below).
The role of these processes in determining \eg a gluon density in the photon
is unique indeed.

As we have already   
mentioned, the jet production in $\gamma \gamma$
collision leads to the complementary information on the photon 
structure to that coming from the $F^{\gamma}_2$;  
moreover, the study of the hadronic final state has become recently
a standard part of the structure function analyses 
performed  at LEP (see discussion in Secs. 2.1 and 2.2).   

In this section we focus on the large $p_T$ jet production
in  the resolved $\gamma \gamma$ and $\gamma p$ collisions, where
photon may interact through its partons.
The main goal here is to extract the individual parton density in the real
photon \footnote{Presently most of the data  deal with the almost real photon;
for the jet production in processes with virtual photon, see next section.}.
Beside the jet production also the inclusive one-particle production
 is  sensitive to the partonic content of photon,
but it  depends on the additional fragmentation 
functions, and  will not be discussed here 
\footnote{with the exception of the newest 
 charged particle production measurements by H1 group,
 where the gluon content has been derived.}.
Also, other hard
processes with resolved photon(s), like the Drell-Yan pair production
may be used  in pinning down   
the parton distributions in the photon.
The lack of data reflects much lower rates for these processes.

The production cross section of jets, which are
the hadronic representations of the
hard partons produced in the basic subprocess,
does depend however on the applied jet 
definition. 
Although this introduces additional 
uncertainty in the description of data, 
relatively high rate for these events  makes
large $p_T$ jet production, apart from the $F^{\gamma}_2$,
the basic source of information
on the partonic structure of $\gamma$.

In the previous section we have discussed single tagged events
with (positive) mass of the photons  
radiated by the tagged electron larger or much larger
than 1 GeV (with exception of one experiment {\bf DELPHI 95}). 
In this section the initial photons may be considered
\underline{real}, as the transverse momenta of the observed jets
are chosen to be much larger than the median mass of the photon(s).

At present, the $\gamma \gamma$ collisions arising
in $e^+e^-$ colliders in no-tag \footnote{The name ``untagged'' is also used.
In some cases antitagged
or single - tagged events in the large $p_T$ jet production
are discussed in ${\gamma}{\gamma}$ and ${\gamma}p$ collisions as well
(see below).}
conditions correspond to the
$\gamma \gamma$ events where  (both) real photons can be described by
 the Weizs\"{a}cker - 
Williams  energy spectra. In the OPAL experiment,
the squared target photon masses
were
estimated to be $<P_{1,2}^2>$=0.06 \g2, 
in the ALEPH experiment around 0.23 \g2. 
In the $\gamma p$ scattering at $ep$ collider 
HERA similar Weizs\"{a}cker - 
Williams  spectrum 
describes the flux of photons coming from the electron
(at HERA at present, \ie in 1996/7, the positron) 
- here $P^2<4$ GeV$^2$ (but with the median 0.001~GeV$^2$) 
or below 0.01-0.02 GeV$^2$,
if the dedicated  detector is used.

\subsubsection{General framework}

Resolved photon processes with  large $p_T$ jets 
can be characterized by the scale of hardness  $\tilde{Q}^2$
which is usually provided by the $p_T$
of the final jets, \ie $\tilde Q^2 \sim p_T^2$.
 (We introduce here the notation
$\tilde{Q}^2$  in order to distinguish 
it from the DIS scale $Q^2$, which is equal to the virtuality 
of the  photon probe.) 
In order to resolve both photons
 the scale $\tilde{Q}^2$ should be much larger than 
$P^2$ --  the virtuality
of the most virtual photon in the process, and moreover 
much larger than $\Lambda_{QCD}^2$ in order to apply the
perturbative QCD. 
(Events where the masses fulfill the relation  
$\Lambda_{QCD}^2\ll P^2 \ll \tilde{Q^2}$
are discussed in the next section,
where the concept and data on the structure of 
\underbar{virtual} photon will be  introduced.) 

\begin{figure}[ht]
\vskip 0cm\relax\noindent\hskip -1.cm
       \relax{\includegraphics{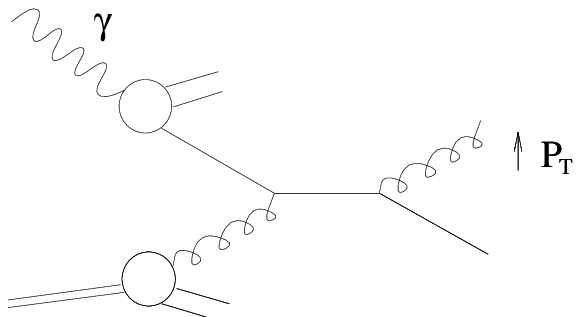}}
\vskip 0.cm\relax\noindent\hskip 6cm
       \relax{\includegraphics{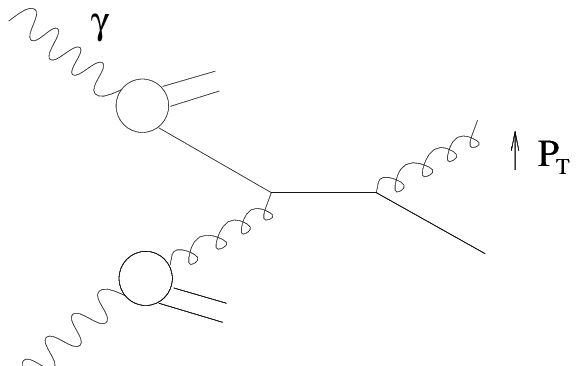}}
\vspace{0ex}
\caption{\small\sl Inclusive large 
$p_T$ jet production in the resolved 
${\gamma}p$ collision (a) and in the (double) resolved
$\gamma\gamma$ collision (b).}
\label{fig:a2}
\end{figure}

In this section initial photon(s) are considered to be real.
Assuming the factorization 
between the hard subprocess and parton densities,
the generic LO cross section 
for the two jet production in  ${\gamma}p$ collision 
(Fig.~\ref{fig:a2}a) is given by 
\be
d\sigma^{{\gamma}p\rightarrow {jet_1}~{jet_2}~X}
=\sum_{i,j}\int \int dx_{\gamma}
dx_{p}f_{i/{\gamma}}(x_{\gamma},\tilde{Q}^2)f_{j/p}(x_p,\tilde{Q}^2) 
\widehat{{\sigma}}^{ij
\rightarrow {jet_1}~{jet_2}} \label{twojets},   
\ee
where $f_{i/{\gamma}}(x_{\gamma},\tilde{Q}^2)$ describes the probability in
the  LL approximation
 of finding a parton of a type ''$i$'' in the photon. 
The $x_{\gamma}(x_p)$ variable is by definition the part 
of the four momentum of the photon (proton)
carried by its  parton. Scale $\tilde{Q}^2$ 
is  taken usually to be equal to  the so called factorization scale.
Note that the precise definition of the  factorization  scale 
can be given in 
 the QCD calculation going beyond the LO approach.

 For the 
two-jet production in ${\gamma}{\gamma}$ collision,
${\gamma}{\gamma} \rightarrow {jet_1}~{jet_2}~X$ 
(Fig.~\ref{fig:a2}b) a similar formula holds: 
\be
d\sigma^{{\gamma} \gamma \rightarrow {jet_1}~{jet_2}~X}
=\sum_{i,j}\int \int dx_{\gamma}^1
dx_{\gamma}^2f_{i/{\gamma}}(x_{\gamma}^1,\tilde{Q}^2)f_{j/{\gamma}}(x_{\gamma}^2,\tilde{Q}^2) 
\widehat{{\sigma}}^{ij
\rightarrow {jet_1}~{jet_2}} \label{twojet}.   
\ee
\vspace*{2.6cm}

Note that, contrary to the DIS$_{\gamma}$ experiments, 
in the resolved photon processes there is no one basic observable,
analogous  to the $F^{\gamma}_2$. 
Usually the $p_T$ or $E_T$  and  pseudorapidity $\eta$
distributions are compared with the QCD prediction using a specific
parton parametrization \footnote{See the collection presented in Appendix.}
and the agreement 
justifies {\it a posteriori}
the correctness of the applied  parton parametrization
and the QCD calculation.
Only in few cases the 
parton distribution is extracted from data (see below).

It is worth noticing the difference between the variables 
describing the produced parton and the corresponding quantities
for the jet, representing 
the considered  parton on the hadronic level. 
The    unfolding procedure is needed in order to
obtain the "true" partonic variables,
\eg transverse momentum of the parton
($p_t$) from transverse momentum or energy 
of the jet ($p_T$ or $E_T$).

\subsubsection{Measurements of the resolved photon processes} 

For the  jet production processes 
where at the Born level (\ie in the LO)
two jets appear, the measurements
of basic distributions in  $E_T$ or the \psr 
$\eta$ are performed 
for single or double jet (dijets) events
as a rule. 

For comparison with  QCD not only the above distributions 
are important but also the
study of the structure of the jet (jet profile or transverse energy flow 
 around the jet axis) and the energy of the underlying event, 
 or the hadronic activity outside the jets
where effects due to accompanying remnant jets may be seen. 

For more detailed study of properties of the resolved photon
processes the separation of the
direct photon events,
where the photon participates directly in the hard subprocess,
 $x_{\gamma}\sim 1$, and 
    resolved photon events
is needed.  Strictly speaking
this separation holds only in the LO
approach. In  the two jet events,
$x_p$ and $x_{\gamma}$ ( or $x_{\gamma}^{1,2}$)
 distributions, sensitive to parton densities,
can be reconstructed, see \eg {\bf ZEUS 95b, 97b}.
The angular distribution (\eg in $\gamma\gamma$ CM system) 
$d\sigma/d\cos\theta^*$, on the 
other hand, is not sensitive to these ingredients.
So this measurement may help to verify the expectation of  
different angular distributions which correspond to
 direct and resolved photon
contributions (\ie due to different subprocesses),
see {\bf ZEUS 96a} for the first results. 

The theoretical predictions based  on the LO 
\cite{l16} - \cite{l1} or NLO(HO)
QCD calculations are available for the inclusive one  and two jet
production for $e^+e^-$ and $ep$ experiments
\cite{l2} - \cite{kkk}. The main MC generators used by the
experimental groups for these kinds of processes are PYTHIA,
 HERWIG and PHOJET (with or without the multiple parton interaction).
 They allow to study different types
of distributions of the final state hadrons.

The following two jet definitions are used in the analyses
reported below. 
The jet cone algorithm with the fixed value of 
the cone variable $R$, defined as $R=\sqrt{(\delta \phi)^2+(\delta\eta)^2}$,
with  $\delta\phi\ 
(\delta \eta)$ 
describing 
the differences between the cone (jet) 
axis and the particle direction in the pseudorapidity and azimuthal angle,
 is  used in PUCELL and EUCELL generators.
The second one corresponds to 
the $k_T$-cluster algorithm on which the KTCLUS generator
is built up.
Beside  $R$ also $R_{sep}$ is included in some analyses,
which corresponds to the additional separation between partons
- note the $R_{sep}=2 ~R$ means no restriction. 
(The discussion of jet definitions can be found in \cite{kkk}).

The basic distributions, \eg the jet $E_T$ distribution,
are in general in agreement with the expectation for both 
single and double
jet production. In the pseudorapidity dependence the discrepancy is
observed both for the jet rates and for the
transverse energy flow around the jet axis, especially for small $E_T$
and a small $x_{\gamma}$.
This is taken as a
hint that the multiple scattering may be important
in the photon induced processes.
Since the direct events should be free from such multiple
interactions, the $\gamma \gamma$ and $\gamma p$ collisions may
offer a unique laboratory to study this problem.  
The strong dependence on the 
proper choice of the jet definition and the parameters
like the cone size $R$, is observed recently in the jet production
at HERA. There is a  possible relation of this effect
to the problem of describing the
 underlying events observed both in $e^+e^-$  and in $ep$
collisions. 
%%%%%%%%%%%%%%%%%%%%%%%%%%%%%%%%%%%%%%%%%%%%%%%%%%
%
%
%%%%%%%%%%%%%%%%%%%%%%%%%%%%%%%%%%%%%%%%%%%%%%%%%%
\newpage
\subsubsection{Jet production in $\gamma \gamma$ collision} 

Here the results for the jet production in ${\gamma}{\gamma}$ collision, 
where  {\it one} or  {\it two} resolved photons may interact,
will be presented. 

As in previous section  we will denote 
the squared (positive) virtualities of two involved photons by $P^2_1$
and $P^2_2$. 
In $e^+e^-$ machine parton momentum fractions -  variables 
$x_{\gamma}^{1,2}=x_{\gamma}^{\pm}$ - are determined from the two final jets 
with the highest $E_T$ according to the formulae: 
\begin{eqnarray}
x_{\gamma}^{\pm}={{{\Sigma}_{jets}(E\pm p_z)}\over{{\Sigma}_{hadrons}
(E\pm p_z)}}.
\label{xpm}
\end{eqnarray}
\newline\newline
\centerline{\huge DATA}
\newline\newline
$\bullet${\bf{DELPHI 94 \cite{delphi94} (LEP 1) }}\\ 
%z.p.C 62(94) 357\\
Data were collected in the 
period 1990-92 in antitagging conditions ($P^2_{max}\sim 0.12 $ 
\g2). Jets defined according to the Lund cluster algorithm 
with $p_T$ greater than 1.75 GeV were observed.

\begin{figure}[hb]
\vskip 0.in\relax\noindent\hskip 1.5cm
       \relax{\includegraphics{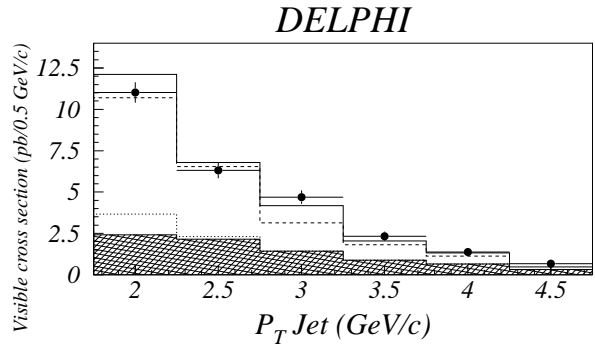}}
\vspace{0ex}
\caption{\small\sl Distribution of jet transverse momentum compared
with MC  predictions (TWOGAM generator). 
Dark area - QPM only; dots - QPM + VDM;
full line - QPM + VDM + DG ($p_T^{0}$ = 1.45 GeV/c);
dashed line - QPM + VDM + DO ($p_T^{0}$ = 1.22 GeV/c);
(from \cite{delphi94}).
}
\label{fig:jet3}
\end{figure}

Results for the $p_T$ of the jet distributions are shown in  
Fig.~\ref{fig:jet3} ($p_T^0$ 
has been introduced to separate the resolved photon contribution
 calculated using  the FPK approach with DG and DO parametrizations).
\vspace*{4.2cm}

\begin{figure}[hb]
\vskip 0.in\relax\noindent\hskip 1.5cm
       \relax{\includegraphics{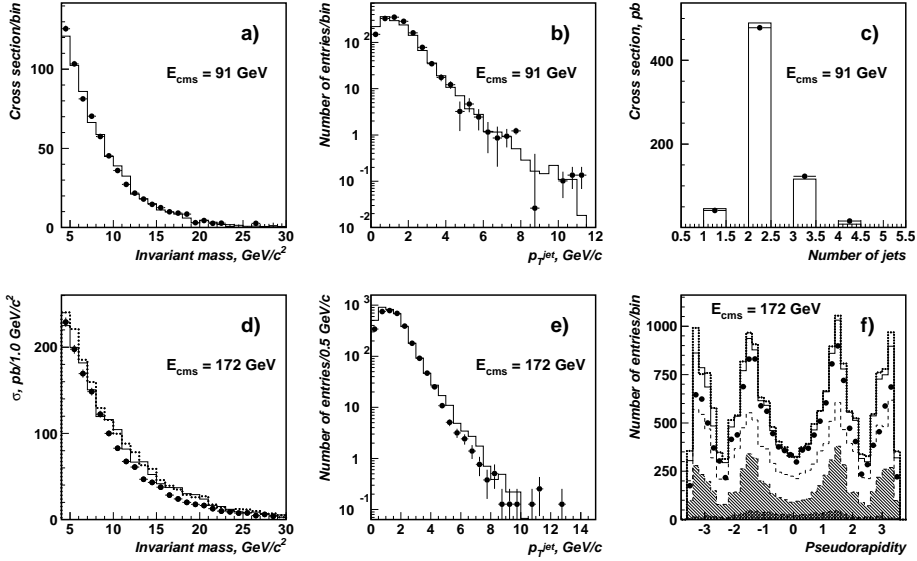}}
\vspace{0ex}
\caption{\small\sl Distributions of: a), d) - invariant mass,
b), e) - $p_T^{jets}$, c) number of reconstructed jets,
f) number of events versus pseudorapidity.
 Solid and dotted lines - VDM + QPM + RPC
with GS2 and GRV parton densities, respectively.
Dashed line - QCD - RPC contribution for GS2; hatched
histogram - VDM part, double hatched - QPM part
(from \cite{zimin}).}
\label{fig:jet4}
\end{figure}

~\newline
$\bullet${\bf{DELPHI 95 \cite{delphi3} (LEP 1) }}\\  
(See also previous section).
The values of  $E_T$ of observed jets were larger than  1.5 GeV, 
while the magnitude of the mass of virtual
photon was equal to $<P_1^2>\approx$ 0.06 GeV$^2$, 
\ie the standard  resolved (almost real) photon
process, from the point of view of both photons. 
~\newline\newline
Comment: {\it This is not typical DIS experiment since the 
photon probe
is almost real. The analysis is not typical, either, 
for the standard large $p_T$
jet study.}
~\newline\newline
$\bullet${\bf{DELPHI 97 \cite{zimin} (LEP 1 and 2) }}\\  
Based on the LEP1 and LEP2 data (1995 and 1996 runs), 
various distributions for the hadronic final state in no-tag events
were measured (Fig.~\ref{fig:jet4}).
\vspace*{7.2cm}

~\newline
Comment:{\it ``MC prediction (based on TWOGAM generator)
 gives perfect agreement 
at CM energy = 91 
\gev and slightly exceeds data at   172 \gev''.} 
~\newline\newline
$\bullet${\bf{OPAL 97f \cite{as4} (LEP 1.5)}}\\    
%ZP C73 (1997)433\\
The inclusive one- and two-jet 
cross sections have been measured
at $\sqrt{s_{ee}}$ = 130 and 136 GeV (based on the 1995 run).
The anti-tagging condition was applied 
to one of the initial photons corresponding 
to maximum photon virtuality $P^2_{max}\approx$ 0.8 GeV$^2$. 
The cone jet finding algorithm with R=1 was 
used for the first time 
in photon - photon collisions at LEP.

The transverse energy of jets $E_T$ is taken to be larger than 3 \gev 
and the \psr lies  within $|\eta |<1.$ 
Analysis of the transverse energy flow around the jet direction
for all jets and in two-jet events  was performed.

\begin{figure}[ht]
\vskip 0.in\relax\noindent\hskip 1.cm
       \relax{\includegraphics{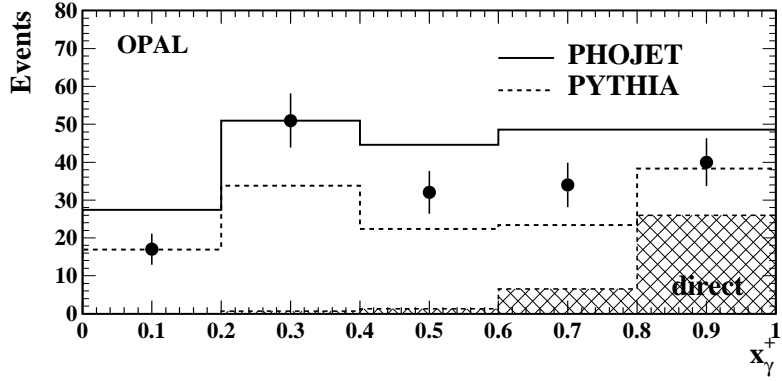}}
\vspace{0ex}
\caption{\small\sl The number of two-jet events as a function of
$x_{\gamma}^+$ compared to PHOJET (solid line) and PYTHIA
(dashed line) simulations. The hatched histogram is the direct contribution
to PYTHIA events (from \cite{as4}).}
\label{fig:jet5}
\end{figure}

\begin{figure}[ht]
\vskip 0.in\relax\noindent\hskip 1.5cm
       \relax{\includegraphics{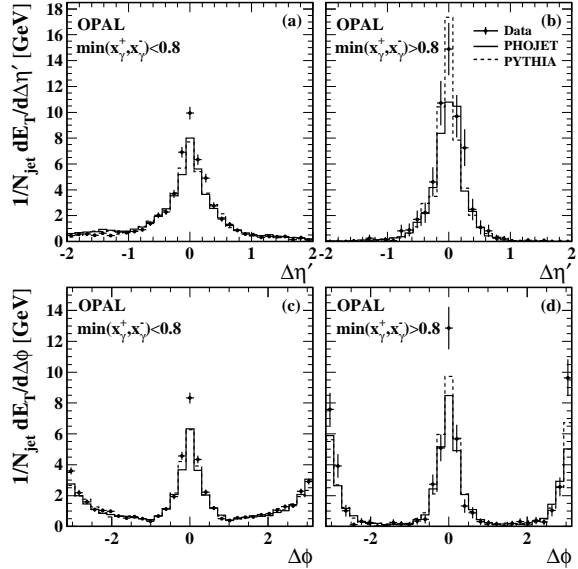}}
\vspace{0ex}
\caption{\small\sl The transverse energy flow around the jets in 
two-jet
events. The data are compared with the PHOJET (solid line)
and PYTHIA (dashed line) simulations (from \cite{as4}).}
\label{fig:jet6}
\end{figure}

For the two-photon events, the $x_{\gamma}^{\pm}$ distribution 
and the transverse energy flow around the jet, studied separately 
for the direct ($x_{\gamma}>0.8$) and the double resolved  
($x_{\gamma}<0.8$) photons  contributions, were obtained 
(see Fig.~\ref{fig:jet5} and Fig.~\ref{fig:jet6},  respectively).\\
\vspace*{4.cm}
\vspace*{6.7cm}

\begin{figure}[ht]
\vskip 0.in\relax\noindent\hskip 1.5cm
       \relax{\includegraphics{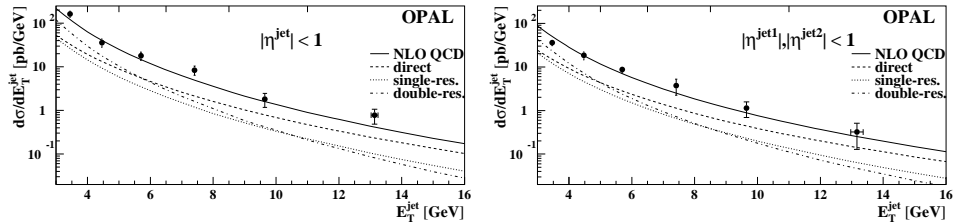}}
\vspace{0ex}
\caption{\small\sl The inclusive one-jet (a) and two-jet 
(b) cross section as a function of $E_T^{jet}$ for the jets with 
$|\eta^{jet}| < 1$ compared to the NLO calculations 
\cite{klek}. The solid line is the sum of direct,
single - resolved and double - resolved contributions shown
separately (from \cite{as4}).}
\label{fig:jet7}
\end{figure}

The one-jet and two-jet cross sections 
${{d{\sigma}}\over {dE^{jet}_T}}$ 
and ${{d{\sigma}}\over{ d{\eta}^{jet}}}$ 
were measured up to $E_T^{jet}$= 16 GeV, 
extending the previous measurement at TRISTAN. 
Results are presented in Fig.~\ref{fig:jet7}.\\
\vspace*{2.1cm}

\begin{figure}[ht]
\vskip 0.in\relax\noindent\hskip 1.5cm
       \relax{\includegraphics{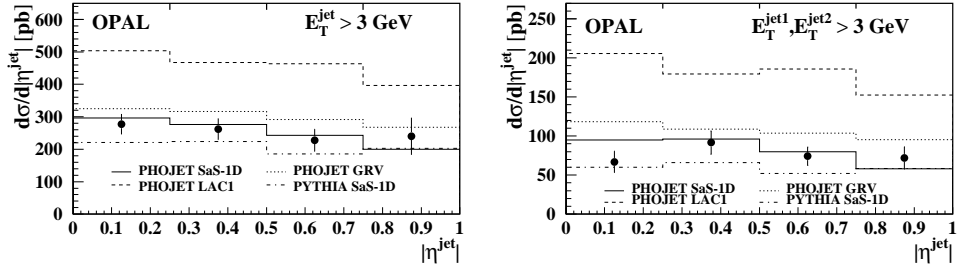}}
\vspace{0ex}
\caption{\sl The inclusive one-jet (a) and two-jet (b)
cross section as a function of $E_T^{jet}$ for the jets with
$E_T^{jet}\ >$ 3 GeV compared to the LO QCD calculations
of PYTHIA and PHOJET generators (from \cite{as4}).}
\label{fig:jet8}
\end{figure}

The $\eta$ distributions for one and two jet events were studied 
as well (see Fig.~\ref{fig:jet8}).\\
\vspace*{2.5cm}

~\newline
Comment: {\it "The data on  $\eta$ distributions  agree well with NLO QCD
calculations based on GRV parametrization.
The GRV-LO and SaS 1D parametrizations describe the data equally well,
the LAC 1, however, gives twice the observed value".}
~\newline\newline
$\bullet${\bf{OPAL 97g  \cite{burgin} (LEP 2)}}\\                    
%burgin\\
The dijet production in two  photon collisions at $e^+e^-$ energy 
161-172 \gev for the average
squared virtuality  $P^2=0.06$ \g2 was measured.
The transverse energy of jets is taken to be  $E_T>3$ \gev 
and the \psr lies within $|\eta |<2.$ 
The cone jet finding algorithm with R=1 was used. 

\begin{figure}[hb]
\vskip 0.in\relax\noindent\hskip 1.5cm
       \relax{\includegraphics{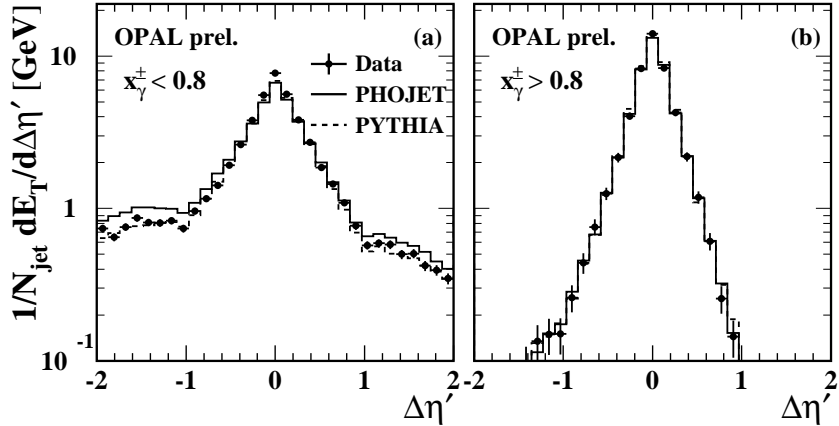}}
\vspace{0ex}
\caption{\small\sl The transverse energy flow around the jet axis
in two - jet events; a) double - resolved events, (b) direct events
(from \cite{burgin}).}
\label{fig:jet9}
\end{figure}

The direct and  resolved photon subprocesses 
($x_{\gamma}^{\pm}\ ^>_<$ 0.8)
were studied. 
The transverse energy flow around the jet axis is presented in 
Fig.~\ref{fig:jet9}
for the double resolved (a) and direct (b) photon processes.
Larger hadronic activity for the resolved photon sample
is observed as expected.\\
\vspace*{4.5cm}

\begin{figure}[ht]
\vskip 0.in\relax\noindent\hskip 1.5cm
       \relax{\includegraphics{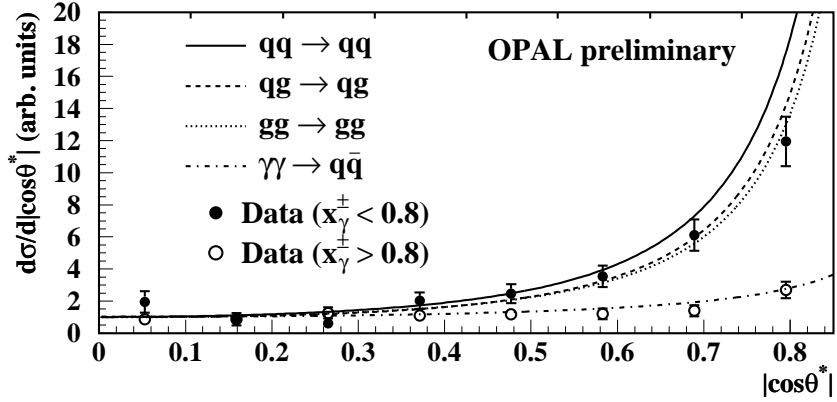}}
\vspace{0ex}
\caption{\small\sl The angular distribution of events separated into 
the direct and the double-resolved contributions, together with the 
QCD expectations (from \cite{burgin}).}
\label{fig:jet10}
\end{figure}

The angular dependence plotted separately for the three basic resolved
photon subprocesses and for the direct events 
is presented in Fig.~\ref{fig:jet10}.\\
\vspace*{4.3cm}

\begin{figure}[ht]
\vskip 0.in\relax\noindent\hskip 1.5cm
       \relax{\includegraphics{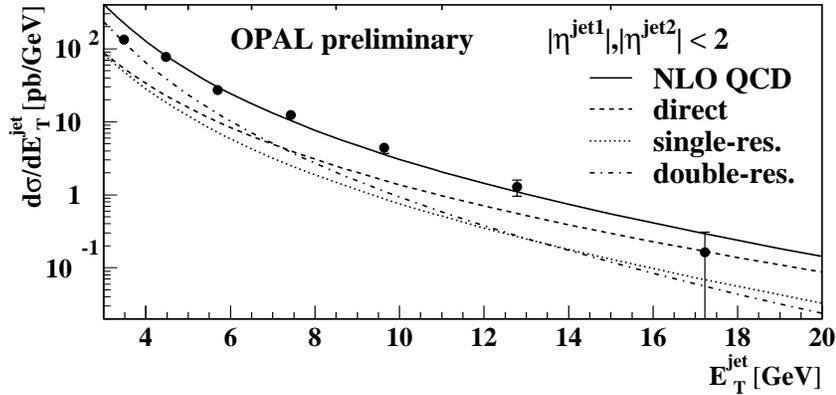}}
\vspace{0ex}
\caption{\small\sl The inclusive two-jet cross section as a function of
$E_T^{jet}$ for jets with $|\eta^{jet}|\ <$ 2, compared to the
NLO calculations \cite{klek}. Solid line is the sum of the direct,
single-resolved and double-resolved cross sections shown
separately (from \cite{burgin}). }
\label{fig:jet11}
\end{figure}

The cross section for the direct, single- resolved and double-resolved
two-jet events versus $E_T$
compared to the NLO calculation with the GRV-HO parton parametrization  
is shown in Fig.~\ref{fig:jet11}.\\
\vspace*{4.2cm}

\begin{figure}[ht]
\vskip 0.in\relax\noindent\hskip 1.5cm
       \relax{\includegraphics{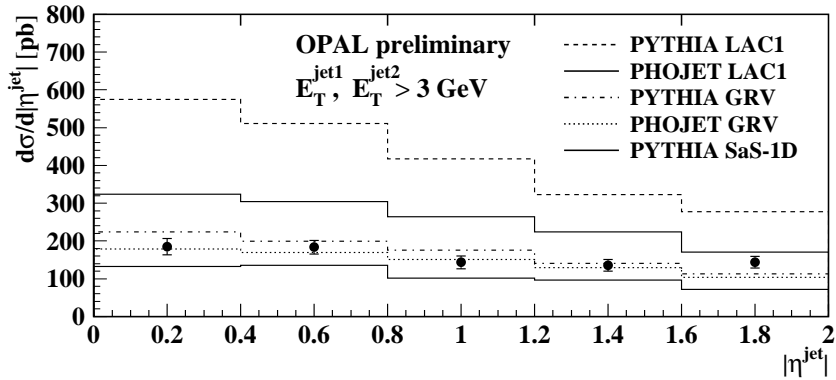}}
\vspace{0ex}
\caption{\small\sl The inclusive two-jet cross sections as a function 
of $|\eta^{jet}|$ for jets with $E_T^{jet}\ >$ 3 GeV
(from \cite{burgin}).}
\label{fig:jet12}
\end{figure}

The jet pseudorapidity  distribution  is presented in 
Fig.~\ref{fig:jet12}    
together with the predictions from PYTHIA and PHOJET generators,
with various parton parametrizations.
\newpage
\vspace*{4.4cm}

\begin{figure}[ht]
\vskip 0.in\relax\noindent\hskip 1.cm
       \relax{\includegraphics{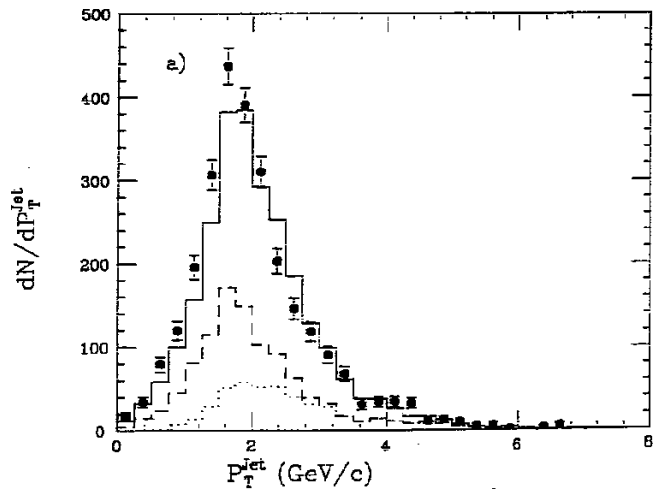}}
\vspace{0ex}
\caption{\small\sl The experimental $p_T^{jet}$ distribution compared 
with the predictions of QPM (dotted histogram), QPM + VMD
(dashed histogram) and QPM + VMD + MJET (solid histogram)
with $p_T^{0}$ = 1.6 GeV
(from \cite{amy3}).}
\label{fig:jet13}
\end{figure}

~\newline
Comment: {\it "The $E_T$ dependent two-jet cross section is in good agreement 
with NLO QCD calculation. The GRV-LO and SaS-1D parametrizations describe 
the inclusive two-jet cross section equally well.
The LAC1 parametrization overestimates the inclusive two-jet cross section 
significantly".}
~\newline\newline
$\bullet${\bf{AMY 92 \cite{amy3} (TRISTAN)}}\\
The measurement of the high $p_T$ hadron production in the quasi-real
${\gamma} {\gamma}$ collision was performed (for the energy between 
55 and 61.4 GeV).
In the observed 3- and 4-jet events there are one
or two  spectator jets coming from the
resolved photon(s). 
Two paramerizations of parton density in the photon were used: 
DG and (DO + VDM)
(Fig.~\ref{fig:jet13})
( The $p_T^0$ has been introduced to separate the QPM+VDM
 contributions from the "MJET" one, which corresponds to 
the RPC contributions).\\
\vspace*{5.6cm}

\begin{figure}[ht]
\vskip 0.in\relax\noindent\hskip 1.cm
       \relax{\includegraphics{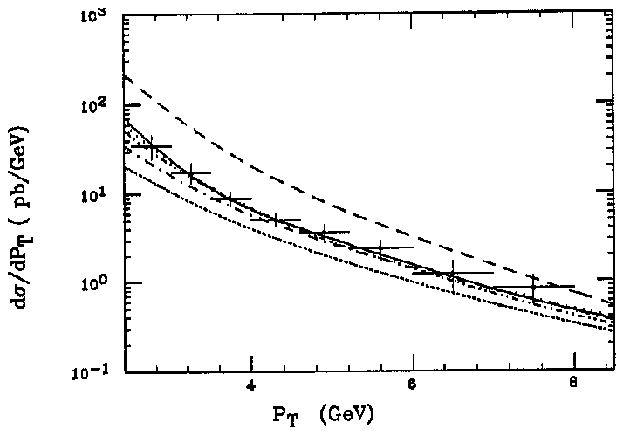}}
\vskip 0.in\relax\noindent\hskip 1.cm
       \relax{\includegraphics{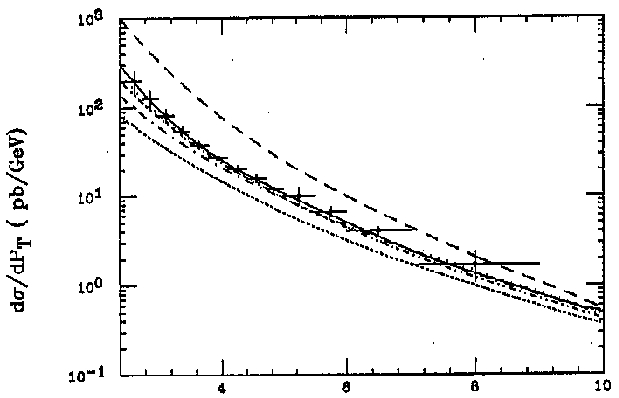}}
\vspace{0ex}
\caption{\small\sl The inclusive jet 
cross section as a function of $p_T$
integrated over $|\eta|\ <$ 1.0 for one-jet (a) and
two-jet events (b). The curves represent the sum of QPM
(direct) and MJET (resolved) cross sections using the LAC1
(full line), GRV-LO (double dot - dashed), DG (dotted line),
LAC3 (dashed line) and LAC1 without the gluon component
(dot - dashed line). The short - dashed curve corresponds 
to the QPM cross section (from \cite{as2}).}
\label{fig:jet14}
\end{figure}

~\newline
Comment: {\it The evidence for hard scattering of hadronic 
constituents of photons in photon - photon collisions at TRISTAN.
``.. without the gluonic component it is impossible
to reproduce the data''.}
~\newline\newline
$\bullet${\bf {AMY 1.5 94 \cite{as2} (TRISTAN)}}\\
AMY 1.5 (the upgraded AMY detector at the energy 60 GeV) measures
the inclusive single- and 
double-jet cross section ${{d{\sigma}}\over{ dp_T}}$ 
in the range 2.63 GeV$<p_T<$ 8.00 GeV (single jet)
and 2.75 GeV$<p_T<$7.50 GeV (two jets).
The cone algorithm with R=1 was used.
In Fig.~\ref{fig:jet14} the $p_T$ 
distributions for one and two jets for $|\eta|< 1.0$,
are presented.
\vspace*{3.3cm}

\begin{figure}[ht]
\vskip 0.in\relax\noindent\hskip 1.cm
       \relax{\includegraphics{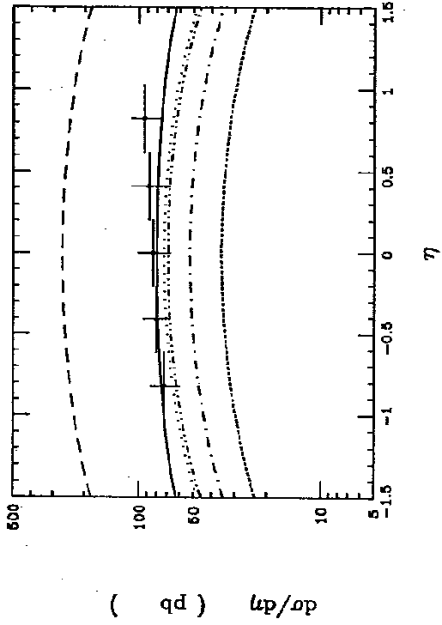}}
\vspace{0ex}
\caption{\small\sl The inclusive jet cross section as a function of
$\eta$ integrated over $|p_T|\geq$ 2.5 GeV. The lines
convention is the same as in Fig.~\ref{fig:jet14}
(from \cite{as2}).}
\label{fig:jet15}
\end{figure}

The $\eta$ dependence for the jet with $p_T$ larger than 2.5 
is presented in Fig.~\ref{fig:jet15}.\\
\vspace*{4.8cm}

~\newline
Comment: {\it ``The data are in good agreement with LO QCD 
calculations based on either
LAC 1, DG or GRV parametrizations of the parton densities in the photon.
The calculation based on LAC 3 disagrees with the data
and we do not see the deviation that is observed in the 
$\eta$ distribution by the H1 experiment'' (here denoted as {\bf H1 93}).}
~\newline\newline
$\bullet${\bf {TOPAZ 93 \cite{as3} (TRISTAN)}}\\
The inclusive jet cross section ${{d{\sigma}}\over {dp_T}}$ was measured 
at the energy of the collision $\sqrt{s_{ee}}$=58 GeV for $<p_T>$
from 2.61 to 7.44 GeV (single jet) and from 2.71 to 7.43 GeV 
for two-jet 
events.  The cone algorithm with $R=1$ was used.

The energy flow within the jet as a function of
 $\delta \eta$ and $\delta \phi$ was also studied
and found in agreement with the MC simulation (not shown). 

\begin{figure}[ht]
\vskip 0.in\relax\noindent\hskip 1cm
       \relax{\includegraphics{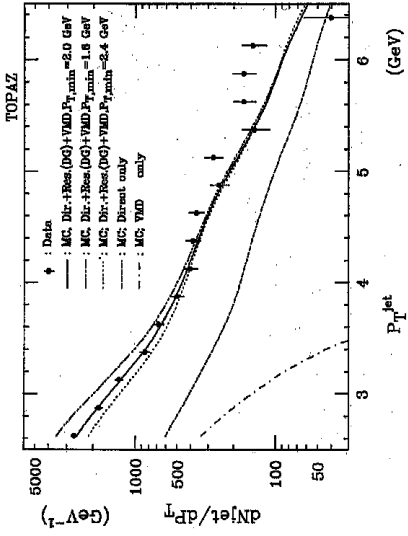}}
\vspace{0ex}
\caption{\small\sl The jet $p_T$ distribution for a no - tag sample.
Predictions are shown for the sum of
direct + resolved (DG) + VDM processes for different $p_T^0$
values. Contributions of the direct
(dashed line) and VDM (dot - dashed line) processes are shown
(from \cite{as3}).}
\label{fig:jet16}
\end{figure}

\begin{figure}[ht]
\vskip 0.in\relax\noindent\hskip 1cm
       \relax{\includegraphics{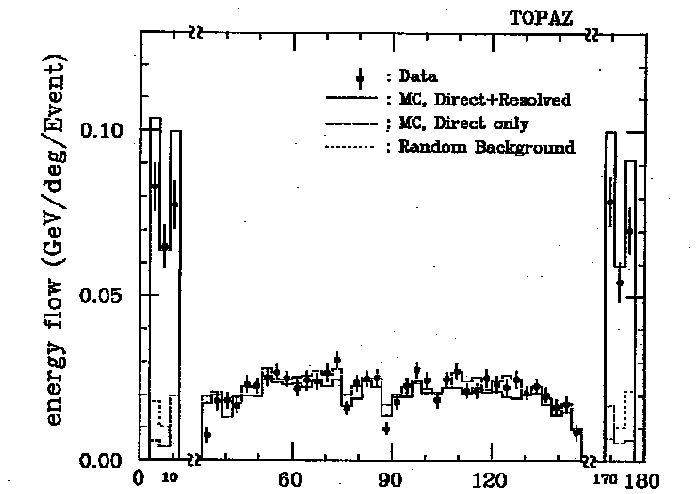}}
\vspace{0ex}
\caption{\small\sl The energy flow versus the polar angle $\theta$ 
for the jet sample. The data are compared with the MC predictions 
for the sum of the direct and the resolved processes (solid histogram)
and the direct process only (dashed histogram)
(from \cite{as3}).}
\label{fig:jet17}
\end{figure}

\begin{figure}[ht]
\vskip 0.in\relax\noindent\hskip 3.5cm
       \relax{\includegraphics{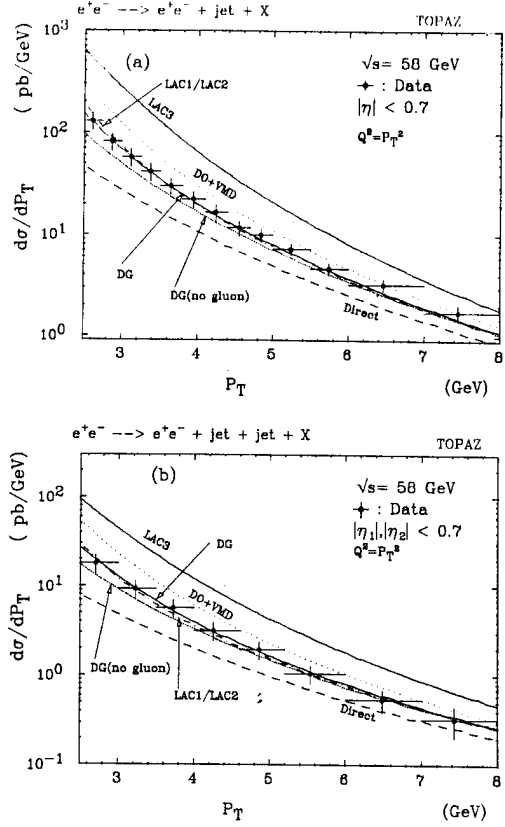}}
\vspace{0.ex}
\caption{ {\small\sl 
Inclusive a) jet and b) two-jet cross sections as 
a function of
 $p_T$ for the pseudorapidity $\mid {\eta}_{jet} \mid \leq0.7$.
TOPAZ Collaboration results from \cite{as3}. 
 }}
\label{fig:rys}
\end{figure}

In Figs.~\ref{fig:jet16} and  \ref{fig:jet17}  
the $p_T$ distribution for the jet and the
energy flow per event
as a function of the polar angle of the jet are presented, respectively.
Fig.~\ref{fig:rys} shows the inclusive one - jet and 
two - jet cross sections.
\vspace*{4.7cm}
\vspace*{4.5cm}
\vspace*{11cm}

~\newline
Comment: {\it 
The polar angle distribution gives ..''direct evidence
of the presence of the resolved processes.
The data exclude the parametrizations of LAC 3 and 
DO+ VMD,
which predict a very large gluon content even at large 
$x_{\gamma}$.''}
%%%%%%%%%%%%%%%%%%%%%%%%%%%%%%%%%%%%%%%%%%%%%%%%%%
%
%
%%%%%%%%%%%%%%%%%%%%%%%%%%%%%%%%%%%%%%%%%%%%%%%%%%
\newpage
\subsubsection{Jet production in the resolved $\gamma p$ scattering}

Here we present the data for the jet production in $\gamma p$ scattering
 taken at HERA collider at $\sqrt {s_{ep}} \sim 300$ GeV \footnote{
 The energy of the electron
was at the beginning of running of the HERA collider equal to 26.7 GeV, 
starting from 1994 it is 27.5 GeV, with the energy of the proton 
820 GeV.}, where only   {\it one}  photon may be resolved
\footnote{The $\gamma \gamma $ events leading to the large $p_T$
jets are rare at HERA.}.
The photoproduction events correspond here to the limit
of virtuality of the initial photon: $P^2 \sim 4$ GeV$^2$ (with median 0.001 
 GeV$^2$)
or (with a special condition) $\sim$ 0.01-0.02 \g2 (see below).
The partonic variables, both related 
to the initial and final states in the hard partonic subprocesses,
are reconstructed from the corresponding quantities 
for final state hadronic jets.
The relation between variables corresponding to  these two levels
depends on the order of the perturbative QCD calculation,
as it was mentioned before.  
%For example   $p_T$ we will denote the corresponding variable for the 
%paton whereas $E_T$ the transverse energy of jet, these two varables  
%should be equal in LO. Similar relation holds for the pair 
%$x_{\gamma}$ and $x_{\gamma}^{vis}$, etc.)

The  parton momentum fraction $x^{jet}_{\gamma}$
 (called also $x_{\gamma}^{vis}$ or 
$x_{\gamma}^{obs}$) is in the LO case  equal to $x_{\gamma}$.
 In practice
it is reconstructed 
 using the two jets  with the highest transverse energy $E_T^{jet}$ 
in the event,using the following relation:
\be
x^{jet}_{\gamma}=
{{E_T^{jet_1}e^{-{\eta}^{jet_1}}+E_T^{jet_2}e^{-{\eta}^{jet_2}}}\over
{2E_{\gamma}}},
\ee
where also jet pseudorapidities,
${\eta}^{jet}=-\ln (\tan(\theta/2))$, and
the energy of the photon $E_{\gamma}$ (= $yE_e$) enter. 
(The positive pseudorapidity
corresponds to the proton direction.)
The scaled energy $y$ of the initial photon (Eq.~(\ref{xy})) 
is measured from the transverse energy $E_T^h$ and pseudorapidity  
of hadrons $\eta^h$ according to the 
 formula
\be
y={{1}\over{2E_e}}\sum_h E^h_Te^{-\eta^h},
\ee
where the sum is over produced hadrons.\newline

In the following discussion the pseudorapidity of the jet 
in the laboratory system is  denoted  by
$\eta$, whereas  $\eta^{*}$ means the corresponding 
variable in the $\gamma p$ CM system.
For the HERA collider typically  $\eta-\eta^{*}  \sim 2$. 
  The difference of the two jets' transverse energy 
  $E_{T\,1}$-$E_{T\,2}$, $\Delta E_T$,
and  analogous difference  for the pseudorapidities, 
$\Delta \eta=|\eta_1-\eta_2|$, are introduced in two-jet events.
(For the analysis of the energy of underlying event the variable
$\delta \eta=\eta_{cell}-\eta$ is used as 
in the $\gamma\gamma$ case.) In the analysis of 
two jet events there appear also the average pseudorapidity 
$\bar \eta=(\eta_1+\eta_2)/2$ and the average 
transverse energy of jets, $\bar {E_T}$.

Below, $\phi$ denotes the azimuthal angle of the particle. 
In the cone jet algorithm, beside $R$, also $R_{sep}$ is introduced both 
in the theoretical calculations and in experimental analysis
of the jet production at HERA.

Note also that the multiple interaction included in the experimental analysis
is modeled as  in the $p {\bar p}$ experiments 
(see \eg {\bf H1 96a}).\\
\newline\newline
\centerline{\bf \huge DATA}
\newline\newline
$\bullet${\bf {H1 92 \cite{h192} (HERA)  }}\\
%T.Ahmed  PL B297 (1992) 205 \\
The evidence for the   hard photoproduction of jet with   
$E_T > 10 $ \gev is reported
(the jet cone algorithm with $R<1$ was used in the analysis).
\newline\newline
$\bullet${\bf {H1 93 \cite{h193} (HERA)  }}\\
%PL B 314 (1993) 436\\
First measurement  of the inclusive jet cross section 
at the $ep$ collider HERA (based on the 1992 data) 
is reported. The events with  
the scaled energy $y$ of the initial photon
between 0.25 and 0.7 and the 
photon virtuality $P^2$   smaller than 0.01 GeV$^2$ were collected.
The photoproduction of jet was studied   
for $E_T$   from 7 to 20 GeV and the \psr 
 ~interval $-1 <{\eta} < 1.5$. The jet cone algorithm with R=1
(PYTHIA 5.6) was used. 

\begin{figure}[ht] 
\vskip 0.in\relax\noindent\hskip 1.5cm
       \relax{\includegraphics{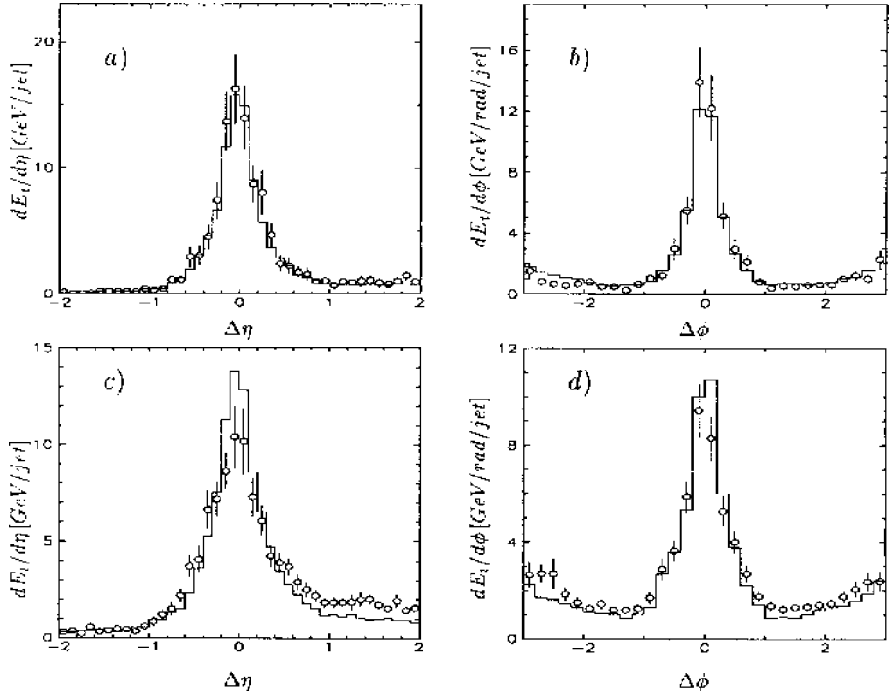}}
\vspace{0ex}
\caption{\small\sl The transverse energy flow as a function of 
$\delta\eta$ (integrated over $|\delta\phi |\ <$ 1.0) (a, c)
and as a function of $\delta\phi$ (integrated over 
$|\Delta\eta |\ <$ 1.0) (b, d). Figs. a) and b) correspond to
-1.0 $<\ \eta\ <$ 0.5, b) and c) to 0.5 $<\ \eta\ <$ 1.5
(from \cite{h193}).}
\label{fig:jet20}
\end{figure}

The transverse energy flow around the jet axis was studied 
 and the discrepency was found
in form of  too large averaged $E_t$ on the forward side of the jet
(see Fig.~\ref{fig:jet20}). 
Among others, multiple parton interactions were
mentioned as a possible  explanation of this effect.\\
\vspace*{8.3cm}

\begin{figure}[hb]
\vskip 0.in\relax\noindent\hskip 4.cm
       \relax{\includegraphics{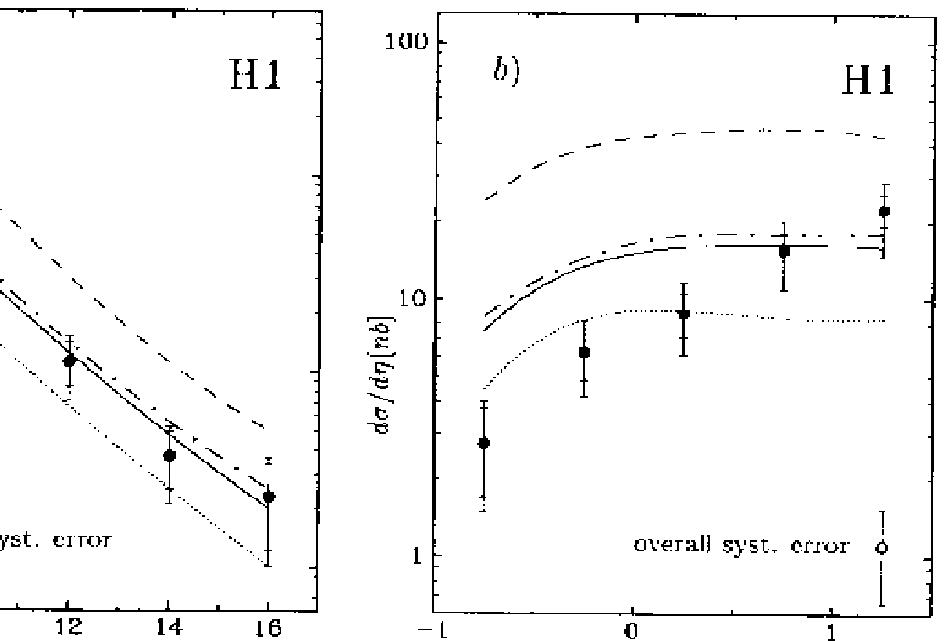}}
\vspace{0ex}
\caption{\small\sl Inclusive jet $E_T$ spectrum integrated over $\eta$
interval -1.0 $<\ \eta\ <$ 1.5 (a) and inclusive $\eta$
spectrum (b) for jets with $E_T\ >$ 7 GeV. LO QCD predictions generated by
the  PYTHIA generator using parametrizations LAC3 (dashed line),
LAC2 (dashed-dotted line), GRV-LO (full line) and GRV-LO (without
gluons, dotted line) (from \cite{h193}).}
\label{fig:jet21}
\end{figure}

The inclusive jet cross sections 
$d\sigma/dE_T$ versus $E_T$ and $d\sigma/d\eta$ as a function of 
$\eta$ integrated over the corresponding range of the $\eta$ and $E_T$, 
respectively, were measured  and compared with the LO calculation
using the following parton parametrizations: for the photon LAC2,
LAC3 and GRV-LO, and GRV for the proton (see Fig.~\ref{fig:jet21}).\\
\vspace*{5.7cm}

The shape of the $d\sigma/dE_T$ is well described in the range 
of $\eta$ between -1 and 1.5. It is a problem to describe
the $\eta$ distribution of jets.

~\newline
Comment: {\it 
"In the (pseudorapidity) range 0.5 to 1.5  the data show 
larger average values of the transverse energy flow
outside the jet cone on the forward side of the 
jet than predicted by the MC."\\
LAC3 gives cross section higher by factor 3 
than data for the  $E_T$ distribution of the jets.
"None of the models describe well the measured 
$\eta$ dependence (for jets)."}
~\newline\newline
$\bullet${\bf {H1 95 \cite{h1} (HERA)  }}\\
%NP B 445(95)195\\
The photoproduction of 2-jet events (the 1993 data) was studied  
for the jet $E_T$  range from 7 to 20 GeV and the \psr 
interval $0\le{\eta}\le2.5$  with $|\Delta \eta|\le1.2$
(between the most energetic jets) (see also {\bf H1 96a}). 
The scaled  energy $y$ of the initial photon
was between 0.25 and 0.7.
Photon virtuality $P^2$ was  smaller than 0.01 GeV$^2$.
The jet cone finding algorithm with R=1 
(and 0.7 for cross checks) was used 
(PYTHIA 5.6 with the GRV-LO for the proton).

For the first time the inclusive (LO) cross sections 
were derived for the  parton level and the gluon density 
in the photon was measured.

The transverse energy flow around the jet direction per event
versus the rapidity distance from the jet direction
was studied for 7 $\le E_T \le $ 8  GeV and  $ 0\le {\eta} \le 1$
and found to be asymmetric and different for the samples with 
$x_{\gamma}> 0.4$ and  $x_{\gamma} <0.4$ (not shown).

\begin{figure}[ht]
\vskip 0.in\relax\noindent\hskip 3.5cm
       \relax{\includegraphics{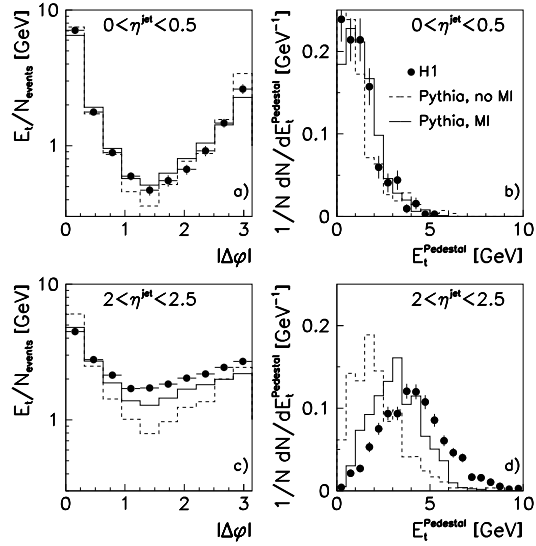}}
\vspace{-.4cm}
\caption{\small\sl (a, c) The transverse energy flow versus the azimuthal
angle with respect to the jet direction in two rapidity bins.
(b, d) Distributions of transverse energy measured outside of the 
jets. Histograms show PYTHIA simulations with (full line)
and without (dashed line) multiple interactions
(from \cite{h1}).}
\label{fig:jet22}
\end{figure}

The transverse  energy flow versus the azimuthal angle around 
the jet direction
 and the transverse energy of the underlying events 
outside the jets (here named $E_t^{pedestal}$, 
another name - the  underlying event)  was studied 
(see Fig.~\ref{fig:jet22}) with the conclusion
``the multiple interaction gives an improved description''.\\
\vspace*{6.5cm}

\begin{figure}[ht]
\vskip 0.in\relax\noindent\hskip 4.3cm
       \relax{\includegraphics{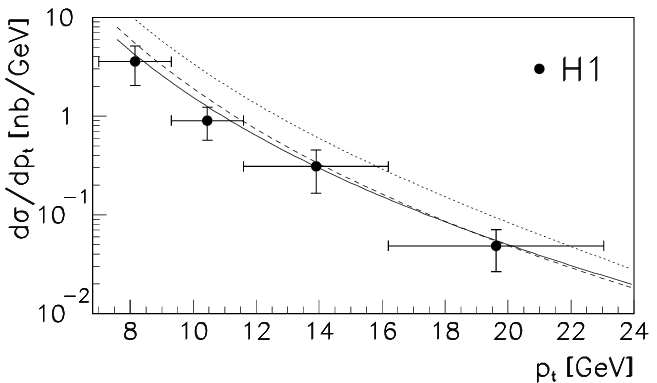}}
\vskip 3.5cm\relax\noindent\hskip 4.3cm
       \relax{\includegraphics{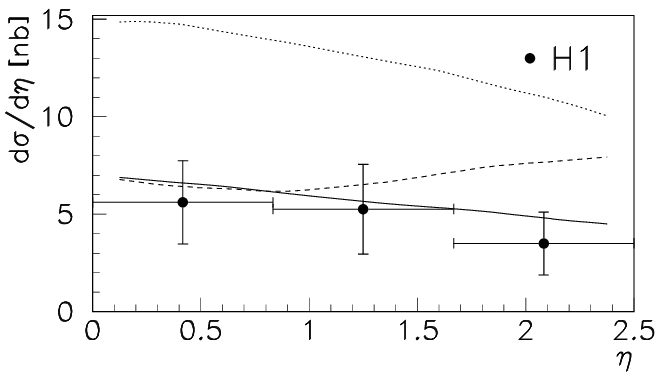}}
\vspace{-0.5cm}
\caption{\small\sl Single parton cross sections: (a) $d\sigma$/$dp_t$
integrated over the pseudorapidity range 0 $<\ \eta\ <$ 2.5,
(b) $d\sigma$/$d\eta$ for $p_t\ >$ 7 GeV. The solid line - the LO QCD
calculation with GRV LO parametrization for partons in the proton
and the photon. The dashed (dotted) line - the same for LAC1 (LAC3)
parametrizations for the photon (GRV-LO for the   proton)
(from \cite{h1}).} 
\label{fig:jet23}
\end{figure}

To achieve the goal which was here the extraction of the 
gluon distribution,
the single {\underline {parton}}  cross section $d\sigma/dp_t$ 
integrated over the parton rapidity range, as well as
the single {\underline {parton}}
 $d\sigma/d\eta$ were studied and compared with the LO
parametrizations GRV and LAC 1 and LAC 3
(see Fig.~\ref{fig:jet23}).\\
\vspace*{3.3cm}

\begin{figure}[ht]
\vskip 0.in\relax\noindent\hskip 4.2cm
       \relax{\includegraphics{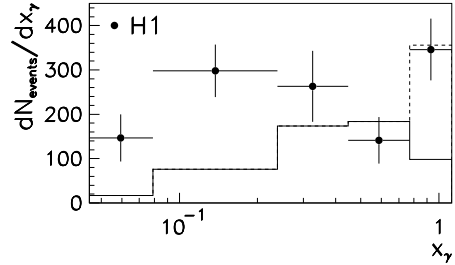}}
\vspace{-0.5cm}
\caption{\small\sl The two-jet event distribution of $x_{\gamma}$ of
the parton from the photon. The solid line - the contribution from 
the quark resolved photon processes; the dashed line - the direct 
photon contribution from the PYTHIA MC (from \cite{h1}).}
\label{fig:jet24}
\end{figure}

To extract the information on the subprocesses, 
the full two-jet kinematics was used and 
the distributions of the $\Delta \eta$, 
$\Delta E_T$, and of  the $x_{\gamma}$ and $x_{p}$ were  studied. 
In Fig.~\ref{fig:jet24} we present the distribution of 
the $x_{\gamma}$.\\
\vspace*{2.9cm}

\begin{figure}[ht]
\vskip -0.8in\relax\noindent\hskip 6.3cm
       \relax{\includegraphics{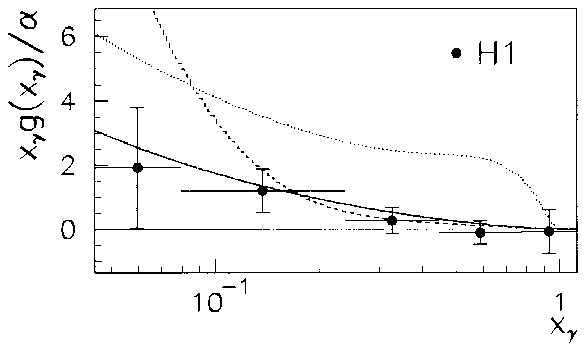}}
\vspace{0.ex}
\caption{\small\sl The gluon distribution 
extracted from the resolved photon processes at
$<$~$\tilde{Q}^2>=<p_T^2>=75~ GeV^2$. 
For comparison GRV-LO (full line), LAC1 (dashed) and LAC3 (dotted)
gluon parametrizations are shown (from \cite{h1}).}
\label{fig:mary13}
\end{figure}

The (LO) gluon distribution in the photon was derived,    
at the average factorization scale  $<\tilde{Q}^2>$=$<p_T^2>$=75 GeV$^2$ 
for 0.04$\leq x_{\gamma} \leq$1, see the table for numbers: 
$$
\begin{array}{|c|c|c|}
\hline
<\tilde{Q}^2>&<x_{\gamma}>&x_{\gamma}g(x_{\gamma})/{\alpha}\\
~[GeV^2]~&&(stat. + syst.) \\
\hline
75&~\,0.059&~~1.92\pm0.87\pm1.68\\
&0.14&~~1.19\pm0.34\pm0.59\\
&0.33&~~0.26\pm0.24\pm0.33\\
&0.59&-0.12\pm0.15\pm0.33\\
&0.93&-0.08\pm0.61\pm0.30\\ 
\hline
\end{array}
$$
The results on $x_{\gamma} g(x_{\gamma})/\alpha$ 
are presented in Fig.~\ref{fig:mary13}.\\
\vspace*{4.9cm}

~\newline
Comment: {\it
 "The multiple interaction option gives an improved description"
 of the jet profiles and pedestal distributions 
(still " deviation from data at large jet rapidities 2 $< \eta < $2.5"). 

In the extracting the gluon density
$q^{\gamma}$ was taken as 
determined by two-photon experiments at LEP and KEK, in 
the form given by the 
GRV-LO parametrization.
"A high gluon density at large parton momenta as suggested 
by the LAC 3 parametrizations is clearly
excluded. The strong rise of the LAC1 parametrization 
below $x_{\gamma}\le 0.08$ is not supported."}
\newline\newline   
$\bullet${\bf{H1 96a \cite{h196} (HERA)}}\\
% desy95-219 ZP C70 (96) 17\\
The single jet production with $E_T\geq 7$ \gev (and -1$<\eta<$2.5)
was  measured   in $ep$ collisions (data from 199?) 
with the scaled photon energy $0.25 < y<0.7$ and $P^2$ below 0.01 GeV$^2$.
The cone algorithm with R=1 was used.

\begin{figure}[ht]
\vskip 0.in\relax\noindent\hskip 3.7cm
       \relax{\includegraphics{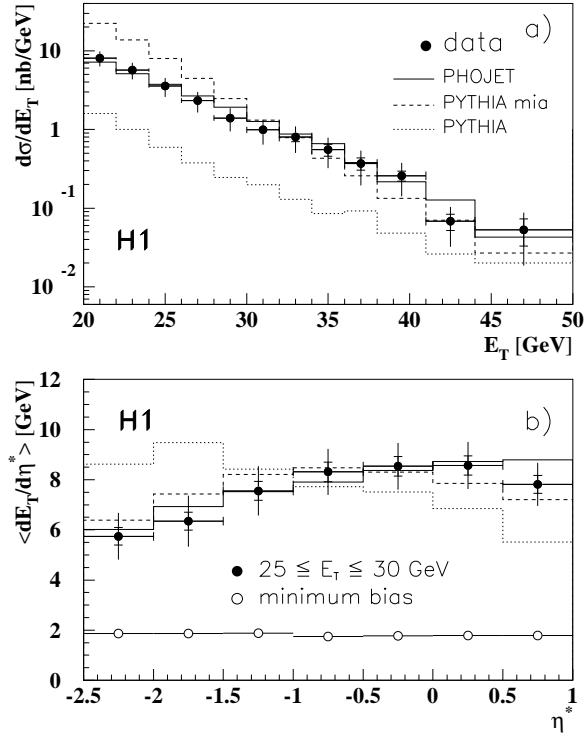}}
\vspace{-0.6cm}
\caption{\small\sl {a) The differential transverse energy cross 
section integrated over the pseudorapidity 
(-2.5 $\leq\eta^*\leq$ 1). Histograms are the simulations with
interactions of the beam remnants (full line - PHOJET,
dashed - PYTHIA) and without them (dotted line - PYTHIA).
b) The corrected transverse energy flow versus
$\eta^*$ ($\eta^*\ >$ 0 corresponds to the proton direction).
The pseudorapidity range and histograms as in a)
(from \cite{h196}).}}
\label{fig:jet26}
\end{figure}

The properties of the hadronic final state 
and the distribution of the transverse energy are studied in detail
(using  PYTHIA, HERWIG and PHOJET generators with 
GRV-LO parton parametrizations for the proton and the photon).
The integration over the $\gamma p$ CM 
system pseudorapidity -2.5$<$\-$\eta^*<$1
leads to the   total transverse event energy  distribution shown in 
Fig.~\ref{fig:jet26}a (here $0.3 < y<$ 0.7).

The average transverse energy flow versus $\eta^*$ 
for the total $E_T$ range between 25 and 30 GeV was also measured 
(see  Fig.~\ref{fig:jet26}b). The shape of both distributions
 may indicate the need of 
the multiple interactions.\\
\vspace*{9.5cm}

\begin{figure}[ht]
\vskip 0.in\relax\noindent\hskip 4.2cm
       \relax{\includegraphics{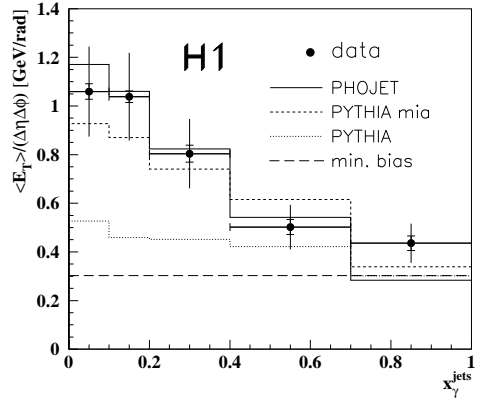}}
\vspace{-0.6cm}
\caption{\small\sl The corrected transverse 
energy density in the region
$|\eta^*|\ <$ 1 outside the jets, as a function of 
$x_{\gamma}^{jets}$. The histograms are as in Fig.~\ref{fig:jet26}
(from \cite{h196}).}
\label{fig:jet28}
\end{figure}

To get an insight into the details of the considered events 
 the transverse energy flow outside of the two jets 
with the highest $E_T$ was studied as a function of  
$x_{\gamma}$  for the $|\eta^*|<1$ and $\Delta \eta <$ 1.2.
Results for the transverse energy density can be found in 
Fig.~\ref{fig:jet28}.\\
\vspace*{4.5cm}

The distribution of the transverse energy around the jet axis was also
 measured as a function of the $\delta \phi$ (not shown).
 The jet width obtained in this analysis is
similar to the corresponding quantity in the $p {\bar p}$ collision.

\begin{figure}[ht]
\vskip 0.in\relax\noindent\hskip 2.5cm
       \relax{\includegraphics{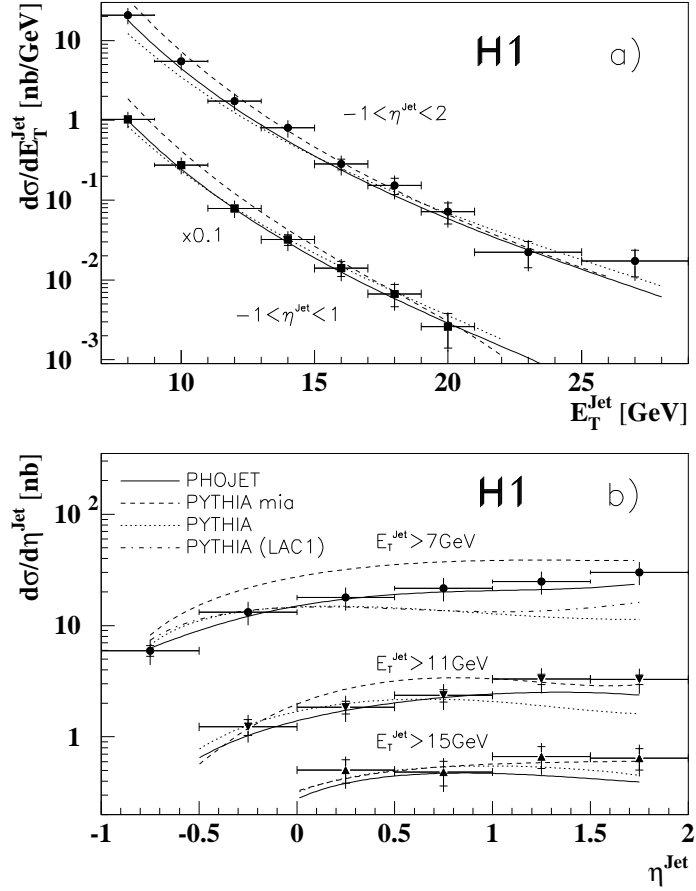}}
\vspace{-0.4cm}
\caption{\small\sl a) Cross section $d\sigma$/$dE_T$ for the jet production
for $E_T^{jet}\ >$ 7 GeV, in two $\eta$ ranges:
-1 $<\ \eta^{jet}\ <$ 2 
and -1 $<\ \eta^{jet}\ <$ 1. The curves 
show the MC simulations with interactions of the beam 
remnants (full line - PHOJET, dashed - PYTHIA) and
without them (dotted - PYTHIA).
b) Cross section $d\sigma$/$d\eta^{jet}$ vs $\eta^{jet}$
for different thresholds in $E_T$: 7, 11 and 15 GeV. The curves
are as in a); additional dash-dotted curve
- PYTHIA with LAC1 parametrization
(from \cite{h196}).}
\label{fig:jet29}
\end{figure}

Further results 
obtained for the jet cross section (here 0.25$ <y<$0.7)
are presented in  Fig.~\ref{fig:jet29}a, 
where  $d\sigma/dE_T$ for the jet production in two $\eta$
regions is shown, and in Fig.~\ref{fig:jet29}b, where 
the distribution $d\sigma/d\eta$ 
for the events with the transverse jet energy 
$E_T>$ 7, 11, 15 \gev is presented.
The  comparison with PHOJET and PYTHIA simulations, with and without 
multiple interactions, was done for both kinds of distributions.
Note that the rapidity distribution is 
  more sensitive to 
the photon structure functions, while the $E_T$ cross section to 
the matrix elements for the hard processes.
Note also that  "for $E_T$ bigger than 7 GeV 
previous measurements ({\bf H1 93}) suffered from a defect
 and are superceded by this new measurement".

~\newline 
Comment: {\it In addition to the primary hard scattering 
process, the interaction between the two beam remnants is included
in the analysis. It  gives ``adequate descriptions of data''
for the transverse energy versus pseudorapidity and the average 
energy flow obtained in this analysis.
``For the first time  the underlying event energy has been measured 
in jet events using direct and resolved photon probes.''

The multiple interaction seems to improve also $d\sigma/dE_T$
and $d\sigma/d\eta$ distributions for jets;
within this approach the low $E_T$ and positive 
$\eta$ range is still not properly described by the considered two
LO parton parametrizations:
LAC1 and  GRV within PYTHIA program. PHOJET describes these data.}
\newpage
\vspace*{12.cm}

~\newline
$\bullet${\bf{H1 96b \cite{as} (HERA)}}\\
%ICHEP'96, Warsaw  , pa02-080\\
The double differential 2-jet cross section $d{\sigma}/dx_{\gamma}
^{jets}/d\log(E_T^2/E_0^2)$ was measured  
as a function of $x_{\gamma}$ 
for different $E_T$ ranges above 8 \gev.
Events (the 1994 data) correspond to: $P^2$ lower than 4 GeV$^2$,
the scaled energy $y$ between 0.2 and 0.83 and the jet pseudorapidity 
 -0.5$< \eta <$2.5, with $\Delta \eta <$ 1.
The cone algorithm with R=0.7 was used. 
The PYTHIA generator with the multiple interaction using GRV-LO
parton parametrizations for the proton and the photon 
 was applied to describe the data.

\begin{figure}[ht]
\vskip 0.in\relax\noindent\hskip .0in
       \relax{\includegraphics{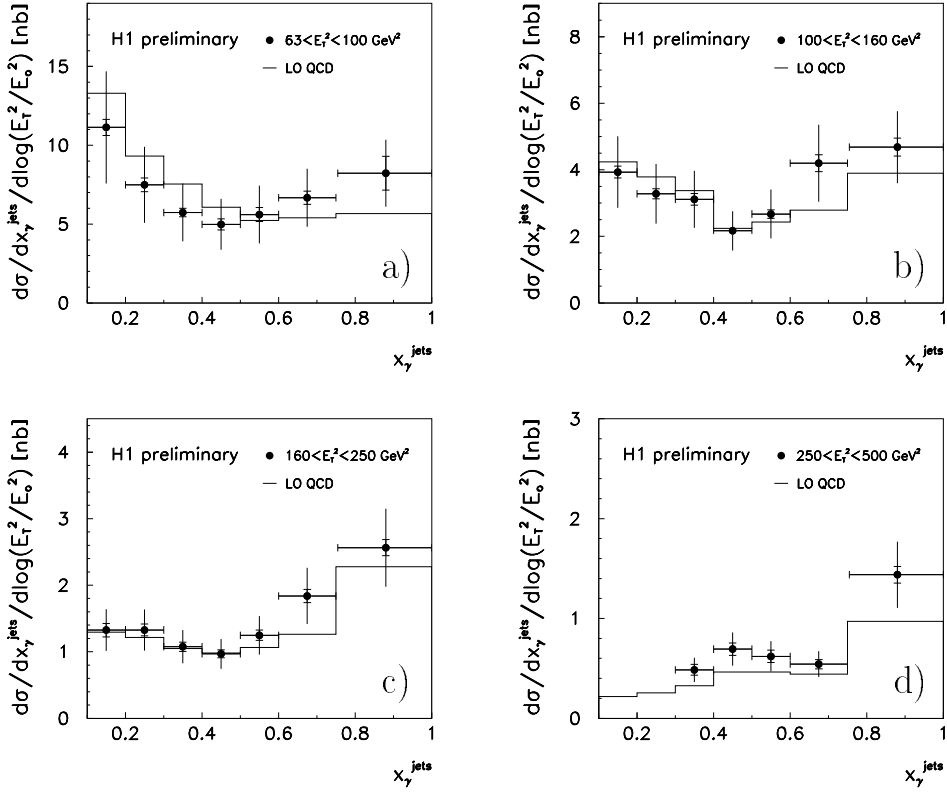}}
\vspace{0ex}
\caption{\small\sl The double differential two-jet cross section
(see the text). The results from the PYTHIA simulation with multiple
interactions and GRV-LO parametrizations are shown
(from \cite{as}).}
\label{fig:jet31}
\end{figure}

In Fig.~\ref{fig:jet31} the cross section is shown 
as a function of the $x_{\gamma}$
for the different $\tilde Q^2$ (=~$E_T^2$) bins (for the 
$E_T^2$ distributions - see {\bf H1 97a}).\\
\vspace*{9.8cm}

For the first time the effective (LL) parton density of the photon
\be
{\alpha}^{-1}x_{\gamma}(\tilde q_{\gamma}+{9\over 4}g_{\gamma}),
\ee 
with $\tilde q_{\gamma} = \sum (g_{\gamma} + {\bar q_{\gamma}}$),
was extracted, for 
63 GeV$^2<p_T^2<$1000 GeV$^2$ and 0.1$<x_{\gamma}<$0.7
(see {\bf {H1 97a}} for the figures). 

~\newline
Comment: {\it "The effective parton distribution grows with 
the scale $p_T^2$, although the increase appears 
slightly steeper than expected from the GRV-LO parametrization."}
\newline\newline
$\bullet${\bf {H1 97a  \cite{t307} (HERA) }}\\
A fixed cone algorithm  with R=0.7 was used to describe
jet events with $P^2<$ 4 GeV$^2$ and $y$ between 0.2 and 0.83.
The pseudorapidity ranges $0<{\bar \eta}<2$ and $\Delta \eta<1$
as well as $\Delta E_T<0.5 {\bar E_T}$  were selected.

\begin{figure}[ht]
\vskip 0.cm\relax\noindent\hskip 1.5cm
       \relax{\includegraphics{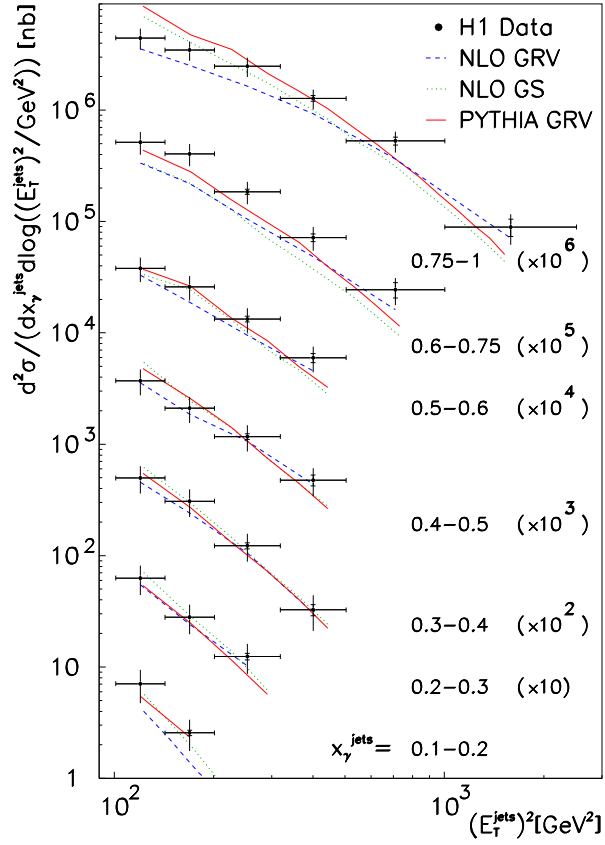}}
\vspace{0.cm}
\caption{\small\sl The double differential cross section 
as a function of the square of the averaged jet transverse energy 
and for the different $x_{\gamma}^{jets}$ (from \cite{t307}).}
\label{fig:t3072}
\end{figure}

The double differential dijet cross section data as a function of 
$\bar E_T^2$ for few ranges of $x_{\gamma}$ are compared with the 
NLO QCD calculation KK \cite{l10} and PYTHIA (GRV) simulation, 
see Fig.~\ref{fig:t3072}.\\
\newpage
\vspace*{10.3cm}
\vspace*{4.2cm}
\begin{figure}[ht]
\vskip 0.cm\relax\noindent\hskip 1.5cm
       \relax{\includegraphics{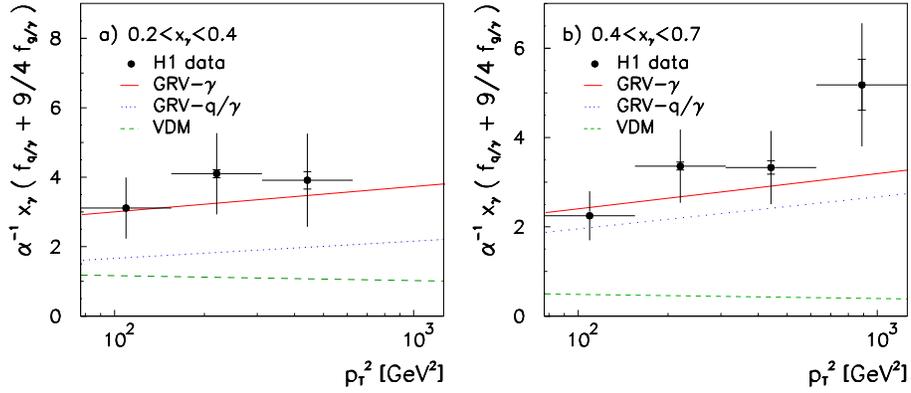}}
\vspace{0.cm}
\caption{\small\sl The effective parton density in the photon
(from \cite{t307}).}
\label{fig:t3073}
\end{figure}

\begin{figure}[ht]
\vskip -1.cm\relax\noindent\hskip -0.2cm
       \relax{\includegraphics{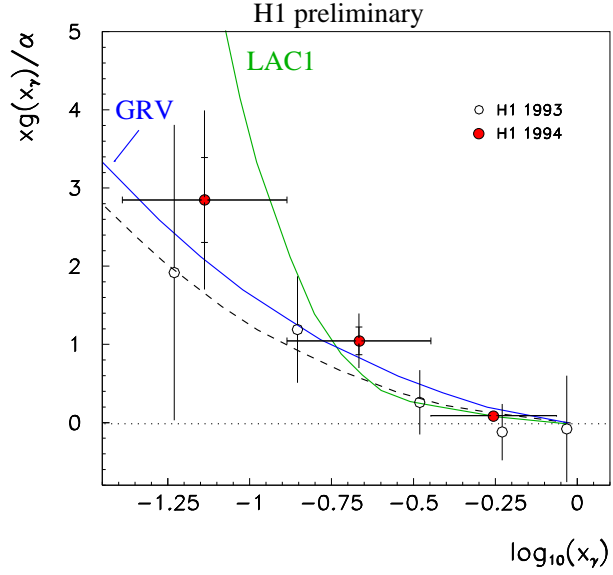}}
\vspace{0.2cm}
\caption{\small\sl The gluon density (from \cite{char}).}
\label{fig:char}
\end{figure}

The effective parton distribution in the photon was extracted for
0.2$<x_{\gamma}<$0.4 and 0.4$<x_{\gamma}<$0.7.
Its dependence on the $\tilde Q^2$ scale (= $p_T^2$) is shown in 
Fig.~\ref{fig:t3073}.
\newline\newline
Comment: {\it "Satisfactory overall description 
(of the double differential cross section for jet) 
except for $x_{\gamma}>0.6$"}.
\newpage
~\newline
$\bullet${\bf {H1 97b \cite{char} (HERA) }}\\
The new method of extracting the gluon density in the photon
from the charged particles is introduced. Events with  
$0.3<y<0.7$, $P^2<0.01$ GeV$^2$ and $|\eta|<1$ were  used.
The  result on the LO gluon density at $\tilde Q^2=<p_T^2>$=38 GeV$^2$
 is presented in Fig.~\ref{fig:char}  and compared with the 
jet data based on the 1993 runs (see also Fig.61).\\
\vspace*{7.6cm}
\newpage
~\newline
$\bullet${\bf {ZEUS 92 \cite{zeus92} (HERA) }}\\
%M. Derrick  PL B 297 (1992) 404\\
The evidence for the hard scattering in the photoproduction
with $E_T>$ 10 GeV at HERA
is reported (the jet cone algorithm with $R=1$ was used).
\newline\newline
$\bullet${\bf {ZEUS 94 \cite{zeus94} (HERA) }}\\
%PL B322(94)287  \\
The measurement was based on the 1992 data
for the single and double jet photoproduction for $P^2$ below 0.02 GeV$^2$
for tagged events, otherwise below 4  GeV$^2$, 
and  for the $y$ between 0.2 and 0.7.
The analysis of the direct and resolved photon
processes was made using the HERWIG generator.
The jet finding cone algorithm with $R$=1 was used.

\begin{figure}[ht]
\vskip 0.in\relax\noindent\hskip 1.7cm
       \relax{\includegraphics{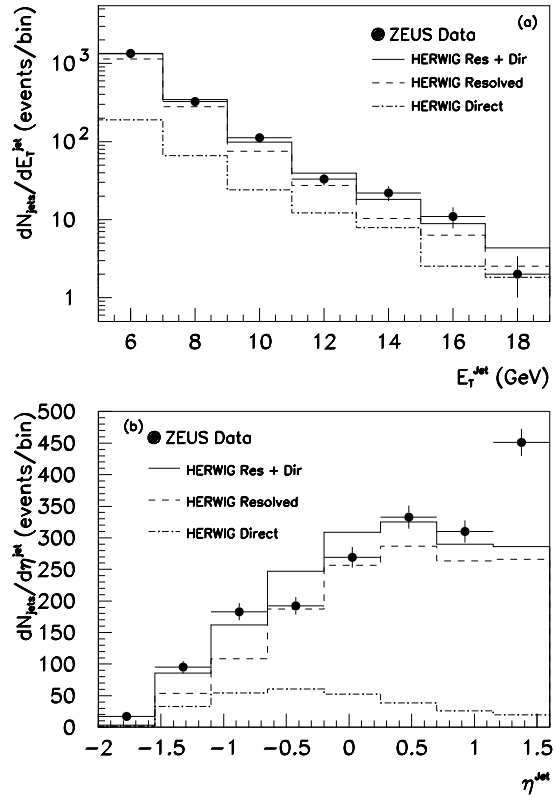}}
\vspace{-0.6cm}
\caption{\small\sl Inclusive jet distributions 
for (a) transverse energy
of jets, (b) pseudorapidity of jets
(from \cite{zeus94}).}
\label{fig:jet34}
\end{figure}

The results for the  $E_T$ distribution  for single jets 
up to $E_T$ = 18 GeV,
integrated over rapidity $ \eta $ below 1.6, are presented in 
Fig.~\ref{fig:jet34}a.
Fig.~\ref{fig:jet34}b shows 
 the $ d\sigma/d\eta$ 
data where the disagreement with MC prediction occurs 
for the positive $\eta$.\\
\vspace*{10.3cm}

\begin{figure}[ht]
\vskip 0.in\relax\noindent\hskip 2.cm
       \relax{\includegraphics{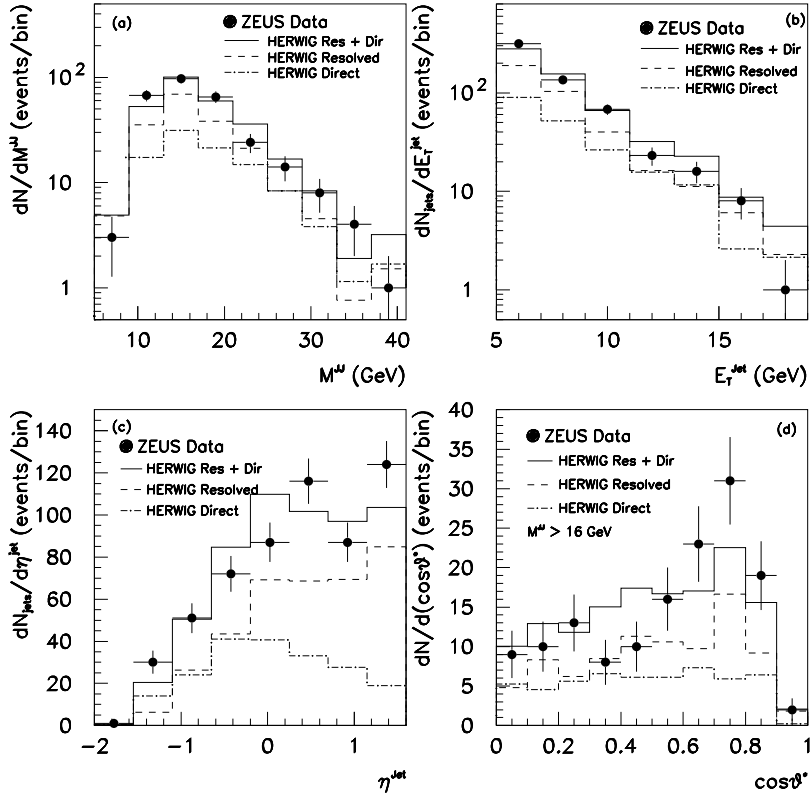}}
\vspace{-0.7cm}
\caption{\small\sl Kinematic distributions for events with two or
more jets: (a) the jet pair invariant mass, (b) the transverse
energy of jets, (c) the pseudorapidity, (d) $\cos\theta^*$
of jet angles in jet-jet CM with respect to the proton momentum
for events with $M_{ij} >$ 16 GeV. The comparison with MC simulations 
is shown (from \cite{zeus94}).}
\label{fig:jet35}
\end{figure}

Di-jet production has been studied by selecting events 
with two or more jets with $E_T> 5$ GeV, for $\eta $ smaller than 1.6.
(Fig.~\ref{fig:jet35}).\\
\vspace*{10.2cm}

The $x_{\gamma}$ and $x_{proton}$ distributions were studied as well 
for events with $| \Delta \eta |< $1.5, $|\Delta \phi|>$120$^o$ and 
the invariant mass of two jets $M_{ij}$ larger than 16 \gev (not shown).
\newline\newline
$\bullet${\bf {ZEUS 95a \cite{zeus95} (HERA) }}\\
%M. Derrick  PL B 342 (1995) 417\\
The 1993 data for the production of at least  one jet  with $E_T>$ 
6 \gev are presented. Events correspond to  $P^2$ below 4 \g2, $y$ 
between 0.2 and 0.85 and the jet pseudorapidity between -1 and 2 (-3 
and 0 for $\eta^*$). PYTHIA 5.6 and HERWIG 5.7 generators (GRV and 
LAC1 parametrizations for the photon and MRSD$_0$ for the proton) 
were used with the cone algorithm for R=1. A wider than before range 
of $E_T$ (up to  41 \gev) and wider $\eta$ range were considered. 

\begin{figure}[ht]
\vskip 0.in\relax\noindent\hskip 1.cm
       \relax{\includegraphics{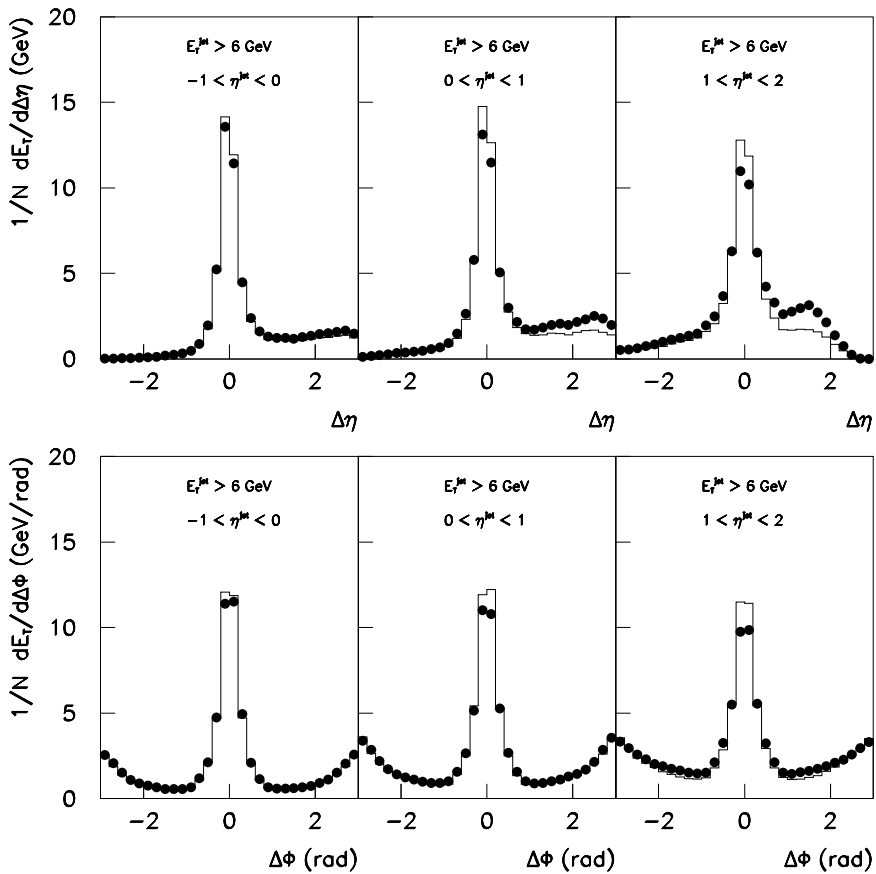}}
\vspace{-0.7cm}
\caption{\small\sl Transverse energy profiles as functions of
$\delta\eta$ (top row) and $\delta\phi$ (bottom row) (see the text). 
Results from the PYTHIA simulation (with both resolved and direct 
processes) are shown (from \cite{zeus95}).}
\label{fig:jet36}
\end{figure}

The transverse energy flow  around jet axis was studied. Results are 
presented in Fig.~\ref{fig:jet36}, where ``there is some discrepancy 
for the forward-going jets in the $\delta \eta>$ 1''.\\ 
\vspace*{11.7cm}

\begin{figure}[ht]
\vskip 0.in\relax\noindent\hskip 2.8cm
       \relax{\includegraphics{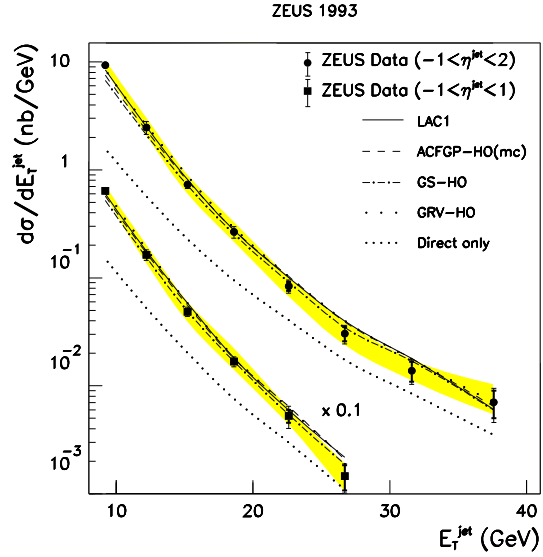}}
\vspace{-1.5cm}
\caption{\small\sl The $E_T$ distributions for jets (see text). 
Results of the PYTHIA simulations with LAC1, ACFGP-HO, GS-HO and 
GRV-HO parametrizations of the parton densities in the photon are 
shown (from \cite{zeus95}).}
\label{fig:jet37}
\end{figure}

\begin{figure}[ht]
\vskip 0.in\relax\noindent\hskip -0.5cm
       \relax{\includegraphics{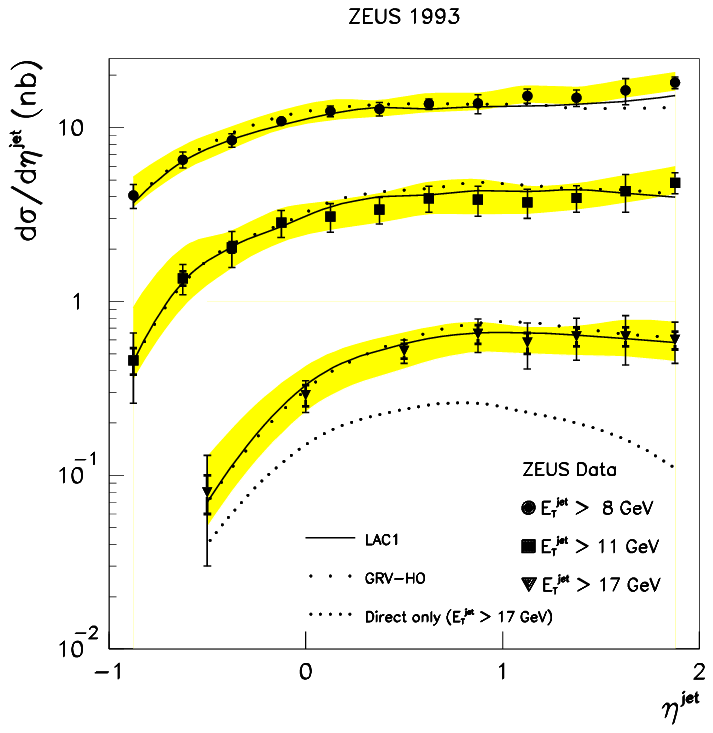}}
\vskip -0.45cm\relax\noindent\hskip 6.6cm
       \relax{\includegraphics{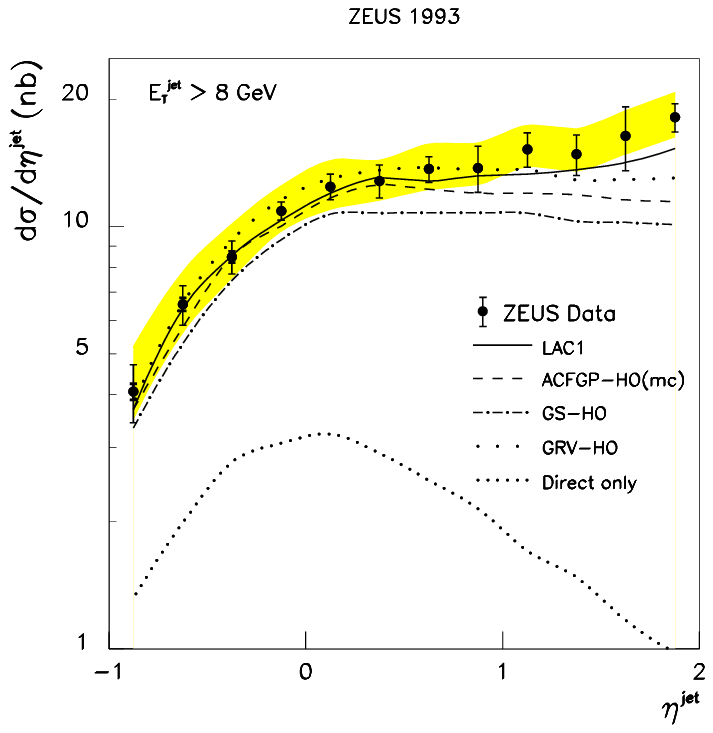}}
\vspace{-1.7cm}
\caption{\small\sl The jet pseudorapidity distributions: (a) 
integrated above three energy thresholds, $E_T\ >$ 8, 11 and 17 GeV 
(see text); the comparison with PYTHIA simulations using LAC1, GRV-HO 
parametrizations is shown; (b) only for $E_T\ >$ 8, and compared in 
addition with ACFGP-HO and GS-HO parametrizations  (from \cite{zeus95}).}
\label{fig:jet38}
\end{figure}

The $E_T$ distributions integrated over two different pseudorapidity 
ranges and $d\sigma/d\eta$ distribution integrated above three $E_T$  
thresholds: 8, 11 and 17 GeV, are presented in  Figs.~\ref{fig:jet37},
\ref{fig:jet38} for different parton parametrizations.

~\newline
Comment: {\it "In the jet profiles, there is a significant 
excess of the transverse energy density in the data with 
respect to the MC expectations
for jets in the region $1<\eta<2$. This excess is located outside 
of the jet in the forward direction $\Delta \eta>1$."\\
`` Except for the region of very forward, low $E_T$ jets,
these measurements  are fully consistent with LO QCD predictions
in new kinematical regime of the structure of the photon'' . 
 The result (for the $d\sigma/d\eta$ for the $E_T>$ 8 \gev
 and the range $-1<\eta<1$) does not support the discrepancy of
 the $d\sigma/d\eta$ with respect to LO QCD calculations 
observed by the H1 Collaboration ({\bf{H1 93}})".}\\
\vspace*{8cm}
\vspace*{8.1cm}

~\newline
$\bullet${\bf {ZEUS 95b \cite{zeus95b} (HERA) }}\\
%derrick pl b348/95/665\\
The photoproduction of dijets, with at least two jets of $E_T$ larger 
than 6 \gev, is considered in the 1993 data. Events corresponding to 
the scaled energy $y$ between 0.2 and 0.8 and $P^2$ lower than 4 \g2 
(for $|\Delta\eta |<0.5$) were grouped in  the resolved and direct 
processes samples. The cone algorithm with R=1 was used within the
HERWIG 5.7 and PYTHIA 5.6 generators with the GRV-LO parametrization
for the photon and the MRSD\_ for the proton.

The important $x_{\gamma}$ distribution was studied. 
(The $x_{\gamma}$ distribution has also been studied in 
{\bf ZEUS 96a, 96b, 97b}.) The cut on the $x_{\gamma}$, equal to
0.75, was introduced later to enhance the resolved or the direct
photon contributions, and a few distributions were studied 
separately for these samples.

\begin{figure}[ht]
\vskip 0.in\relax\noindent\hskip 1.5cm
       \relax{\includegraphics{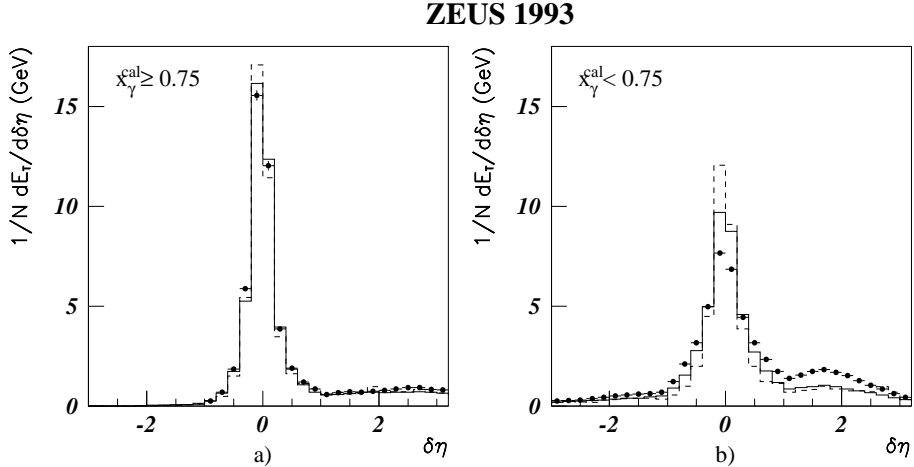}}
\vspace{0ex}
\caption{\small\sl Transverse energy flow around the jet axis versus
$\delta\eta$: (a) $x_{\gamma}\ >$ 0.75 (direct events);  
(b) $x_{\gamma}\ <$ 0.75 (resolved events). Solid (dashed) line - 
PYTHIA (HERWIG) simulation 
%direct contribution to the HERWIG distribution-shaded histogram.
(from \cite{zeus95b}).}
\label{fig:jet39}
\end{figure}

In Fig.~\ref{fig:jet39} the transverse energy flow around the jet 
axis versus $\delta \eta$ is shown, for the first time separately 
for the resolved photon and direct photon contributions ( ...with 
the failure to describe low $x_{\gamma}$ data).\\
\vspace*{5.2cm}

The $d\sigma/d{\bar {\eta}}$ was also measured 
for the direct and resolved photon events (not shown, 
see below for comments and new data in {\bf ZEUS 96a}).
\newline\newline
Comment: {\it 
"Both simulations fail to describe the transverse energy flow 
in the forward region (see also {\bf ZEUS 95a} and {\bf H1 93})."

"The  LO QCD predictions (with DG, GRV and GS2
parton parametrizations) 
 lie below  the dijet cross section 
$d\sigma/d{\bar {\eta}}$  data by factor 
1.5-2". The importance of the NLO calculation is stressed.}  
\newline\newline
$\bullet${\bf {ZEUS 96a  \cite{zeus96} (HERA) }}\\
%derrick pl b 384/96/401 (desy 96-094)\\
Analysis of the 1994 data for dijets (2 jets or more) for $E_T$ above 6 GeV and
with the jet pair invariant  mass above 23 GeV was performed.
Events correspond to the range of  
$y$ between 0.25 and 0.8 and $P^2$  below 4 \g2.
The cone algorithm with R=1 was used within 
 PYTHIA 5.7 and HERWIG 5.8 generators
(with the MRSA parton parametrization for the proton and the GRV-LO for 
the photon).

To obtain the scattering angle $\cos \theta^*$ 
distribution, sensitive to the parton dynamics and 
not parton densities as in analysis above, the cut not on 
$\Delta \eta $ (as in previous analysis) but on $\bar {\eta}$
was introduced.

\begin{figure}[ht]
\vskip 0.in\relax\noindent\hskip 0.cm
       \relax{\includegraphics{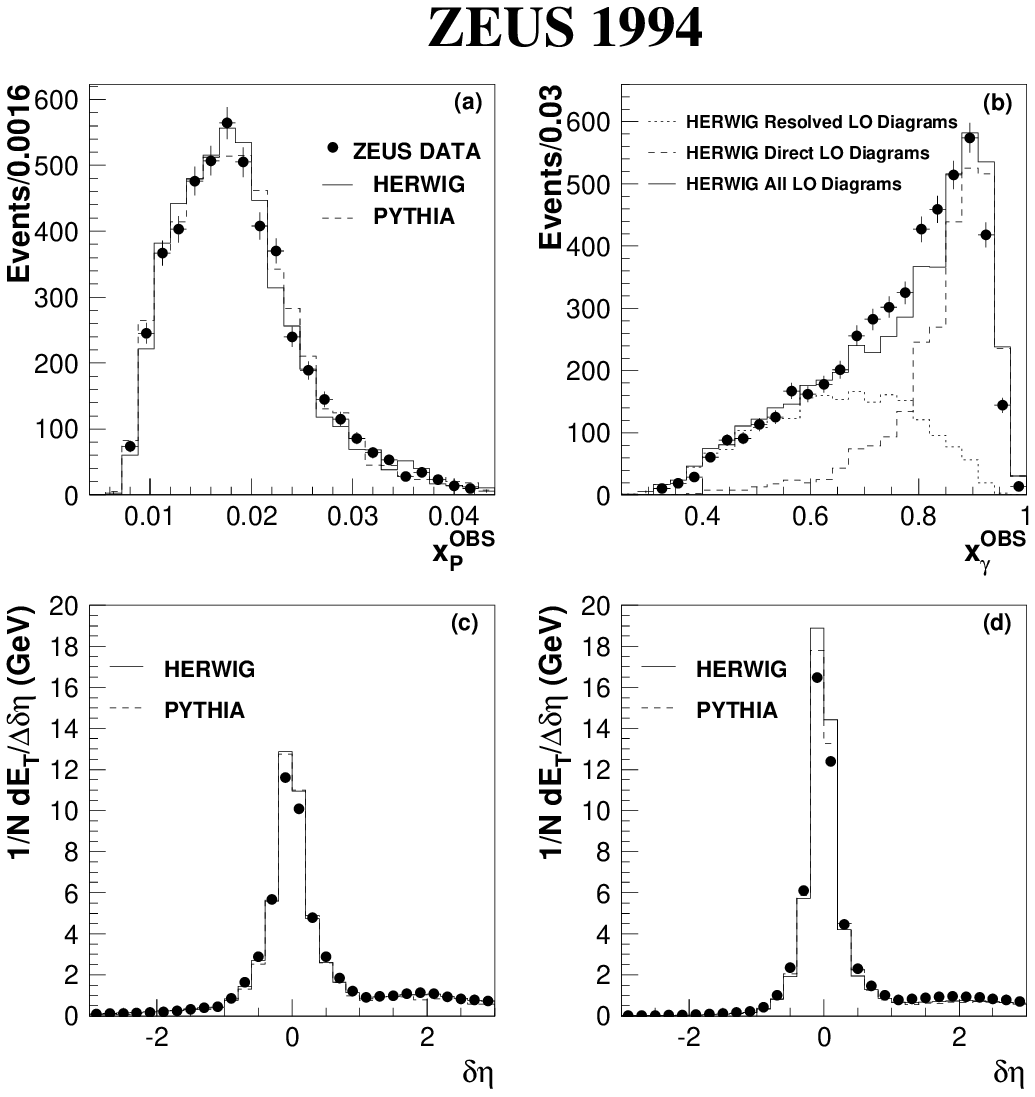}}
\vspace{0.4cm}
\caption{\small\sl Distributions  in $x_p$ (a),
$x_{\gamma}$ (b) of the transverse energy flow. The resolved
($x_{\gamma} <$ 0.75) and the direct ($x_{\gamma} >$ 0.75)
events as a function of $\delta\eta$ are presented separately 
(c and d, respectively).
The MC results are also shown
(from \cite{zeus96}).}
\label{fig:jet40}
\end{figure}

The results for the $x_{\gamma}$, $x_p$ and $\delta \eta$ distributions
are shown in Fig.~\ref{fig:jet40}. 
Due to the  cut on  $\bar {\eta}$ the absolute
value of $\eta$ is restricted to be below 1.8.
Note that the applied cut on the
invariant mass suppresses events with low $x_{\gamma}$.\\ 
\vspace*{10.cm}

\begin{figure}[ht]
\vskip 0.in\relax\noindent\hskip 1.5cm
       \relax{\includegraphics{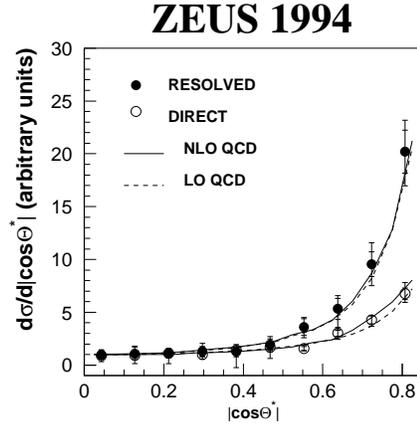}}
\vspace{0ex}
\caption{\small\sl The dijet angular distributions for the resolved
and the direct contributions (see text)
(from \cite{zeus96}).}
\label{fig:jet41}
\end{figure}

Very important results concerning the angular distribution 
due to various partonic subprocesses were obtained for the first 
time in the large $p_T$ resolved photon processes.
The angular distributions $d\sigma/ d\cos \theta^*$
for the  resolved and direct processes are  presented 
together with the LO and NLO calculation based on the CTEQ3M 
parametrization for the proton and the GRV for the photon
in Fig.~\ref{fig:jet41}. The comparison was also made with the 
HERWIG and PYTHIA simulations (not shown).

~\newline
Comment: {\it The transverse energy flow is described properly,
the " requirements of high mass and small boost remove the 
disagreement in the forward flow between data and the simulations 
which has been reported elsewhere in hard photoproduction at HERA."

The dijet angular dependence  is well described by the  LO and NLO
QCD calculations, and also by HERWIG and PYTHIA models.}\\
\vspace*{4.7cm}

~\newline
$\bullet${\bf {ZEUS 96b \cite{zeus96b} (HERA) }}\\
%Warsaw 02-040\\
This is the extension of previous analyses ({\bf {ZEUS 95, 96a}}) based 
also on the 1994 data 
 on the dijet production with $P^2$ lower than 4 \gev 
 and  $y$  between 0.2 and 0.8. 
 Results for the production of at least two jets with the \psr 
between -1.375 and 1.875
and for $E_T^{min}$ = 6, 8, 11 and 15 \gev are presented,  
 assuming $|\Delta \eta|<0.5$.
In the data analysis different jet finding algorithms were
applied: the
cone algorithms EUCELL  and PUCELL (both with R=1) and the $k_T$ -
cluster algorithm        
KTCLUS. 

\begin{figure}[ht]
\vskip 0.in\relax\noindent\hskip -1.3cm
       \relax{\includegraphics{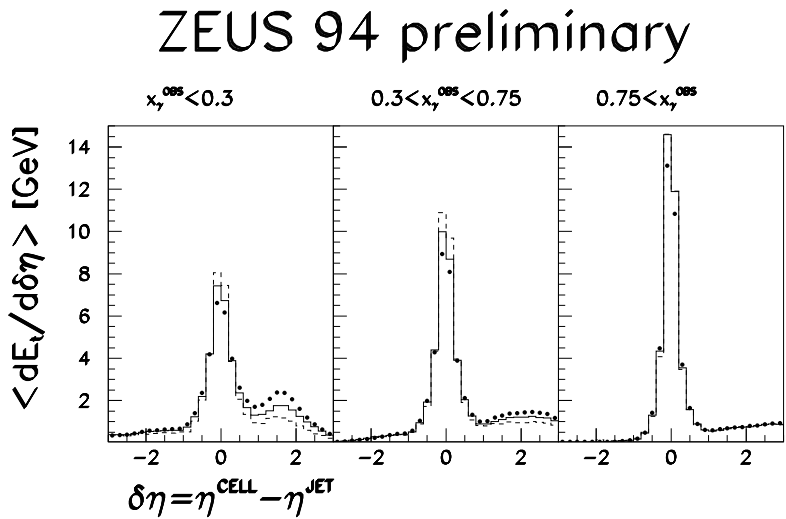}}
\vspace{0.2cm}
\caption{\small\sl Jet profiles $<dE_T$/$d\delta\eta>$ in different
$x_{\gamma}$ bins. Histograms based on the HERWIG generator with
(solid line) and without (dashed line) multiparton interactions  
(from \cite{zeus96b}).}
\label{fig:jet42}
\end{figure}

The resolved cross section was measured in the range 
$0.3<x_{\gamma}<0.75$ and the direct one - for $x_{\gamma}>0.75$.
Analysis of the event distribution versus
$x_\gamma$ (see {\bf {ZEUS 97b}} for the figures),
jet profiles in form of the transverse energy flow around the jet axis
(presented in Fig.~\ref{fig:jet42}) and the $d\sigma/d{\bar \eta}$ 
for the various jet definition and transverse energy thresholds
(see {\bf {ZEUS 97a,b}} for the new results)
were performed.
\vspace*{6.2cm}

~\newline
Comment:{\it ``inclusion of multiparton interactions improves 
the description (of the jet profiles) significantly.''}
~\newline\newline
$\bullet${\bf {ZEUS 96c \cite{zeus96c} (HERA) }}\\
%Warsaw 02-041\\
The inclusive single jet  
cross section for the $P^2$ below 4 \g2 
with $y$ between 0.2 and 0.8 and for three energy regions for 
$W$ between 134 and 277 GeV was studied. 
The measurement  of the transverse energy of the jets bigger 
than 14 \gev and  the pseudorapidity range from  -1 to 2
was performed. The
HERWIG generator (using the jet cone algorithm with R=1) 
including the multiparton interaction was introduced 
in the analysis. The MRSA parton parametrization was used to
describe  the proton structure.

\begin{figure}[ht]
\vskip 0.in\relax\noindent\hskip -2.cm
       \relax{\includegraphics{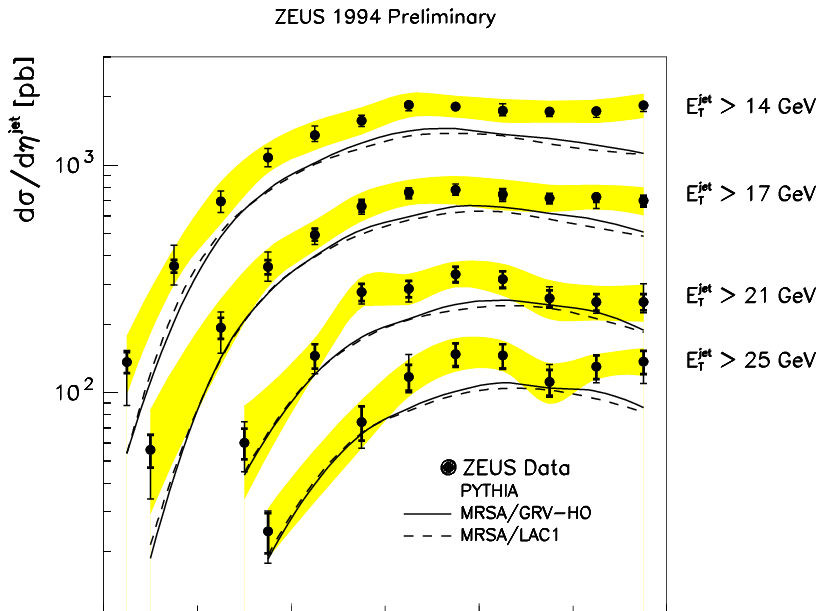}}
\vskip 0.in\relax\noindent\hskip 6.5cm
       \relax{\includegraphics{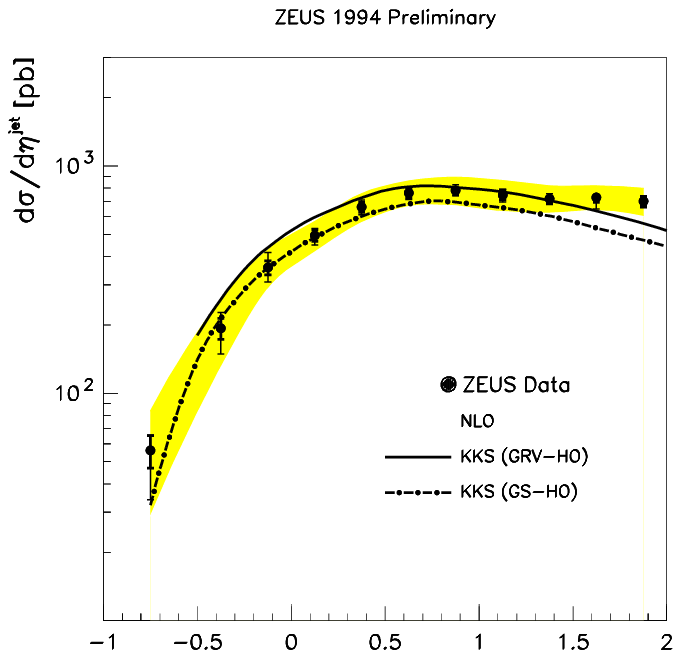}}
\vspace{0ex}
\caption{\small\sl a) The differential cross section
$d\sigma$/$d\eta^{jet}$ integrated over the $E_T^{jet}$
from four thresholds: $E_T^{jet}\ >$ 14, 17, 21 and 25 GeV;
PYTHIA results with the LO parametrizations MRSA for the proton 
and the GRV-HO and LAC1 for the photon are shown.
(b) The same for $E_T^{jet}\ >$ 17 GeV only; curves based on
the NLO calculations (KKS \cite{l8},
using the GRV-HO and the GS-HO parton parametrizations)
are displayed (from \cite{zeus96c}).}
\label{fig:jet43}
\end{figure}

Fig.~\ref{fig:jet43} shows the $\eta$ distributions with the 
comparison to the LO (PYTHIA)
and the NLO QCD predictions (KKS \cite{l8})
(Fig.~\ref{fig:jet43} a and b, respectively).\\
\vspace*{5.7cm}

\begin{figure}[ht]
\vskip 0.in\relax\noindent\hskip 0.7cm
       \relax{\includegraphics{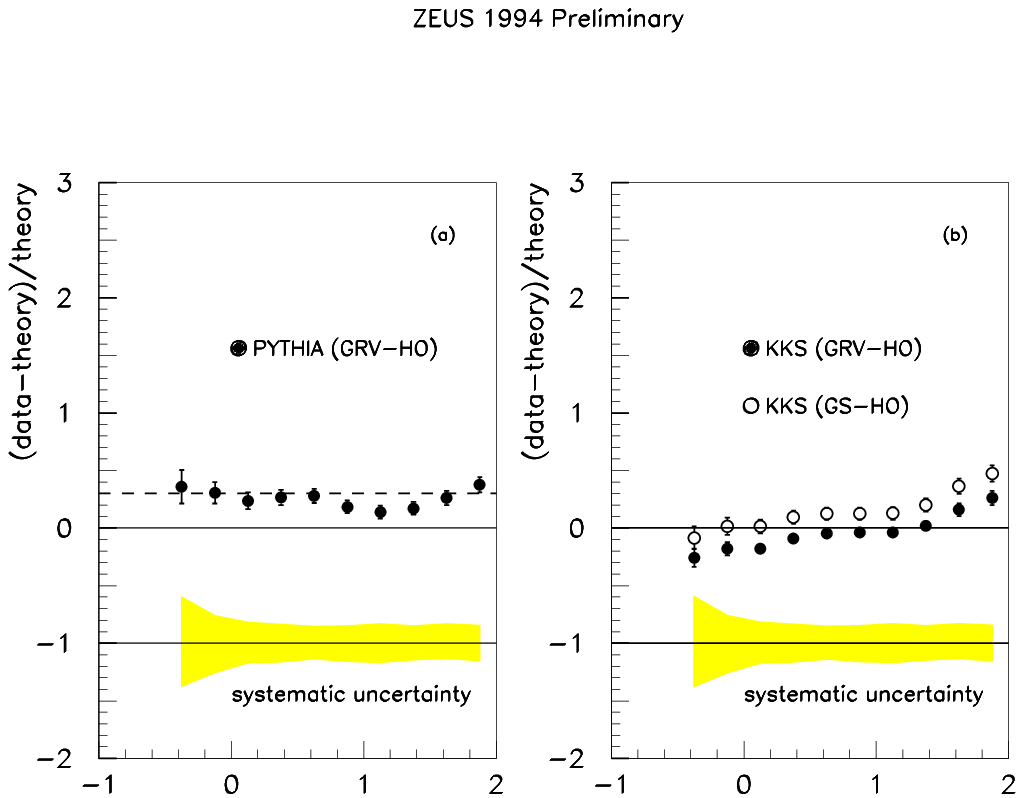}}
\vspace{0.3cm}
\caption{\small\sl The data as in Fig.~\ref{fig:jet43}b 
with the prediction of the PYTHIA 
simulation (GRV-HO) and the KKS\cite{l8} calculation (GRV-HO and 
GS-HO parton parametrizations)
(from \cite{zeus96c}).}
\label{fig:jet44}
\end{figure}

The same is shown (for the NLO approach) 
in the form of the (data-theory)/the\-ory plot
for $E_T>$ 17 \gev in Fig.~\ref{fig:jet44}.

\begin{figure}[ht]
\vskip 0.in\relax\noindent\hskip 0.7cm
       \relax{\includegraphics{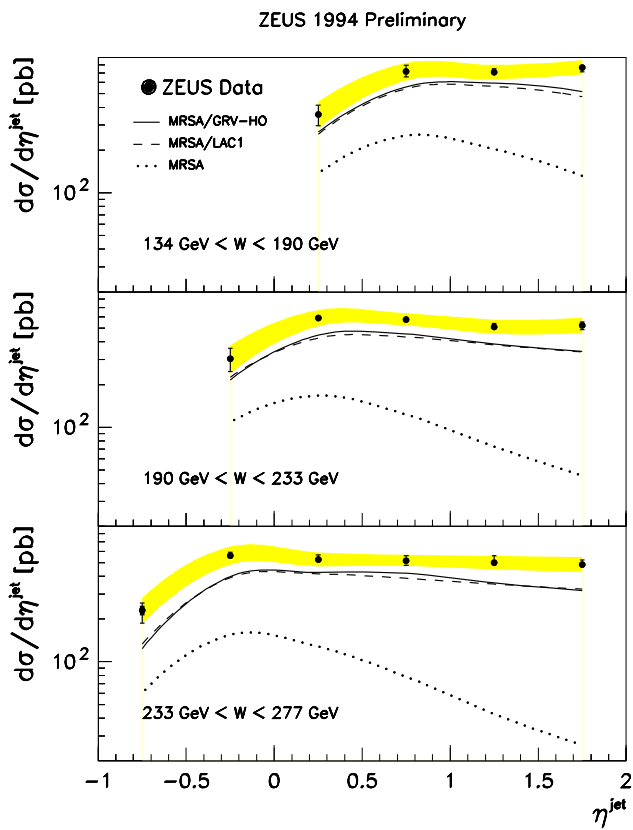}}
\vspace{0.3cm}
\caption{\small\sl The differential cross 
section $d\sigma$/$d\eta^{jet}$
for three regions of the energy W. The PYTHIA results with the MRSA 
for the proton and with  
parton distributions in the photon: GRV-HO and LAC1 are shown
(from \cite{zeus96c}).}
\label{fig:jet45}
\end{figure}

For  events with $E_T>$ 14 \gev  the $\eta$ distribution 
for  different $\gamma-p$ CM energy 
ranges is plotted in Fig.~\ref{fig:jet45}.

~\newline
Comment: {\it ``In the region $\eta >$ 1.5 the data show 
a flattening which is not described by the calculations.''}
\newpage
\vspace*{7cm}
\vspace*{9.8cm}

\newpage
\vspace*{-1cm}

~\newline
$\bullet${\bf {ZEUS 97a \cite{zeus650} (HERA) }}\\
The inclusive jet production $d\sigma/d\eta$
was studied with the iterative cone
algorithm
for events with $P^2<$ 4 GeV$^2$ and 0.2$<y<0.85$. 
In the analysis  $R_{sep}=R$ or 2$R$ was 
\vspace*{18.4cm}
\begin{figure}[ht]
\vskip -9.cm\relax\noindent\hskip 2.7cm
       \relax{\includegraphics{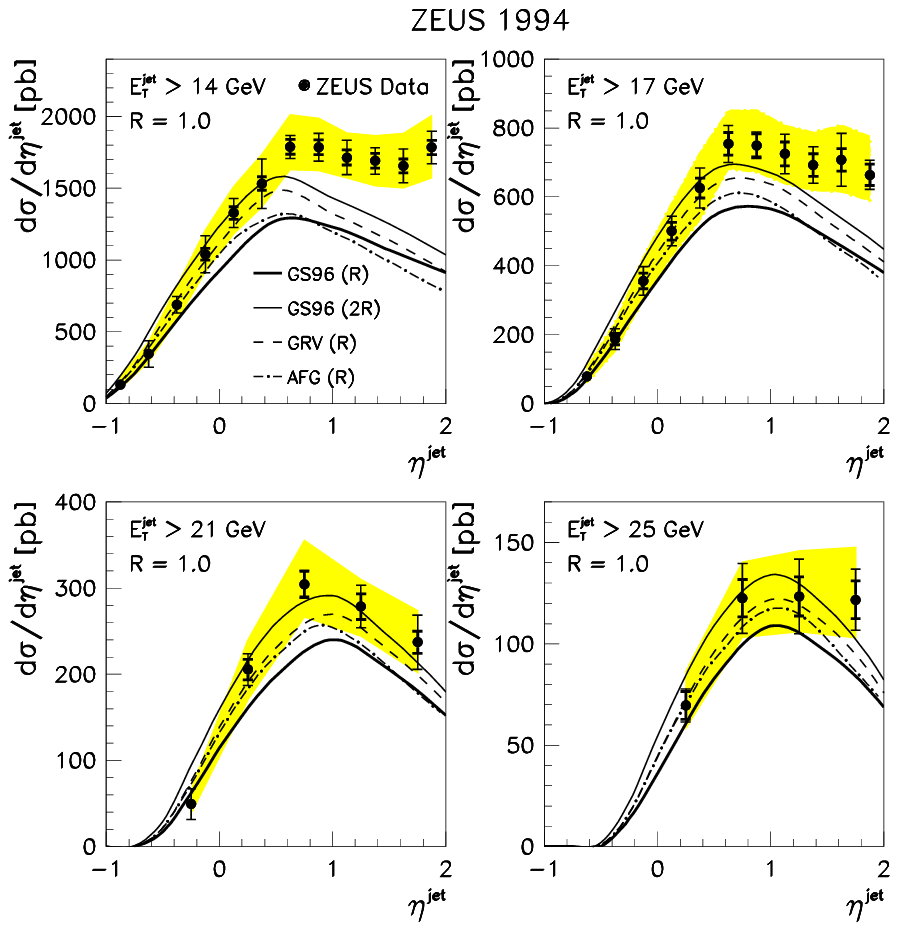}}
\vskip 9.cm\relax\noindent\hskip 2.7cm
       \relax{\includegraphics{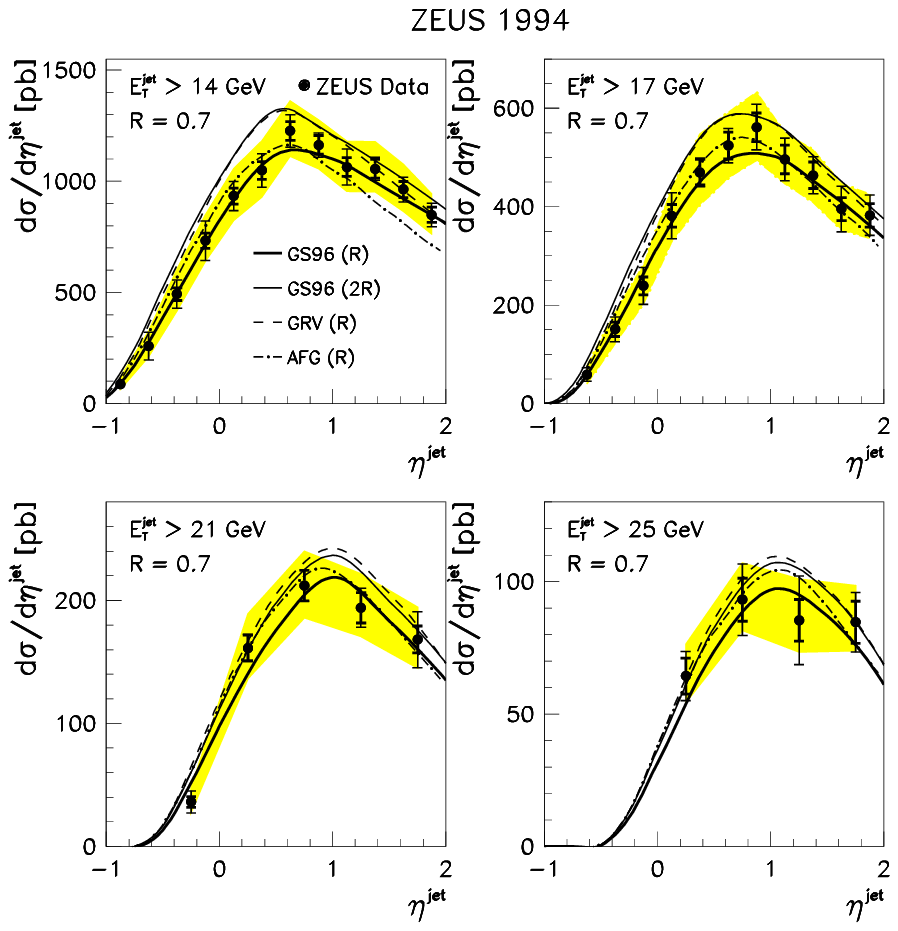}}
\vspace{-0.6cm}
\caption{\small\sl The differential cross section
$d\sigma$/$d\eta^{jet}$ integrated over $E_T^{jet}$
from four thresholds: $E_T^{jet}\ >$ 14, 17, 21 and 25 GeV;
curves based on the
NLO calculations KK\cite{l10} 
 using GRV-HO and GS-HO parton parametrizations
for the photon and the CTEQ4M for the proton with a) $R$=1, b) $R$=0.7
are shown (from \cite{zeus650}b).}
\label{fig:jet6501}
\end{figure}
applied. The distributions of jets  
with $E_T^{min}$=14,17,21 and 25 GeV as a
function of $\eta$ (-1$<\eta<2$) were measured. 
They are not properly 
 described in the forward low $E_T$ region by the NLO calculation
KK \cite{l10} if $R$=1 is used,  for $R$=0.7 the agreement
is obtained (Figs.~\ref{fig:jet6501} a,b).\\

\begin{figure}[ht]
\vskip 0.cm\relax\noindent\hskip 2.7cm
       \relax{\includegraphics{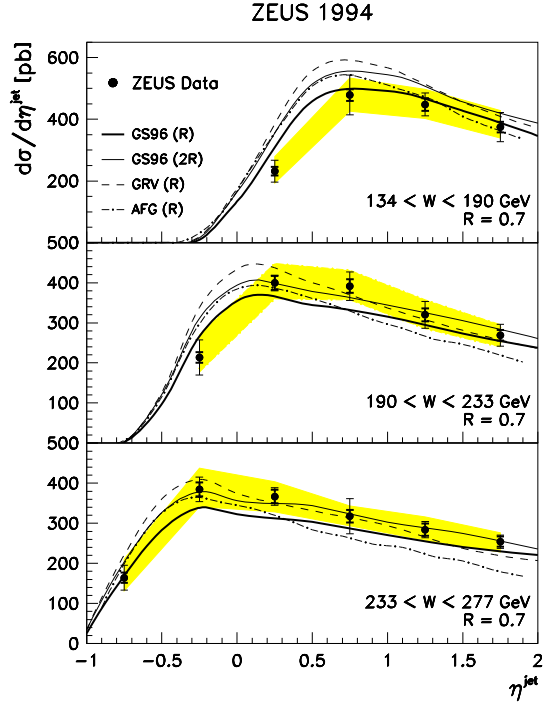}}
\vspace{-0.6cm}
\caption{\small\sl The differential cross 
section $d\sigma$/$d\eta^{jet}$
for three regions of the energy W. The PYTHIA results with the MRSA 
for the proton  and   with the 
  parton distributions in the photon: GRV-HO and LAC1 are shown
(from \cite{zeus650}b).}
\label{fig:jet6504}
\end{figure}

The $\eta$ distributions in three regions of energy W
are also in agreement with the NLO calculation for the $R$=0.7,
 what can be seen in Fig.~\ref{fig:jet6504}.
The data for $R=1$ (not shown) are not in agreement with a QCD 
calculation, as in Fig.~\ref{fig:jet45}.\\
\vspace*{9cm}
~\newline
Comment: {\sl "The measured cross sections for jets with $R$=0.7 are 
well described by the (NLL QCD) calculations in the entire range of $\eta$.}"
~\newline\newline
$\bullet${\bf {ZEUS 97b \cite{zeus18v2} (HERA) }}\\
The dijet cross section based  on the 1994 data for  
$d\sigma/d\bar {\eta}$ for jets with $E_T>$ 6 GeV, 1.375 $<\eta<$1.875, 
and with the $|\Delta \eta|<$ 0.5 was measured. The
different $E_T$ thresholds and the jet definitions 
(as in {\bf {ZEUS 96b}}) were implemented.
 The resolved and the direct photon events
(0.3$<x_{\gamma}<0.75$ and $x_{\gamma}>0.75$, respectively) 
were separated in the analysis.
The results together with the predictions from a NLO
QCD calculation with the additional parameter 
describing the separation of jets $R_{sep}=R$ or 2$R$,
as well as corresponding MC simulations (HERWIG 5.8 and PYTHIA 5.7
with or without the multiple interaction) are presented. 

\begin{figure}[ht]
\vskip 0.in\relax\noindent\hskip 1.cm
       \relax{\includegraphics{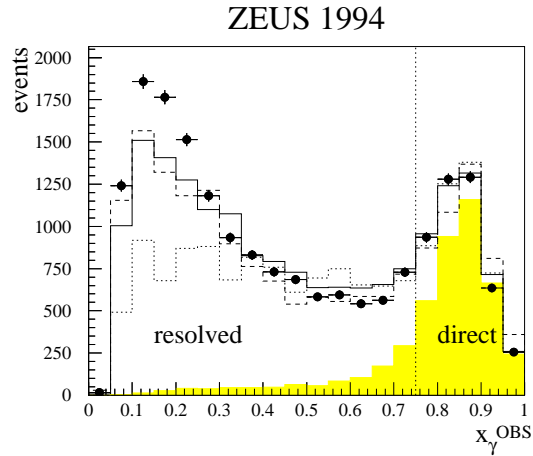}}
\vspace{0.cm}
\caption{\small \sl The corrected $x_{\gamma}^{OBS}$ distribution, 
solid line - HERWIG with the multiple interaction, dashed line -
PYTHIA with the multiple interaction, dotted line - HERWIG without 
the multiple interaction (from \cite{zeus18v2}).}
\label{fig:z18v2.1}
\end{figure}

The  $x_{\gamma} $ distribution obtained 
from the two-jet events 
indicates a need for the resolved photon contribution
(for the first time this kind of measurement 
was performed in 1995, see {\bf ZEUS 95b} and {\bf 96a,b}).
This analysis was based on the KTCLUS algorithm, for the result
see Fig.~\ref{fig:z18v2.1}.
The small $x_{\gamma} $ region is not properly described by the MC
simulations both with and without the multiple interaction.\\
\vspace*{5.cm}
\vspace*{11.7cm}
\begin{figure}[ht]
\vskip 0.in\relax\noindent\hskip 1.3cm
       \relax{\includegraphics{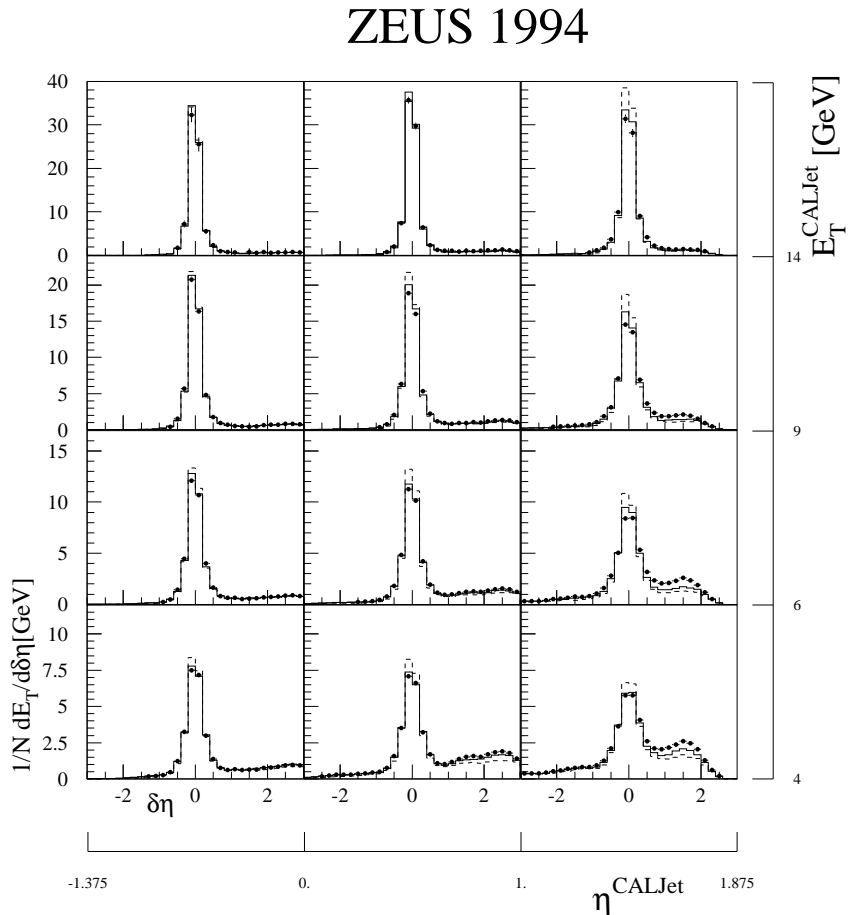}}
\vspace{-0.8cm}
\caption{\small\sl The uncorrected transverse energy flow around the 
jet: solid line - HERWIG with the multiple interaction, 
dashed line  - HERWIG without the multiple interaction
(from \cite{zeus18v2}).}
\label{fig:z18v2.2}
\end{figure}

The transverse energy flow 
obtained using the KTCLUS algorithm is presented 
for different $E_T$ in Fig.~\ref{fig:z18v2.2}, with a similar discrepancy seen in the forward direction, as in previous measurements 
of the same quantity.
 
\begin{figure}[ht]
\vskip  0.in\relax\noindent\hskip 0.cm
       \relax{\includegraphics{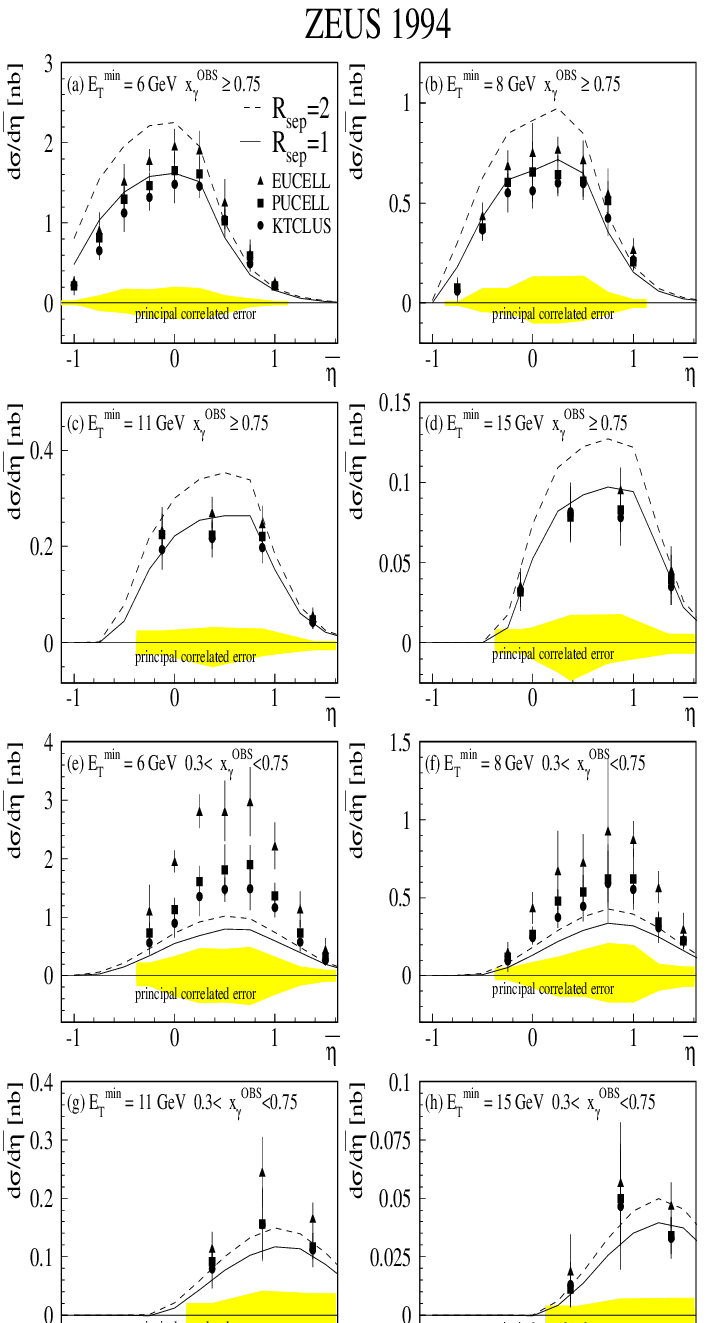}}
\vskip 0.in\relax\noindent\hskip  8.2cm
       \relax{\includegraphics{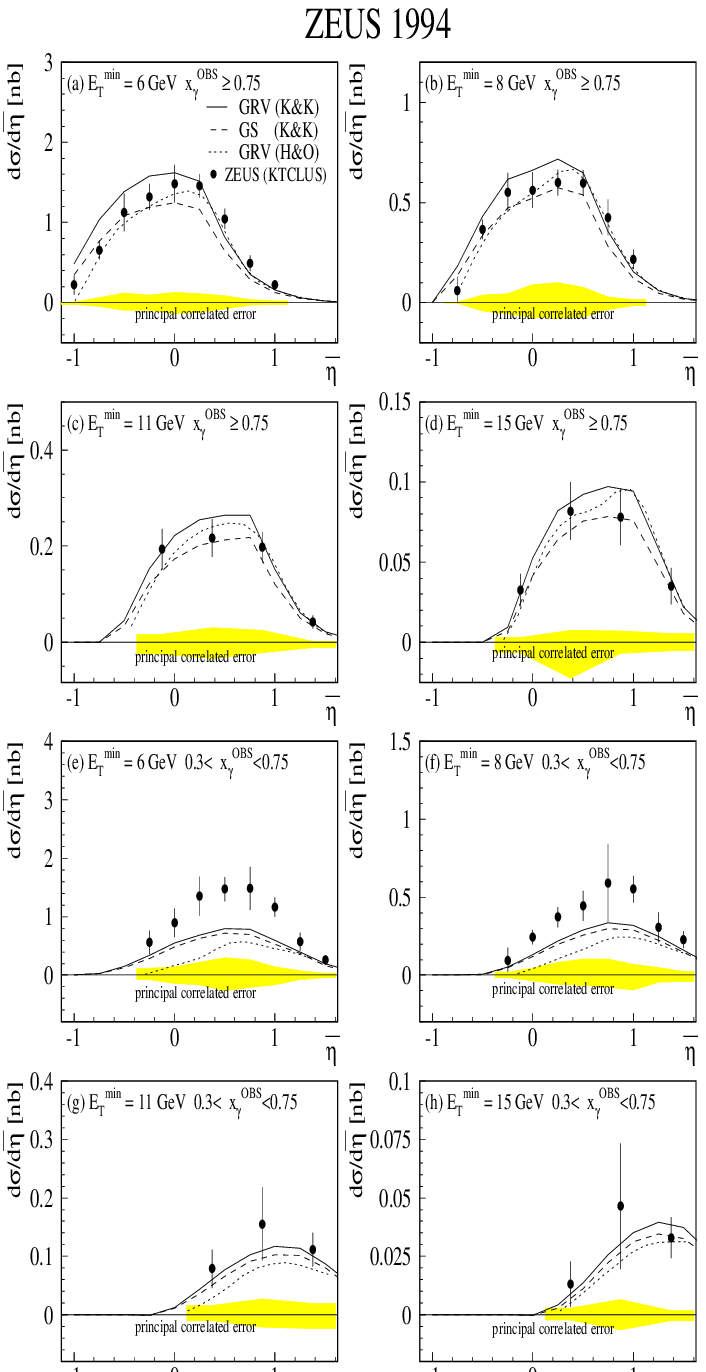}}
\vspace{0cm}
\caption{\small\sl a) Dijet cross sections obtained using the 
different jet algorithms. The curves show the results of the NLO 
calculation \cite{l10} with $R_{sep}$ = 1 (solid line) and 
$R_{sep}$ = 2 (dashed) (GS?).
b) Dijet cross sections obtained using the definite (KTCLUS) 
jet algorithm. The curves show the results of the NLO calculation
\cite{l10} with $R_{sep}$ = 1 (solid line) and 
$R_{sep}$ = 2 (dashed) for different parton parametrizations 
(from \cite{zeus18v2}).}
\label{fig:z18v2.4}
\end{figure}

The dijet cross section 
$d\sigma/d\bar {\eta}$ obtained under condition  $ | \Delta \eta | < $ 0.5,
for different $E_T$ thresholds and with  different jet definitions 
was studied and the results are  
 plotted in Fig.~\ref{fig:z18v2.4}a. The predictions of     
the KK NLO QCD approach \cite{l10}
are also presented.

\begin{figure}[ht]
\vskip 0.in\relax\noindent\hskip -2.5cm
       \relax{\includegraphics{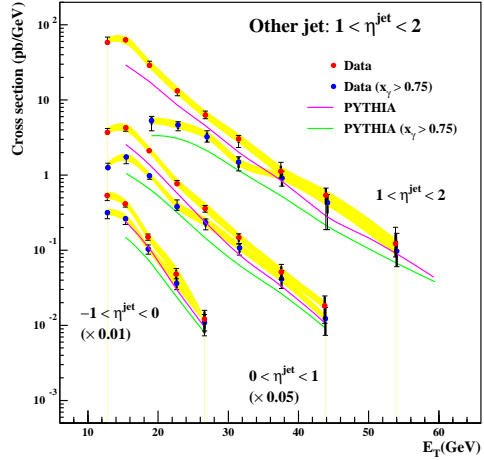}}
\vspace{0ex}
\caption{\small\sl The dijet cross section 
$d\sigma$/$dE_T$ with one jet
within the rapidity range 1~$<\ \eta_1\ <$~2 and the other jet in 
three rapidity ranges, described  in the figure 
(from \cite{hayes}).}
\label{fig:jet49}
\end{figure}

The same cross section, 
now with the KTCLUS jet definition,  
is  plotted in Fig.~\ref{fig:z18v2.4}b. The predictions of the    
KK NLO QCD approach \cite{l10}
with the different parton pa\-ra\-me\-tri\-za\-tions (GS and GRV)
are compared with the data.\\
\vspace*{12.7cm}
~\newline
Comment: {\it The discrepancy in the $\bar {\eta}$ distributions for
resolved photon contributions
was found for events with $E_T^{min}>6 $ GeV,
as discussed in {\bf {ZEUS 96b}}.}
~\newline\newline
$\bullet${\bf {ZEUS 97c  \cite{hayes} (HERA) }}\\
The new 1995 data for the photoproduction of dijets 
($P^2<$ 4 \gev$^2$) and $y$
between 0.2 and 0.85
  are reported for the first time. 
The cone algorithm was used in the PYTHIA simulation; 
both jets have  $E_T>$ 11 \gev  
and are in the rapidity range $-1<\eta<2$.
The preliminary data for $d\sigma/dE_T$ (symmetrized in $\eta$; $E_T$ refers 
to the highest $E_T^{jet}$) are shown in Fig.~\ref{fig:jet49}.\\
\vspace*{5.2cm}

\begin{figure}[hc]
\vskip 0.in\relax\noindent\hskip -5.5cm
       \relax{\includegraphics{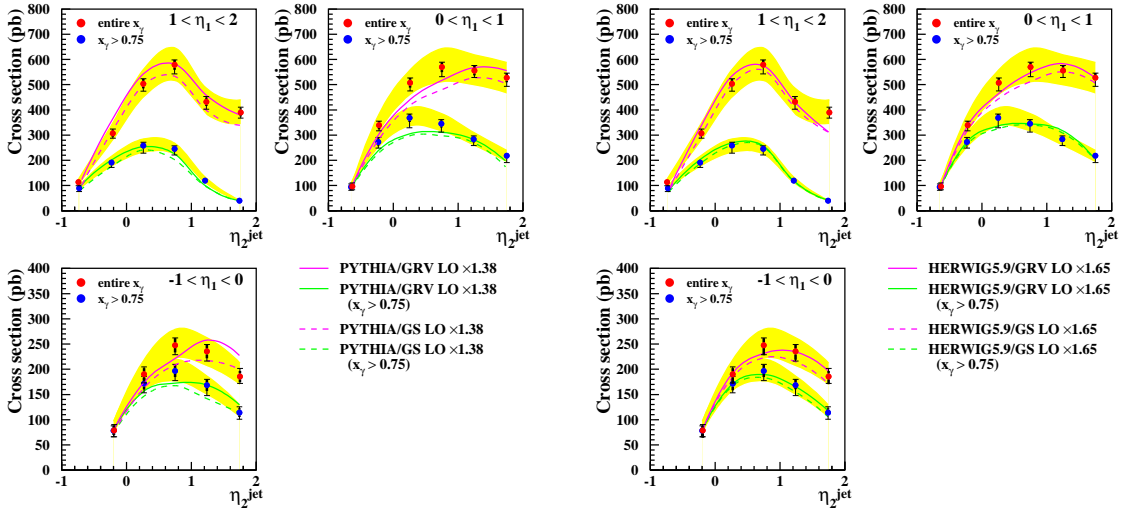}}
\vspace{0ex}
\caption{\small\sl The rapidity distribution $d\sigma$/$d\eta_2$.
The extra cut for the leading jet, $E_T\ >$ 14 GeV.
Three $\eta_1$ jet rapidity ranges were considered
(as in Fig.~\ref{fig:jet49}). The comparison: PYTHIA and 
HERWIG predictions with the GRV-LO and the GS-LO parametrizations
(from \cite{hayes}).}
\label{fig:jet50}
\end{figure}

The preliminary 
results for the \psr  distributions are presented in 
Fig.~\ref{fig:jet50}
for the direct and resolved photon samples.
Below HERWIG and PYTHIA generators were used with the LL parton 
parametrizations for the photon.
Similar analysis with the comparison to the KK \cite{l10} prediction
is discussed in {\bf {ZEUS 97d}}.  
\vspace*{6.3cm}

~\newline
$\bullet${\bf {ZEUS 97d  \cite{zeus654} (HERA) }}\\
The $k_T$- cluster algorithm was used to study the dijet cross
sections, \newline $d^3\sigma/dE_T d \eta_1 d \eta_2$,
for $E_{T,1}>$ 14 GeV and $E_{T,2}>$ 11 GeV and the pseudorapidity
range -1$<\eta_1,\eta_2<$ 2 (see also {\bf ZEUS 97c}). 
Note that $y$ is between 0.2 and 0.85 as before, but $P^2<$ 1 GeV$^2$.
The $E_T$ dependence was also studied and compared with the KK \cite{l10}
predictions 
(with the CTEQ4M and the GS96 parametrization for the proton 
and the photon, respectively).

\begin{figure}[ht]
\vskip 0.cm\relax\noindent\hskip 1.5cm
       \relax{\includegraphics{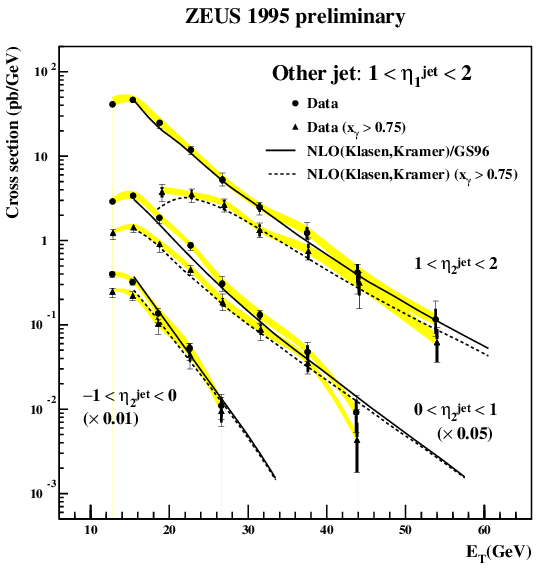}}
\vskip -0.48cm\relax\noindent\hskip 8.5cm
       \relax{\includegraphics{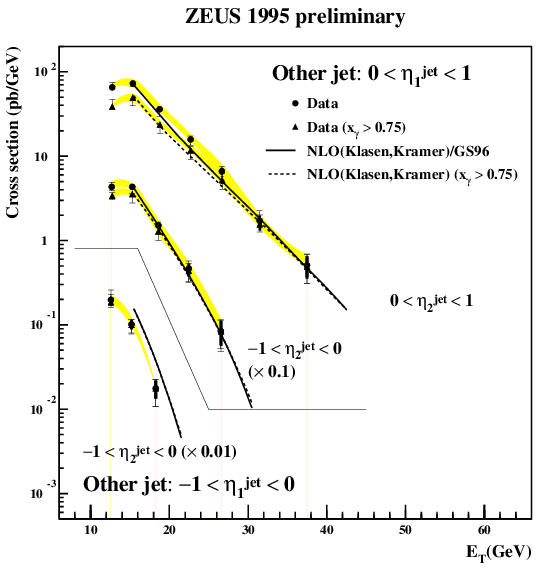}}
\vspace{-0.6cm}
\caption{\small\sl a) Dijet cross section 
$d\sigma$/$dE_T$ with one jet
within the rapidity range 1~$<\ \eta_1\ <$~2 and the other jet in 
three rapidity ranges, described  in the figure.
b) Dijet cross section 
$d\sigma$/$dE_T$ with one jet
within rapidity range 0~$<\ \eta_1\ <$~1 and the other jet in 
three rapidity ranges, described  in the figure.
Comparison with the NLO calculation KK \cite{l10}
- solid line (dashed line shows the direct contribution)
(from \cite{zeus654}).}
\label{fig:6541}
\end{figure}

Below the results for the $E_T$ distribution in different regions of 
$\eta_{1,2}$ are presented (Figs. \ref{fig:6541} a, b). 
\vspace*{5.7cm}

\begin{figure}[ht]
\vskip 0.in\relax\noindent\hskip 3.3cm
       \relax{\includegraphics{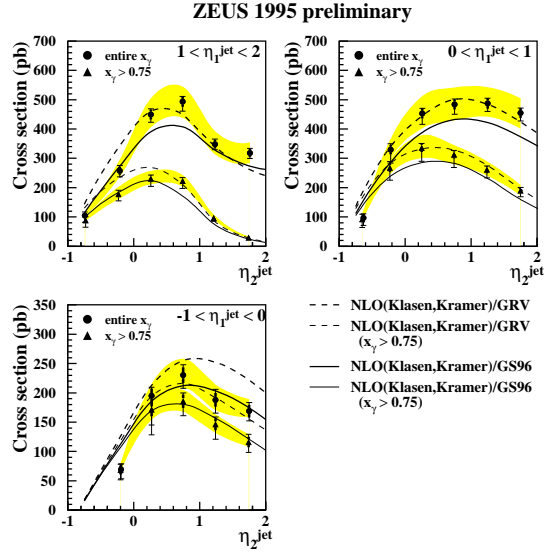}}
\vspace{0ex}
\caption{\small\sl The rapidity distribution $d\sigma$/$d\eta_2$.
The extra cut for the leading jet, $E_T\ >$ 14 GeV.
Three $\eta_1$ jet rapidity ranges were considered
(as in Fig.~\ref{fig:jet49}). The comparison with the NLO calculation,
see text
(from \cite{zeus654}).}
\label{fig:6543}
\end{figure}

The rapidity distribution $d\sigma$/$d\eta_2$
integrated over  $E_T\ >$ 14 GeV is presented in Fig. \ref{fig:6543}.
Three $\eta_1$ jet rapidity ranges were considered
(as in Fig.~\ref{fig:jet50}). Here the  comparison
was made with the NLO calculation by KK \cite{l10}
 with GRV-HO and GS96 parton parametrization for the photon
(CTEQ4M parametrization for the proton).
The comparison with the HERWIG simulation was also made (not shown, see {\bf 
ZEUS 97c}).

\begin{figure}[ht]
\vskip 0.in\relax\noindent\hskip 1.5cm
       \relax{\includegraphics{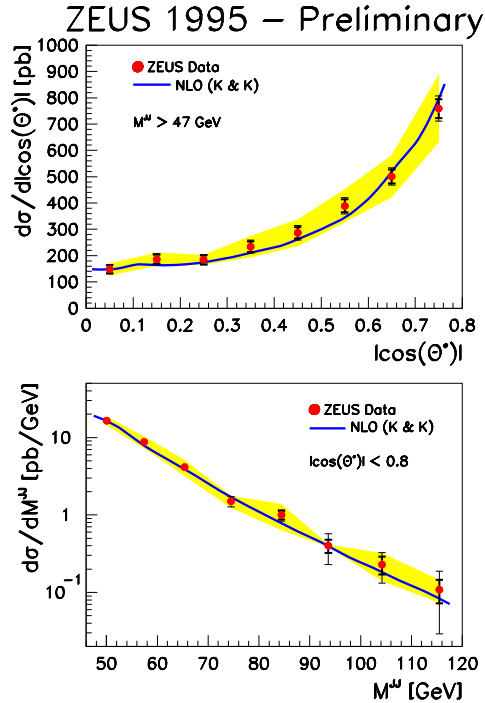}}
\vspace{0ex}
\caption{\small\sl 
The dijet distribution $d\sigma$/$d \cos \theta^*$ for $M_{ij}>$ 47 GeV
(upper plot) and $d\sigma$/$d M_{ij}$ for the $| \cos \theta^* |< 0.8$
(lower plot).
 The comparison with the NLO calculation KK \cite{l10} with 
the CTEQ4M (GS96) parametrization for the proton (photon)
is shown (from \cite{zeus654}).}
\label{fig:jet6545}
\end{figure}

The high mass ($M_{ij}>$ 47 GeV) dijet cross section was also measured. 
Events with  $P^2$ smaller than 4 GeV$^2$ were collected 
and compared with the
same NLO  calculation (KK\cite{l10}) as above. In this particular
analysis 
the cone algorithm was used with $R$=1. Results are presented in Fig.
\ref{fig:jet6545}.\\
\vspace*{6.cm}
\vspace*{9.cm}
%%%%%%%%%%%%%%%%%%%%%%%%%%%%%%%%%%%%%%%%%%%%%%%%%%
%
%
%%%%%%%%%%%%%%%%%%%%%%%%%%%%%%%%%%%%%%%%%%%%%%%%%%
\newpage
\section{Partonic content of the virtual  photon}
The notion of partonic content of the virtual  photon
has appeared in  high energy interactions soon after
the related concept for the real one \cite{rev,uw}. 
Not only  the DIS$_{\gamma}$ formalism discussed before 
can be extended to the scattering on
the virtual photon, but  also  the 
\underbar{resolved virtual photon} processes
can be measured in this case.  

Note that for the resolved virtual photon processes
the flux (convention  dependent)
of the virtual photon in the initial electron is  
introduced in the analysis
{\cite{www}} (see also discussion below).
The flux is usually taken (\eg {\bf H1 97})
in a form integrated over the 
relevant range of $y$, the scaled energy of the initial photon,
and over the squared mass of photon, $P^2$:
\begin{equation}
F_{\gamma/e}=\int^{y_{max}}_{y_{min}}dy \int^{P^2_{max}}_{P^2_{min}}d P^2
f_{\gamma/e}(y,P^2) 
\end{equation}
with
\begin{eqnarray}
f_{\gamma/e}(y,P^2)={{\alpha}\over {2 \pi P^2}} [ {{1+(1-y)^2}\over {y}} 
- {{2 (1-y) P^2_{min}}\over {y P^2}} ].
\label{26}
\end{eqnarray}
For the photoproduction $P^2_{min}=m_e^2 {y^2} / (1-y)$, 
where $m_e$ is  
the electron mass.
The assumption of the factorization of the cross section
for the $ep$ scattering and for the $\gamma^* p$
process remains valid as long as $p_T^2\gg P^2$.

\subsection{Theoretical framework}
The structure function of the virtual photon  can 
be obtained in the Parton Model assuming   
the production of the $q {\bar q}$ pairs. In the  
mass parameter range
\begin{eqnarray}
m_q^2\ll P^2\ll Q^2
\label{27}
\end{eqnarray}
it has a   form  (to be compared with  the Eq.~(\ref{f2})):
\begin{eqnarray}
\nonumber{F_2^{\gamma}(x,Q^2,P^2)=N_c N_f<Q^4>
{{{\alpha}}\over{\pi}}}
x\{ [x^2+(1-x)^2]\ln{{Q^2}\over{P^2x^2}}+6x(1-x)-2\},
\end{eqnarray}
where
\be
<Q^4>={1\over{N_f}}\sum^{N_f}_{i=1}Q_{i}^4.
\ee
One can see clearly that the scale of the probe
has to differ from the $P^2$ in order to test the structure of 
the virtual photon (the so called '$P^2$ suppression' 
compared to the real photon case).
The QCD  
evolution equations for the virtual photon are analogous to those
for the real photon.
Moreover in the case of the virtual photon there is a hope  
that the initial conditions are not needed,
since for $Q^2\gg P^2\gg \Lambda^2_{QCD}$ the nonperturbative effects
should be absent (see Ref.\cite{uw}).
The virtual photon 
may play therefore a unique   role in testing the QCD.
(The existing parton parametrizations for 
the virtual photon are shortly discussed in the Appendix.)

The jet production in the processes involving one or 
two virtual photons 
can be studied in a similar way as for the real photons. 
The factorization between the emission of virtual photons
and the jet production cross section for
 $\gamma^*\gamma$, $ \gamma^*\gamma^*$ and 
$\gamma^* p$ scattering 
is usually assumed. 
Note that these cross sections involving virtual initial photons 
do depend on the convention used 
for the definition of the flux of virtual photons, since
$\sigma^{\gamma^*}\sim 1/\Gamma_{\gamma^*/e}$. 
At the same time the $ee$ or $ep$ cross 
sections are free from such ambiguity,
as it cancels in the corresponding  
cross sections  according to the (symbolical) relation: 
$\sigma^e\sim \Gamma_{\gamma^*/e} ~\sigma ^{\gamma^*}$.
The notion of the \underbar{resolved electron} 
may happen to be very useful here (see the next section).
 
\subsection{DIS$_{\gamma^*}$ for virtual photons}
Measurements of the virtual photon structure function in the 
deep inelastic electron scattering are performed using double-tag events.
Usually these  events were selected in the kinematic 
region where one of the virtual photons (the probe) 
has, on the average, a large virtuality $Q^2$ and the other, 
the target, a small one, $P^2\ll Q^2$. 

For the DIS$_{\gamma^*}$ measurements, the quantity 
$$x_{vis}=Q^2/(Q^2+P^2+W^2_{vis})$$  
needs to be converted to the true $x_{Bj}=Q^{2}/2p q$ variable
\footnote {At finite $P^2$ a modified variable $x_{Bj}$,
which extends over the whole range between 0 and 1 may be introduced.
For small ratio $P^2/Q^2 \ll1$ they coincide.}.
We 
start with the old data from PLUTO, the only  measurements of  
``DIS'' type for the virtual photon so far.

\begin{figure}[ht]
\vskip 0.in\relax\noindent\hskip 2.cm
       \relax{\includegraphics{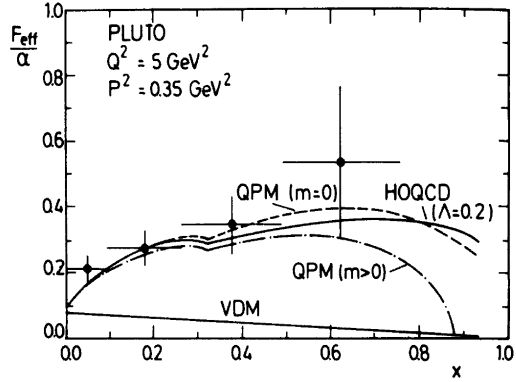}}
\vspace{-2cm}
\caption{\small\sl The PLUTO Collaboration data for the effective 
structure function for the virtual photon.
For the fixed averaged $Q^2=5~ GeV^2$ and $P^2=0.35~ GeV^2$,
a dependence on $x_{Bj}$ is shown (from \cite{pluto}).}
\label{fig:marysia2}
\end{figure}

\begin{figure}[hc]
\vskip 0.in\relax\noindent\hskip 1.5cm
       \relax{\includegraphics{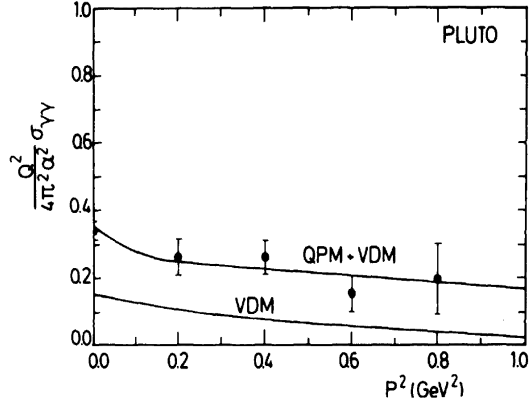}}
\vspace{-2.3cm}
\caption{ {\small\sl The PLUTO Collaboration results for 
${{Q^2}\over{4{\pi}^2{\alpha}^2}}{\sigma}_{{\gamma}{\gamma}}$
for virtual photon as a function of its virtuality $P^2$
(averaged over $x_{Bj}$ and $Q^2$ ranges)(from \cite{pluto}). }}
\label{fig:marysia1}
\end{figure}

~\newline
\centerline{\bf \huge DATA}
~\newline
$\bullet${\bf {PLUTO 84 \cite{pluto} (PETRA) }}\\
The double - tag events   were measured 
where  one of the virtual photons (the probe) 
had, on the  average, virtuality  $<Q^2>=$5 GeV$^2$ and the other 
(the target)  $<P^2>=$0.35 GeV$^2$. 
(The energy of the beam was here 17.3 GeV.) 
The experiment was sensitive to the following combination of  
the virtual photon structure functions:\\
\centerline{$F_{eff}\equiv F_2+(3/2)F_L$.}
$F_{eff}$ was extracted 
for $x_{Bj}$ range between 0.05 and 0.6,
 see Fig.~\ref{fig:marysia2}.
In Fig.~\ref{fig:marysia1}  the  quantity
$Q^2{\sigma}_{{\gamma}{\gamma}}/4{\pi}^2{\alpha}^2$,
averaged over both $x_{Bj}$ and $Q^2$,
is shown as a function of the measured $P^2$ (0.2 - 0.8 GeV$^2$).\\
\vspace*{6.8cm}
\vspace*{7.3cm}

\subsection{Jet production in resolved virtual photon(s)
 processes in   $\gamma \gamma$ and $\gamma p$ collisions}
As for the real photon, the large $p_T$ jets may resolve 
the virtual photon(s).
Provided the corresponding mass relation
$\tilde Q^2\sim p_T^2 \gg P^2_1 ( P^2_2 ) \gg \Lambda^2_{QCD}$ occurs,
%where the $P^2_1$ represents squared mass of the most virtual photon,
one may use the QCD improved parton model as in the case of the real photon
(Eqs.~(\ref{twojets}, \ref{xpm})).

Recently the LO and the NLO QCD calculations have appeared for the 
jet production by virtual photon(s) in the considered processes
$\gamma\gamma$ and $\gamma p$.
% (\cite{l14}, \cite{l15}).

The transition region 
between the interaction of an almost real  photon and of a virtual photon
with the proton 
is studied by  H1 and ZEUS collaborations in the $ep$ collision at HERA.
In such analysis the MC generators used to describe the photoproduction
and the DIS events as well as the rapidity gap events at HERA are used.
As far as the flux of virtual photons is concerned,
it  is  integrated over the corresponding range of the virtuality 
(see Eqs.~(\ref{26}, \ref{27})).

Some kinematical variables are defined  in the $\gamma^*p$
CM system, and are denoted below by  a star, \eg $E_T^*$. 
Note that we use the notation $P^2$ for the squared virtuality
of the photon although in the context of the DIS events (on the 
{\underline{proton}}) at HERA it plays the role of the $Q^2$.

~\newline
\centerline{\bf \huge DATA}
~\newline
$\bullet${\bf {TOPAZ 94 \cite{topaz} (TRISTAN)}}\\ 
The jet production {\bf TOPAZ 94} data (see discussion in Sec.2.2) 
are related to the resolved virtual ${\gamma}(P^2_1)$+real ${\gamma}
(P^2_2)$ process. 
The jet production (one and two jets) has been studied  
with 3.0 GeV$^2<P_1^2<$30 GeV$^2$, 
and $p_T$ between 2 and 8 GeV (corresponding to  4$< \tilde {Q^2}<$64 GeV$^2$).
To what extent one can describe these events using 
a partonic language for the virtual photon
is not clear since in some cases one is probing
the region $\tilde Q^2/P^2_1$ smaller than 1. (See also Sec.2.2.)
\newline\newline
$\bullet${\bf {H1 97 \cite{rick} (HERA)}}\\
%DESY 97-179; hep-ph/9709017
The single jet cross section for the events with $E_T^*>$ 4-5 GeV 
 and with 0.3 $<y<$0.8 is studied in  the transition between
the photoproduction and the standard DIS$_p$ regime at 
\vspace*{10.3cm}
\begin{figure}[ht]
\vskip 0.in\relax\noindent\hskip 0.cm
%%%%%       \relax{\special{psfile = rick3.ps}}
       \relax{\includegraphics{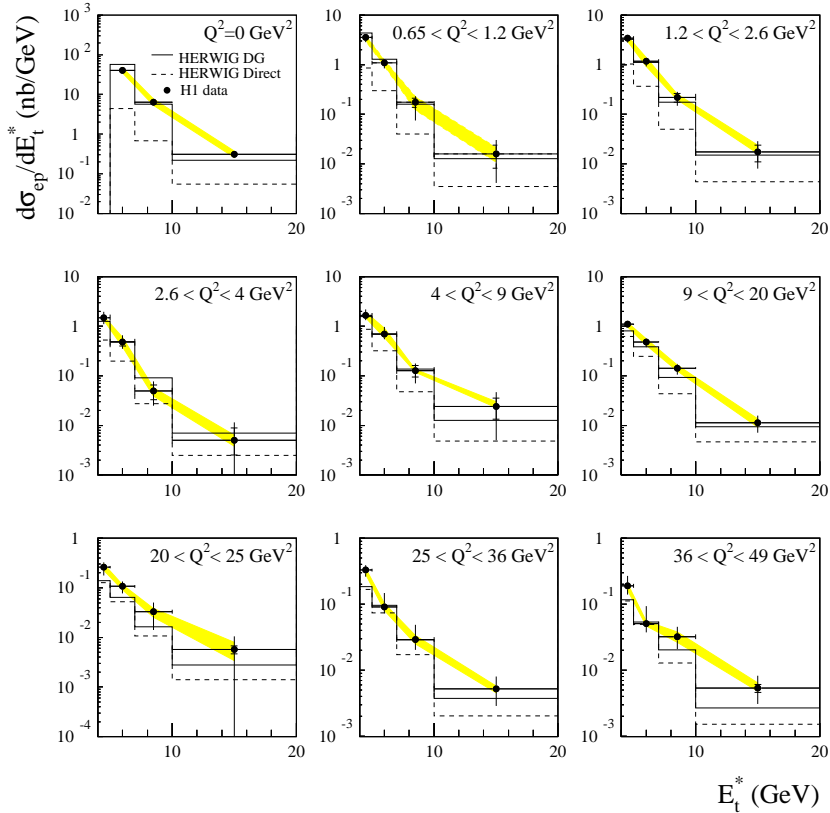}}
\vspace{0ex}
\caption{\small\sl The inclusive $d\sigma_{ep}/dE_T^*$ jet cross section
as a function of the transverse energy $E_T^*$ for
various initial photon virtuality $P^2$ ranges and for the 
$-2.5<\eta^*<-0.5$. The  HERWIG prediction is denoted by 
the solid line, the dashed line corresponds to the direct contribution to 
this model (from \cite{rick}).}
\label{fig:97091} 
\end{figure}
HERA. The data from the years 1994 and 1995, in  three ranges
of the squared mass of the virtual photon $P^2<$10$^{-2}$ GeV$^2$,
0.65 $<P^2<$20 and 9 $<P^2<$49 GeV$^2$ were collected.
The jet pseudorapidity   was studied in the range
$-2.5<\eta^*<-0.5$,
using the $k_T$- clustering algorithm and 
the PHOJET 1.03, LEPTO 6.5 and ARIADNE 4.08, 
RAPGAP and HERWIG 5.9 generators
(with the JETSET  used for the hadronization). The GRV 94-HO
parton parametrization
 for the proton and the Drees-Godbole, GRV-HO, SaS-2D 
parametrizations of the virtual photon were used.

\begin{figure}[ht]
\vskip 0.in\relax\noindent\hskip 0.cm
       \relax{\includegraphics{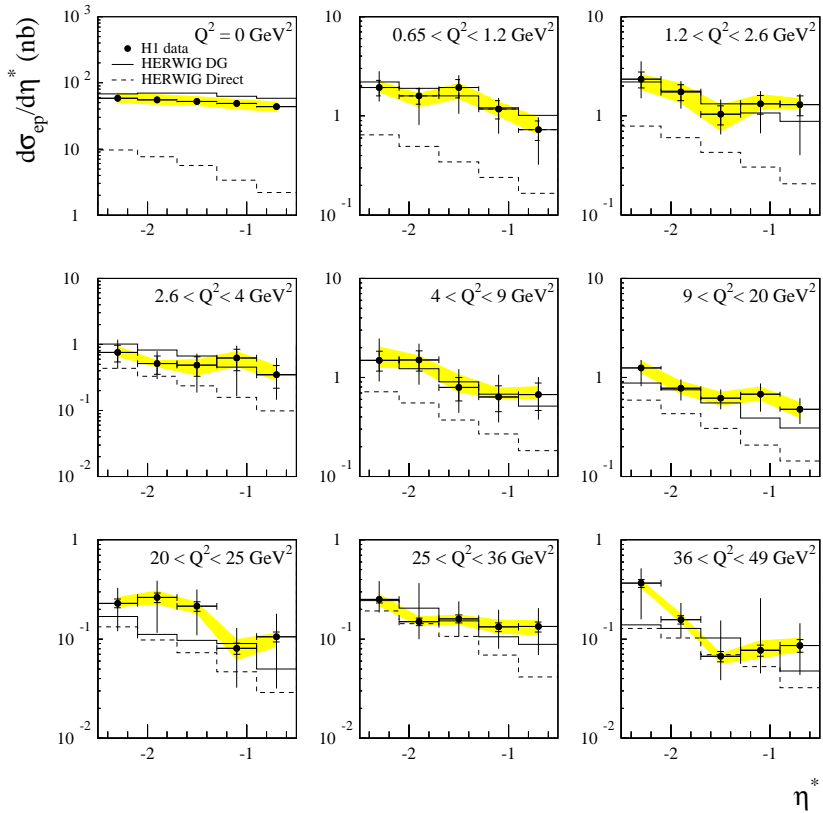}}
\vspace{0ex}
\caption{\small\sl The inclusive $d\sigma_{ep}/d\eta^*$ jet cross section
as a function of the rapidity $\eta^*$ for
various initial photon virtuality $P^2$ ranges for the $E_T^*>$ 5 GeV. 
The  HERWIG (DG) prediction is denoted by 
the solid line, the dashed line denotes the direct contribution 
(from \cite{rick}).}
\label{fig:97092} 
\end{figure}

The measured $d\sigma_{ep}/dE_T^*$ as a function of the transverse 
energy of the jet for various $P^2$ ranges, integrated over $y$
between 0.3 and 0.6,
is presented in Fig.~\ref{fig:97091}. 
It was found to be in agreement with the HERWIG (DG) model.
In Fig.~\ref{fig:97092}   the  corresponding 
data for the rapidity distribution
for jets with $E_T^*>$ 5 GeV are shown for the various virtuality 
ranges.\\
\vspace*{9.8cm}

\begin{figure}[ht]
\vskip 0.in\relax\noindent\hskip 0.cm
       \relax{\includegraphics{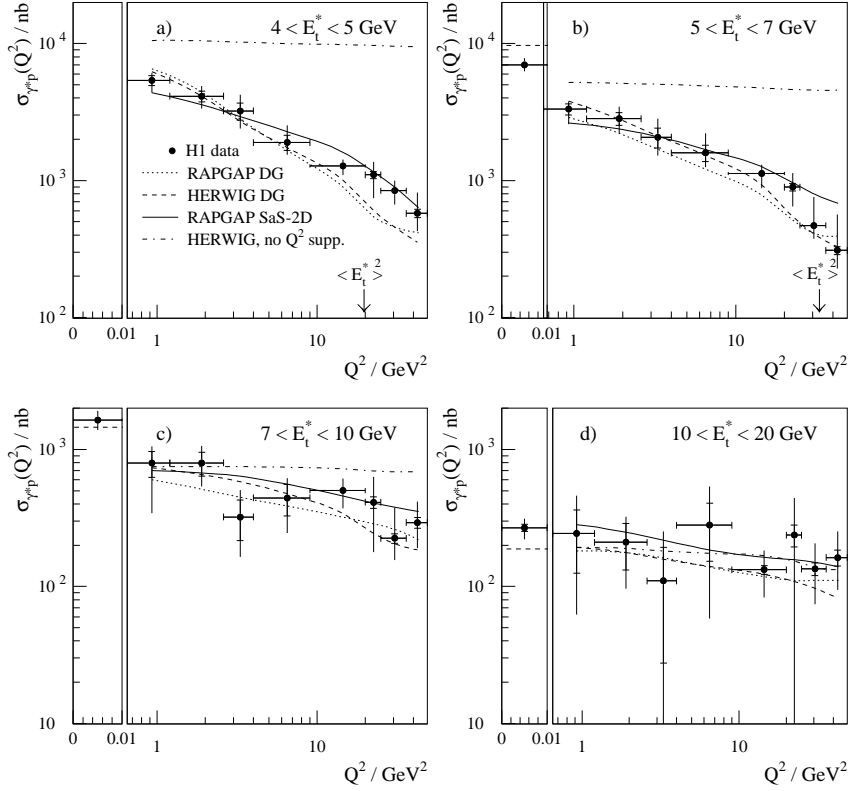}}
\vspace{-0.1cm}
\caption{\small\sl The inclusive $\gamma^* p$ jet cross section
as a function of $P^2(=Q^2)$ for
various ranges of transverse jet energy 
$E_T^*2=(\tilde Q^2)$ for the 
$-2.5<\eta^*<-0.5$. The  HERWIG (DG) prediction is denoted by 
the dashed line, RAPGAP (DG) - the dotted line, RAPGAP (SaS-2D) - 
the solid line and the dot-dashed line corresponds to the HERWIG 
with the GRV-HO parametrization as for the real photon 
("no $P^2$ suppression") (from \cite{rick}).}
\label{fig:97093} 
\end{figure}

The results together with the HERWIG and RAPGAP predictions based on the
DG and SaS-2D parton parametrizations in the virtual photon
are shown in Fig. \ref{fig:97093}. 
To study the dependence of the virtuality of the photon 
the cross section $\sigma_{\gamma^* p}$ is introduced,
\be
\sigma_{\gamma^*p\rightarrow jet+X}=  {{\sigma_{ep\rightarrow jet+X}
\over{F_{\gamma/e}}}},
\ee
although it is not certain that the above factorization really holds
for the whole range of kinematical variables.\\
\vspace*{9.4cm}
~\newline
Comment: {\sl The HERWIG (DG) model gives " a good description of the 
data except of jets in the lowest $E_t^*$ range when 9$<Q^2<49$ GeV$^2$".}
~\newline\newline
$\bullet${\bf {ZEUS 95c \cite{zeus} (HERA) }}\\
The measurement of the direct and resolved photoproduction 
at HERA with the virtual and the quasi - real photons was performed.
In the 1994 run two samples of
events were collected: with the photons of virtualities 0.1 GeV$^2<P^2<$ 0.55
GeV$^2$ and with the quasi - real photons ($P^2 < $ 0.02 GeV$^2$). 
For each dijet event the fraction of the photon momentum 
$x_{\gamma}^{obs}$, manifest in the two highest $E_T$ 
jets ($E_T>$ 4 GeV),  was calculated.
The events associated with the direct photon process (high $x_{\gamma}^{obs}$)
and with the resolved photon processes (low $x_{\gamma}^{obs}$) were found both
in the virtual and quasi - real photon samples. The ratio $N_{res}/N_{dir}$
was calculated as a function of $P^2$ and it seems to decrease with 
an increasing photon virtuality, see {\bf ZEUS 97c} for new results.
~\newline\newline
$\bullet${\bf {ZEUS 97c \cite{zeus657} (HERA) }}\\
A study of the ZEUS dijet production data taken during 1995 
for the transition region between the photoproduction and the
DIS$_p$ is reported. In particular two samples were studied:
0.1 $<P^2<$ 0.7 GeV$^2$ and $P^2<$ 1.0 GeV$^2$. 
The range of the scaled energy 
was 0.2$<y<$0.55.\\
\vspace*{7.cm}
\begin{figure}[hc]
\vskip 0.cm\relax\noindent\hskip 2.5cm
%       \relax{\special{psfile=t307_6.ps}}
       \relax{\includegraphics{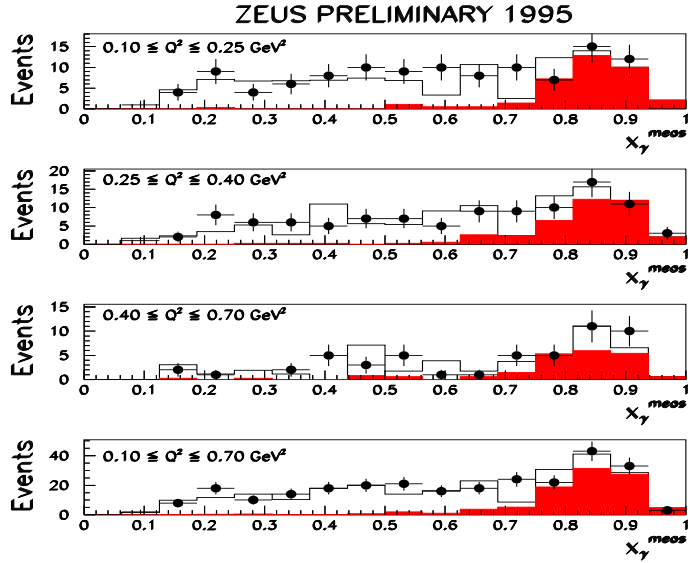}}
\vspace{-0.5cm}
\caption{ {\small\sl The 
$x_{\gamma}$ distributions for the different squared masses 
of the initial photon (from \cite{zeus657}).
}}
\label{fig:n6572}
\end{figure}

The jets with $E_T>$ 6.5 GeV and -1.125$<\eta<$1.875 were used.
The HERWIG 5.9 generator with or without the multiple interaction 
was applied (the MRSA parametrization was used for the proton and 
the GRV parametrization for the photon).
The events corresponding to the direct processes, with
$x_{\gamma} >$ 0.75, and to the resolved ones for $x_{\gamma} <$ 0.75,
were studied.

\begin{figure}[ht]
\vskip 0.cm\relax\noindent\hskip 3.4cm
%       \relax{\special{psfile=t307_5.ps}} 
       \relax{\includegraphics{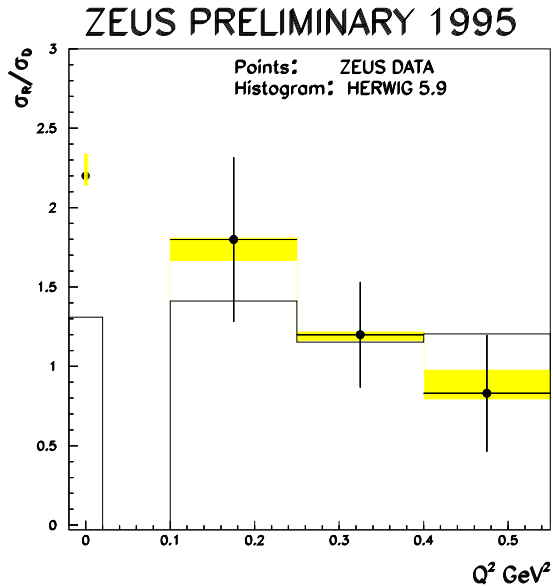}} 
\vspace{-0.3cm}
\caption{\small\sl The ZEUS Collaboration results for the ratio
of the events associated with the resolved photon to the events due to 
the direct photon processes as a function of the squared virtuality of the initial
photon (from \cite{zeus657}).}
\label{fig:n6571}
\end{figure}

The results for the $x_{\gamma}$
distribution for the different $P^2$ ranges are presented 
in Fig. \ref{fig:n6572},
and the 
ratio $\sigma_{resolved}/\sigma_{direct}$ versus $P^2$ is plotted in
Fig. \ref{fig:n6571}.\\
\vspace*{7.3cm}
%%%%%%%%%%%%%%%%%%%%%%%%%%%%%%%%%%%%%%%%%%%%%%%%%%
%
%
%%%%%%%%%%%%%%%%%%%%%%%%%%%%%%%%%%%%%%%%%%%%%%%%%%
\newpage
\section{Related topics} 
\subsection{On leptonic structure functions of the photon}
The lepton pair production by two photons in e$^+$e$^-$
scattering, being one of the basic QED 
processes,\footnote{The earliest experiments on 
e$^+$e$^-\rightarrow $leptons are summed up in 
\cite{fraz,cold}.} is also a test for experimental 
methods applied to more complicated two-photon
reactions involving hadrons. This 
applies especially to the process e$^+$e$^-\rightarrow hadrons$
where the hadronic 
structure functions of the photon are measured 
and the unfolding and tagging methods have to be tested 
(Secs. 2.1 and 2.2).

The lepton pair production in single tagged events 
\be
e^+e^-\ra e^+e^- l^+l^- \label{eell}
\ee
can be described by introducing the QED structure functions for photon:
$F_2^{\gamma (QED)}$, $F_1^{\gamma (QED)}$, $F_L^{\gamma (QED)}$,
etc (see \eg \cite{fraz,cold}).
Unlike in the hadronic case (\ie with the hadronic final 
state), these photon
structure functions  
can be reliably calculated in QED.
Moreover all the particles in the 
final state can be directly observed.

Although the final state e$^+$e$^-\ra $e$^+$e$^-$e$^+$e$^-$ has
been also measured \cite{cello}, for technical reasons only
muonic structure function, denoted below as $F_2^{\gamma (QED)}$,
 could be extracted
(Eq.~(\ref{eell}) with $l=\mu$).

$F_L^{\gamma (QED)}$ structure function is much harder to
measure, because its contribution is  
weighted by the small factor $y^2$ (as for hadronic
$F_L^{\gamma}$, see Eq.~(\ref{si2})).
However, this longitudinal structure function
is not the only structure function that
contains additional information. It has been shown that
there are azimuthal correlations in the final state particles
from two-photon collisions which are sensitive to
additional structure functions \cite{aret,aure}. 

If, instead of measuring the cross section
${d\sigma^{e\gamma\ra eX}\over dx_{Bj}dy}$
(Eqs.~(\ref{si}), (\ref{si2})), one measures in two-body 
$\gamma^*\gamma$ collision also one 
final state particle $a$, additional structure functions 
$F_A^{\gamma}$ and $F_B^{\gamma}$ appear
\cite{aure}:
\begin{eqnarray}
\nonumber {d\sigma (e\gamma\ra eaX)\over dx_{Bj}dyd\Omega_a/4\pi}=
{2\pi\alpha^2\over Q^2}{1+(1-y)^2\over x_{Bj}y}
[(2x_{Bj}\tilde{F}_T^{\gamma}+\epsilon (y)\tilde{F}_L^{\gamma})\\
-\rho (y)\tilde{F}_A^{\gamma}\cos 
\phi_a+{1\over 2}\epsilon (y)\tilde{F}_B^{\gamma}\cos 2\phi_a]. 
\label{ea}
\end{eqnarray}
Here $\Omega_a$ describes the direction of particle $a$ in the
$\gamma^*\gamma$ rest frame, and $\phi_a$ is its 
azimuthal angle around the $\gamma^*\gamma$ axis, relative to
the electron (tag) plane. The functions $\epsilon (y)$ and
$\rho (y)$ are very close to 1. The standard functions
$F_T^{\gamma}$ and $F_L^{\gamma}$ are 
obtained from the corresponding 
$\tilde{F}_i^{\gamma}$
by integration over the solid angle $\Omega_a$. Note that the
formula (\ref{ea}) holds for two leptons or \underbar{two partons}
produced in the final state.

The function $F_B^{\gamma (QED)}$ is, 
in the LL approximation and zero muon
mass limit, equal to $F_L^{\gamma (QED)}$, 
although it involves quite 
different photon helicity structures. Thus extracting 
$F_B^{\gamma (QED)}$ can give us indirectly
information on $F_L^{\gamma (QED)}$ (in LLA).
\newline\newline
\centerline{\bf \huge DATA (Early experiments)}
~\newline
$\bullet${\bf CELLO 83  \cite{cello} (PETRA)}
\newline 
Here both $ee$ and $\mu \mu$  pairs were observed. At
$<Q^2>$=9.5 GeV$^2$  the ``muonic'' structure function 
$F_2^{\gamma (QED)}$ was
measured \footnote{Results for all $F_2^{\gamma (QED)}$
measurements discussed here will be shown together in Fig. 107.}
\label{fotn} and was found to be in good 
agreement with QED calculations.
\newline\newline
$\bullet${\bf PEP-9/TPC 84 \cite{pep} (PEP)}
\newline
$F_2^{\gamma (QED)}$ was measured at $<Q^2>$=0.3 GeV$^2$. 
Results are shown in Fig.105. The quantity
2$x_{Bj} F_1^{\gamma (QED)}$ was also 
extracted in this experiment, from the $y$ region where its
contribution was comparable to that of $F_2^{\gamma (QED)}$.
~\newline\newline
Comment: {\it The importance of $\mu^+\mu^-$ bremsstrahlung background
subtraction has been shown in extracting $F_1^{\gamma (QED)}$. 
Reasonable
agreement with QED predictions was obtained, especially for
$x_{Bj}\leq 0.25$.}\\
\newline
$\bullet${\bf PLUTO 85 \cite{pl} (PETRA)}\\
$F_2^{\gamma (QED)}$ was measured at $<Q^2>$=5.5 and 40  GeV$^2$ 
in the full $x_{Bj}$ range$^{18}$.
Results are in agreement with QED calculations.
\newline\newline
\centerline{\bf \huge DATA(Recent results)}
~\newline
In the last few years new measurements of the leptonic structure function 
 $F_2^{\gamma (QED)}$ have been performed at LEP by the four collaborations; 
the additional structure functions $F_{A,B}^{\gamma (QED)}$  
have also been measured by ALEPH, L3 and OPAL groups.

Since the invariant mass of the $\mu^+ \mu^-$ pair, and hence 
$x_{Bj}$, can be determined very accurately, the measurements
 of $F_2^{\gamma (QED)}$ are only statistically limited.
This is in contrast to the hadronic final states, where the extraction of 
$x_{Bj}$ introduces significant uncertainties.
Thus the investigation of QED structure functions
 is no longer treated only as a test of QED, 
but rather as a clean experimental procedure 
meant for testing and refining the experimental procedures to be 
used in much more complex case of hadronic final states \cite{tnis}.

\begin{figure}[hb]
\vskip 0.in\relax\noindent\hskip 0.cm
       \relax{\includegraphics{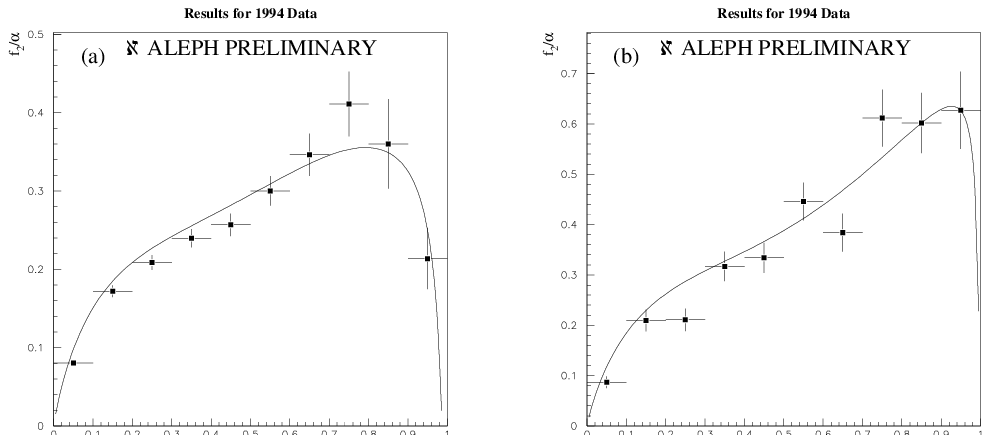}}
\vspace{0cm}
\caption{\small\sl Structure functions 
$F_2^{\gamma (QED)}$/$\alpha$ measured in the ALEPH
experiment \cite{brew} at $<Q^2>$=2.79 GeV$^2$, $<P^2>$=0.153
GeV$^2$ (a) and  $<Q^2>$=14.65 GeV$^2$, $<P^2>$=0.225
GeV$^2$ (b).
(from \cite{brew}).}
\label{fig:brew1}
\end{figure}

Because of the precision of LEP data it is also possible to study the
 effect of the (small) virtuality $P^2$ of the quasi-real (target) photon.
\newline\newline
$\bullet${\bf ALEPH 97d \cite{brew} (LEP 1)}
\newline
$F_2^{\gamma (QED)}$ has been measured (data from 1994) for
0.6$<Q^2<$6.3 GeV$^2$
($<Q^2>$=2.8 GeV$^2$) and 3.0$<Q^2<$60.0 GeV$^2$ 
($<Q^2>$= 14.6 GeV$^2$) (Fig.~\ref{fig:brew1}).
\newpage
\vspace*{4.2cm}

\begin{figure}[ht]
\vskip 0.in\relax\noindent\hskip 0.cm
       \relax{\includegraphics{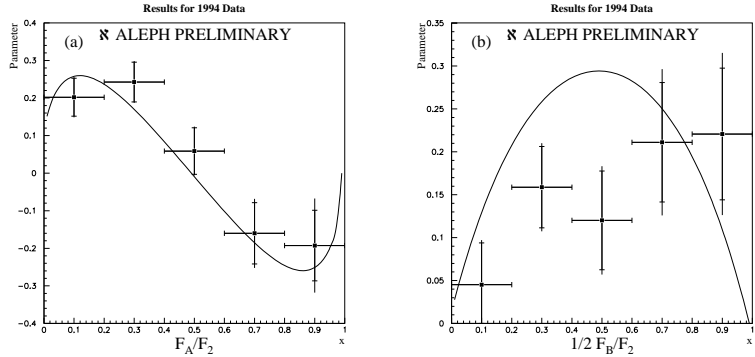}}
\vspace{0.ex}
\caption{\small\sl The structure functions 
$F_A^{\gamma (QED)}/F_2^{\gamma (QED)}$ (a) and
${1\over 2}F_B^{\gamma (QED)}/F_2^{\gamma (QED)}$ (b) measured 
in the ALEPH experiment for $<Q^2>$=8.8 GeV$^2$ \cite{brew}; the solid lines 
are the QED expectations.
(from \cite{brew}).}
\label{fig:brew2}
\end{figure}

In this experiment also the azimuthal angle distributions have been 
measured and functions $F_A^{\gamma (QED)}$ and 
$F_B^{\gamma (QED)}$ extracted 
(Fig.~\ref{fig:brew2}).
\vspace*{4.2cm}

\begin{figure}[ht]
\vskip -0.5in\relax\noindent\hskip -2.6cm
       \relax{\includegraphics{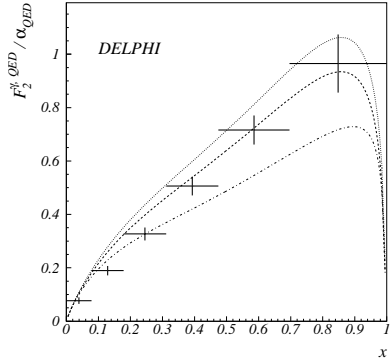}}
\vspace{0.ex}
\caption{\small\sl
Unfolded $F_2^{\gamma (QED)}/\alpha$
compared with QED predictions obtained for different masses
of the target photon: zero mass (upper curve), 
$<P^2>$ = 0.04 GeV$^2$
(middle curve), and average value $<P^2>$ = 0.13 GeV$^2$
(lower curve)
(from \cite{delphi2}).}
\label{fig:delphi4}
\end{figure}

\begin{figure}[ht]
\vskip 0.in\relax\noindent\hskip 1.7cm
       \relax{\includegraphics{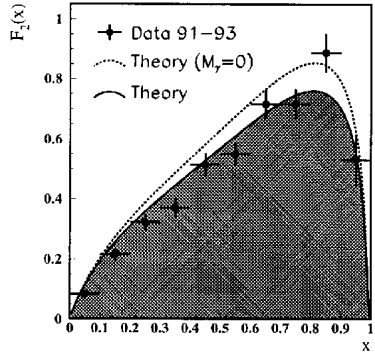}}
\vspace{-4.9cm}
\caption{\small\sl  $F_2^{\gamma (QED)}/\alpha$ measured
 at $<Q^2>$=1.7 GeV$^2$. The effect of nonzero 
target photon mass is also shown (from \cite{l3qed}).}
\label{fig:scan4}
\end{figure}

~\newline
Comment: {\it Both $F_2^{\gamma (QED)}$ and $F_A^{\gamma (QED)}$ 
agree very well with QED predictions. Poorer quality of 
$F_B^{\gamma (QED)}$ data can be helped with better statistics.}
~\newline\newline
$\bullet${\bf DELPHI 96a \cite{delphi2} (LEP 1)}
\newline
$F_2^{\gamma (QED)}$ has been measured at $<Q^2>$=12 GeV$^2$,
as a test for the unfolding and tagging methods in extraction 
of the hadronic $F_2^{\gamma}$ in DELPHI experimental environment
(see Sec. 2.1.2). The effect of non - zero
target virtuality has been studied 
(see Fig.~\ref{fig:delphi4}). A satisfactory fit to the
measured $F_2^{\gamma (QED)}$ is obtained for the fixed value
of $P^2$=0.04 GeV$^2$.\\
\vspace*{5.cm}
~\newline
$\bullet${\bf L3 95 \cite{l3qed} (LEP 1)} 
\newline
The data collected by L3 detector in the years 1991-93 were used to 
extract the $F_2^{\gamma (QED)}$ structure function for  
1.4$<Q^2<$7.6 GeV$^2$ ($<Q^2>$=1.7 GeV$^2$) 
(see Fig.~\ref{fig:scan4}). 
In addition, the angular distribution in the
azimuthal angle $\phi_a$ was measured and some information on 
previously unmeasured $F_A^{\gamma (QED)}$ 
and $F_B^{\gamma (QED)}$ was obtained.\\
\vspace*{8.9cm}
~\newline
$\bullet${\bf L3 97 \cite{l3jer} (LEP 1)}\\
The L3 Collaboration has performed a new measurement of
QED photon structure functions: $F_2^{\gamma (QED)}$, 
$F_A^{\gamma (QED)}$ and $F_B^{\gamma (QED)}$ in
the $e^+e^-\rightarrow e^+e^-\mu^+\mu^-$ process \cite{tnis}.

\begin{figure}[ht]
\vskip -0.5in\relax\noindent\hskip 2.1cm
       \relax{\includegraphics{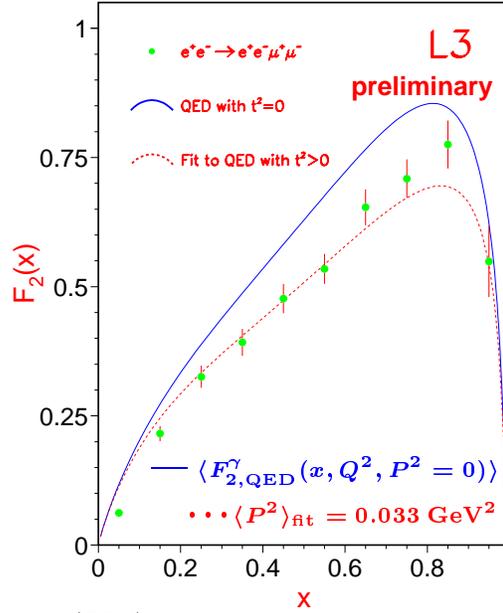}}
\vspace{0.ex}
\caption{\small\sl L3 data on $F_2^{\gamma (QED)}(x,Q^2,P^2)/\alpha $
compared to QED predictions at $P^2$=0 (solid line) and
$<P^2>_{fit}=0.033$ GeV$^2$ (dotted line)
(from \cite{l3jer}).}
\label{fig:tnis_6c}
\end{figure}

\begin{figure}[hc]
\vskip 0.2in\relax\noindent\hskip -0.1cm
       \relax{\includegraphics{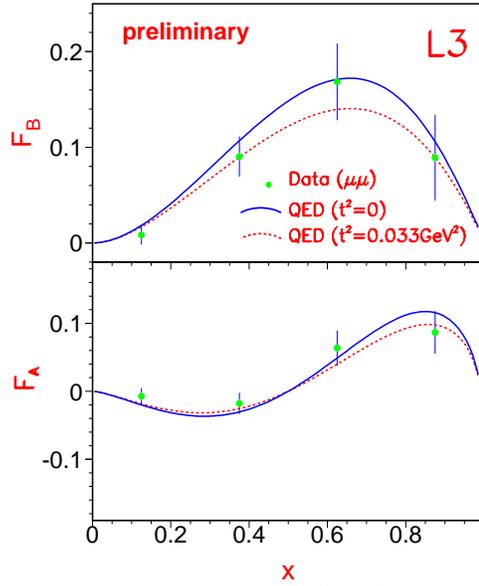}}
\vspace{0.ex}
\caption{\small\sl L3 data on structure functions 
$F_B^{\gamma (QED)}/\alpha$ and $F_A^{\gamma (QED)}/\alpha$ 
measured at 1.4 $<Q^2<$ 7.6 GeV $^2$. The curves
are the QED predictions for $P^2$=0 (solid) and 
$<P^2>_{fit}=0.033$ GeV$^2$ (dashed), respectively
(from \cite{l3jer}).}
\label{fig:tnis_8a}
\end{figure}

\begin{figure}[ht]
\vskip 0.in\relax\noindent\hskip 0.7cm
       \relax{\includegraphics{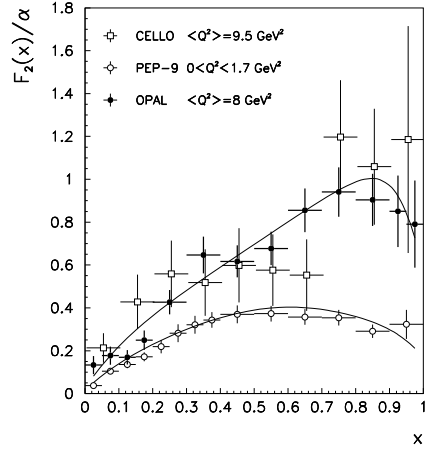}}
\vspace{0.1cm}
\caption{\small\sl $F_2^{\gamma (QED)}/\alpha$ measured at
CELLO \cite{cello} (open squares), PEP-9 \cite{pep} (open circles)
and OPAL \cite{opal} (full circles). The solid lines are the QED 
expectations (identical for CELLO and OPAL within the systematic
errors) (from \cite{opal}).}
\label{fig:cepeop}
\end{figure}

The effect of the photon target virtuality has been studied 
and is clearly seen in all the three structure functions (see
Figs.~\ref{fig:tnis_6c} and \ref{fig:tnis_8a}).\\
\vspace*{8.5cm}
\vspace*{6.1cm}
~\newline
$\bullet${\bf OPAL 93 \cite{opal} (LEP 1)}
\newline 
The QED structure function $F_2^{\gamma (QED)}$ for $<Q^2>$=8.0 GeV$^2$
was extracted from single-tag events at CM energy $\sim M_Z$. 
In Fig.~\ref{fig:cepeop} the data and QED expectations are presented 
together with the earlier measurements at {\bf CELLO 83} \cite{cello}
and {\bf PEP-9/TPC 84} \cite{pep}.
\newpage
\vspace*{5.5cm}
~\newline 
$\bullet${\bf OPAL 97h \cite{opalqed} (LEP 1)}
\newline
Here the extraction of $F_B^{\gamma (QED)}$ was performed for
0.85$<Q^2<$31 GeV$^2$ ($<Q^2>$ =5.2 GeV$^2$).
$F_A^{\gamma (QED)}$ was not extracted due to the partial integration
of the cross section. 

\begin{figure}[hb]
\vskip 0.in\relax\noindent\hskip -3.cm
       \relax{\includegraphics{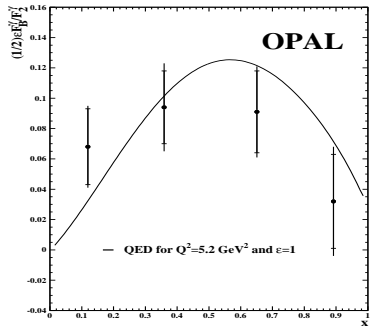}}
\vspace{-0.6cm}
\caption{\small\sl 
The values of 
${1\over 2}F_B^{\gamma (QED)}/F_2^{\gamma (QED)}$ obtained 
from the azimuthal angle distributions (corrected
for the effects of the detector). The solid line is the QED
prediction for $Q^2$=5.2 GeV$^2$ and $\epsilon $=1  
(from \cite{opalqed}).}
\label{fig:doucet2}
\end{figure}

The measured value of
${1\over 2}\epsilon F_B^{\gamma (QED)}/F_2^{\gamma (QED)}$ 
is significantly different from zero and its variation with 
$x_{Bj}$ is consistent with QED (see Fig.~\ref{fig:doucet2}).\\
\vspace*{4.cm}

\vspace*{0.1cm}
\centerline{*****}
\vspace*{0.4cm}

\begin{figure}[ht]
\vskip 0.in\relax\noindent\hskip 0.2cm
       \relax{\includegraphics{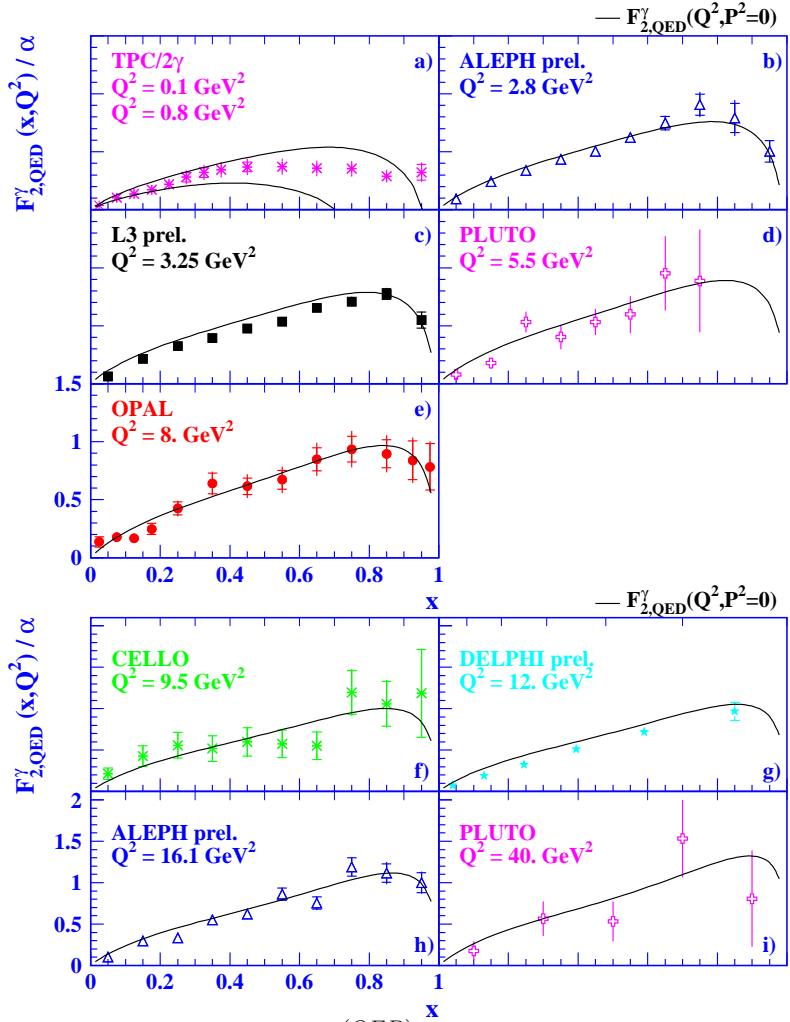}}
\vspace{-0.ex}
\caption{\small\sl Summary of existing $F_2^{\gamma (QED)}/\alpha$
data for broad $Q^2$ range shown with QED predictions for $P^2$=0
(from \cite{tnis}).}
\label{fig:tnis_5}
\end{figure}

A compilation of results for the QED structure function 
$F_2^{\gamma (QED)}$ can be found in Fig.~\ref{fig:tnis_5} 
(from \cite{tnis}). Data from old and new experiments, compared to 
QED calculations of $F_2^{\gamma (QED)}(x,Q^2,P^2=0)$, are presented 
as a function of $x_{Bj}$ for $Q^2$ from 0.1 - 40 GeV$^2$.\\ 
\vspace*{12.3cm}

\subsection{On the structure function of the electron}
In the DIS experiments in $e^+e^-$ collisions 
the inclusive hadron production can be ascribed  not to the 
photon - target but rather to the parent
electron or positron - target. This in some
cases may be more straightforward 
than pinning down the structure of the virtual photon,
as was mentioned before.

This topic is discussed in \eg Ref.~\cite{rev}f,~\cite{mr}. 
See also the results in \cite{ws}, where the structure function of the electron
(in general - lepton) is related not only to the structure function of photon 
but also  to the electroweak gauge bosons $W$ and $Z$.
The {\em structure of weak bosons} which appears in this approach 
was introduced and discussed in Ref.~\cite{ws1}.
%%%%%%%%%%%%%%%%%%%%%%%%%%%%%%%%%%%%%%%%%%%%%%%%%%
%
%
%%%%%%%%%%%%%%%%%%%%%%%%%%%%%%%%%%%%%%%%%%%%%%%%%%
\newpage
\section{Summary and outlook}
The status of recent measurements ($\sim$ 1990 and later) of
the ``structure'' of unpolarized  real photon in 
``DIS$_{\gamma}$'' experiments as well as in large $p_T$ jet
production processes involving resolved photon(s) is presented.
A qualitative change has appeared in both types of measurements.
Final results based on few years' runs at LEP 1 are being published
with the higher  statistics and improved unfolding methods.
The new data from LEP 1.5 and 2 appeared during  last year. 
On the other hand impressive progress has been obtained in 
pinning down the individual parton contributions in large $p_{T}$
resolved photon processes both in $e^+e^-$ and in $ep$ collisions.

The existing discrepancies in describing the final hadronic states in the 
DIS$_{\gamma}$ experiments as well as in the resolved photon processes
in $\gamma \gamma$ and $\gamma p$ collisions seem to have the 
common origin. The need of additional $p_T$
in the distribution of produced hadrons and jets
may suggest an extra interaction involving the constituents of 
photon(s). One of the possible explanations could
 be a multiple interaction described in a similar way as
 introduced in the $p {\bar p}$ processes.   

As far as the data related to the "structure" of the virtual photon 
are concerned, they have just appeared from the hard photon-proton 
and photon-photon collisions with resolved photon(s). The 
corresponding new DIS$_{\gamma^*}$ measurement of the structure 
functions of virtual photon will come probably only with the new 
generation of accelerators. The interesting extension of the idea
to other gauge boson structure functions and the related concept of 
the structure function of the electron might also be tested there.

The new measurements of the partonic content of  the  photon are 
accompanied by the impressive progress  made in the NLO QCD 
calculations for the resolved real and for virtual photon processes.

The future high-energy linear $e^+e^-$ colliders LC as well as
related $e{\gamma}$ and ${\gamma}{\gamma}$ colliders, based on 
the backward Compton scattering on the laser light, will offer 
a unique opportunity to measure the structure of photon in a new 
kinematical regime. Moreover, at these colliders measurements of  
structure functions for photon with a definite polarization should 
become feasible with a good accuracy (see e.g.\cite{NLC}).

\vskip 2cm
\noindent{\large \bf {Acknowledgments.}} %\noindent
\newline
One of us (MK) would like to thank P. Zerwas, G. Kramer,
G. Ingelman, W. Buchm\"{u}ller for  enlightening discussions,
and the whole Theory Group  for the  hospitality 
during her stay at DESY.
She  would like to thank Halina Abramowicz  and Aharon Levy 
 for  figures on $F_2^{\gamma}$ , important discussions and useful comments.
She is also very 
grateful to Stefan S\"{o}ldner-Rembold for  
the compilation of the newest
results on $F_2^{\gamma}$ and to  
A. Caldwell, M. Erdmann and P.J. Bussey
 for discussions on  HERA results.

We express our special thanks to Aharon Levy for his 
encouragement, critical reading of the manuscript and
very important comments.

We want to thank 
Zygmunt Ajduk for his help in preparation of  this review,
 we appreciate also fruitful   
collaboration with Jerzy Rowicki, Jan \.Zochowski and Pawe\l ~Jankowski.
We would like to thank Mike Whalley for sending us
copies of their "Compilation..." \cite{wal}.\newline\newline
Supported in part by Polish Committee for Research; grant number
2P03B18209 (1996-1997).
%%%%%%%%%%%%%%%%%%%%%%%%%%%%%%%%%%%%%%%%%%%%%%%%%%
%
%
%%%%%%%%%%%%%%%%%%%%%%%%%%%%%%%%%%%%%%%%%%%%%%%%%%
\newpage
\section{Appendix}

\subsection{Parton parametrizations for the  real photon}
{\bf{Duke - Owens (DO)}} \cite{do}
\newline
A leading logarithmic parametrization of the parton distributions 
in an asymptotic form. Quarks with equal charges have the same
distribution functions: $f_u = f_c$, $f_d = f_s$ ($N_f=4$).
\newline
{\bf{Drees - Grassie (DG)}} \cite{dg}
\newline
A parametrization for a full solution of the leading order evolution
equations. The input parton distributions with free parameters
assumed at $Q_0^2 = 1$ GeV$^2$ and fitted to the only data on
$F_2^{\gamma}$ existing at that
time, at $Q^2 = 5.9$ GeV$^2$, from PLUTO.
\newline
{\bf{Field - Kapusta - Poggioli (FKP)}} \cite{fkp}
\newline
In this approach $F_2^{\gamma}$ is divided into the hadronic part 
($F_2^{HAD}$) and the point-like one ($F_2^{PL}$). The $F_2^{PL}$ 
arises from the basic ${\gamma}^*\rightarrow q\bar{q}$ coupling and 
higher order QCD corrections, if the final jet $p_T$ is greater 
than $p_T^0$. If $p_T$ is smaller than $p_T^0$ then the $q\bar{q}$ 
pair creates a bound state. $F_2^{HAD}$ in this non-perturbative 
case is taken from the VMD model. The perturbative $F_2^{PL}$ is
calculated using the first order splitting functions and the 
one-loop $\alpha_S$. 
\newline
{\bf{Levy - Abramowicz - Charchu\l a (LAC)}} \cite{lac}
\newline
A parametrization for a full solution of the leading order 
evolution equation fitted to all available in 1991 measurements of 
$F_2^{\gamma}$ for $Q^2\geq Q_0^2$. Three sets are provided with the
different choices of an input scale $Q_0^2$, and the $x\ra 0$ 
behaviour of a gluon distribution $G(x)$, namely:\newline
$\bullet$ LAC1: $Q_0^2 = 4$ GeV$^2$\newline
$\bullet$ LAC2: $Q_0^2 = 4$ GeV$^2$, $xG(x)\ra const.$\newline
$\bullet$ LAC3: $Q_0^2 = 1$ GeV$^2$.\newline 
{\bf{Gl\"{u}ck - Reya - Vogt (GRV)}} \cite{grv}
\newline
The LO and NLO parametrizations of the parton distributions generated 
dynamically from the valence-like VMD input. The low initial scale 
$Q_0^2 = 0.3$ GeV$^2$ is universal for the proton, the pion and the 
photon structure functions. The DIS$_{\gamma}$ scheme is introduced 
to avoid the large-$x_{Bj}$ instability problems. The one free 
parameter, which is a VMD input normalization, is fixed by the data. 
\newline
{\bf{Gordon - Storrow (GS)}} \cite{gs}\newline
The LO and NLO parametrizations. The input structure function at 
scale $Q_0^2 = 5.3$ GeV$^2$ \cite{gs}a and $Q_0^2 = 3$ GeV$^2$ 
\cite{gs}b in the LO analysis is chosen as a sum of a hadronic part 
from the VMD model and of a point-like part based on the Parton Model. 
Free parameters (also light quarks masses) are fitted to the data for 
$Q^2\geq Q_0^2$. The NLO distributions in the $\overline{\rm MS}$ are 
obtained by matching of the $F_2$ in the LO and the NLO approaches 
at the $Q_0^2$ scale.
\newline
{\bf{Aurenche - Chiapetta - Fontannaz - Guillet - Pilon (ACFGP)}} \cite{agf}a
\newline
A solution of the NLO evolution equation with the boundary condition
taken at $Q_0^2$ = 0.25~GeV$^2$. The input parton distributions was 
obtained from the VMD model at $Q^2$ = 2~GeV$^2$ and evaluated down 
to $Q_0^2$.
\newline
{\bf{Aurenche - Guillet - Fontannaz}} \cite{agf}b
\newline
The NLO parton distributions obtained with the input distributions 
(shown to be scheme-dependent) at $Q_0^2 = 0.5$ GeV$^2$. The input 
distributions are based on the VMD model modified to agree with the 
$\overline{\rm MS}$ scheme used in this analysis. 
\newline
{\bf{Watanabe - Hagiwara - Tanaka - Izubuchi (WHIT)}} 
\cite{whit}\newline
A set of six the LO parametrizations obtained by fitting the input 
distributions to all available data for $4$ GeV$^2\leq Q^2\leq 100$ 
GeV$^2$. The parametrizations WHIT1-WHIT6 are based on different 
input gluon distributions. A massive charm contribution is calculated 
from the quark parton model and for $Q^2>100$ GeV$^2$ from the 
massive quark evolution equations. 
\newline
{\bf{Schuler - Sj\"{o}strand (SaS)}} 
\cite{sas}\newline
Four sets of the LO parametrizations. The non-perturbative input 
distributions at $Q_0 = 0.6$ GeV in the SaS1D, SaS1M and $Q_0 = 2$ 
GeV in the SaS2D, SaS2M sets are based on the VMD model (their 
normalization is fixed and the $x$ - dependence is obtained from 
the fits to the data). The fully calculable point-like contribution 
to the $F_2^{\gamma}$ is expressed as an integral of the ``state'' 
distributions over the virtuality $k^2$ of the 
$\gamma^*\rightarrow q\bar{q}$ state.\newline 
The non leading term $C_{\gamma}$ is included into the $F_2^{\gamma}$, 
leading to the $\overline{\rm MS}$ distributions (SaS1M, SaS2M). 
$C_{\gamma} = 0$ gives distributions in the DIS scheme (SaS1D, SaS2D). 

\subsection{Parton parametrizations for the virtual photon}
The $Q^2$-dependence of the parton distributions of the virtual 
photon ($P^2\neq 0$) for $\Lambda^2\ll P^2\ll Q^2$ follows 
from the corresponding evolution equations \cite{uw}, as for
the real photon ($P^2=0$) case.
Experimentally important is, however, the low-$P^2$ region 
$\Lambda^2\lsim P^2$. The parametrizations \cite{rev}f,~\cite{grs,ss},
valid for $0\leq P^2$, are constructed in such a way
that they reproduce the distributions of the real photon in the
limit $P^2\rightarrow 0$ and obey the exact $Q^2$-evolution 
equation in the region $\Lambda^2\ll P^2$.
\newline
{\bf{Schuler - Sj\"{o}strand}} \cite{ss}\newline
An extention of the SaS parton distributions in the real photon
to the virtual photon case. In the point-like contribution, the
integral over the virtuality of $\gamma^*\rightarrow q\bar{q}$
state, $k^2$, is modified by a factor $({k^2\over k^2+P^2})^2$. In the
hadronic contribution a factor $({m_V^2\over m_V^2+P^2})^2$
is introduced, where $m_V$ is a vector-meson mass.\newline
%{\bf{Ioffe - Oganesian}} \cite{io}\newline
{\bf{Drees - Godbole}} \cite{rev}f\newline
Sets of the parton distributions obtained from the corresponding
distributions for the real photon (DG) by including multiplicative 
factors to include properly the $e\rightarrow e\gamma^*$ vertex. 
\newline
{\bf{Gl\"{u}ck - Reya - Stratmann (GRS)}} \cite{grs}\newline
The LO and NLO distributions obtained by solving the $Q^2$-evolution 
equation with the boundary conditions being a smooth interpolation
betwen the boundary conditions valid at $P^2=0$ and for 
$P^2\gg\Lambda^2$. The applicalibility is assumed for the ranges:
$P^2\leq 10$ GeV$^2$, 10$^{-4}\leq x \leq 1$, 0.6$\leq Q^2\leq 
5\cdot 10^4$ GeV$^2$ {\underline {and}} $Q^2\gsim 5 P^2$.
\newline
%%%%%%%%%%%%%%%%%%%%%%%%%%%%%%%%%%%%%%%%%%%%%%%%%%
%
%
%%%%%%%%%%%%%%%%%%%%%%%%%%%%%%%%%%%%%%%%%%%%%%%%%%

\end{document}